\documentclass[12pt,preprint]{aastex}

\newcommand{\msun}{{~{\rm M}_\odot}}

\newcommand{\etal}{{et~al.}~}
\newcommand{\bgc}{B$_{\rm{gc,red}}$}

\shorttitle{X-ray Observations of High Redshift Clusters}
\shortauthors{Hicks et al.}

\begin{document}

\title{{\it{Chandra}} X-ray Observations of the $0.6<z<1.1$ Red-Sequence Cluster Survey Sample}

\author{A.K. Hicks\email{ahicks@alum.mit.edu}
\affil{Department of Astronomy, University of Virginia, P.O. Box 400325,
Charlottesville, VA 22904}}

\author{E. Ellingson\email{elling@casa.colorado.edu} 
\affil{Center for Astrophysics and Space Astronomy, University of Colorado at Boulder, Campus Box 389, Boulder, CO 80309}}

\author{M. Bautz\email{mwb@space.mit.edu}}
\affil{MIT Kavli Institute for Astrophysics and Space Research, 77 Massachusetts Ave., Cambridge, MA 02139, USA}

\author{B. Cain\email{bcain@mit.edu}}
\affil{MIT Kavli Institute for Astrophysics and Space Research, 77 Massachusetts Ave., Cambridge, MA 02139, USA}

\author{D.G. Gilbank\email{dgilbank@astro.uwaterloo.ca}}
\affil{Department of Astronomy and Astrophysics, University of Toronto, 50 St. George St., Toronto, ON, M5S 3H4, Canada}
%\affil{Department of Physics and Astronomy, University of Waterloo, 200 University Avenue West, Waterloo, Ontario, Canada N2L 3G1}

\author{M.G. Gladders\email{gladders@uchicago.edu}}
\affil{Department of Astonomy and Astrophysics, University of Chicago, 5640 S. Ellis Ave, Chicago, IL 60637, USA}

\author{H. Hoekstra\email{hoekstra@uvic.ca}}
\affil{Alfred P. Sloan Research Fellow, Department of Physics \& Astronomy, University of Victoria, Elliott Building, 3800 Finnerty Rd, Victoria, BC, V8P 5C2}

\author{H.K.C. Yee\email{hyee@astro.utoronto.ca}}
\affil{Department of Astronomy and Astrophysics, University of Toronto, 50 St. George St., Toronto, ON, M5S 3H4, Canada}

\and

\author{G. Garmire\email{garmire@astro.psu.edu}}
\affil{Department of Astronomy and Astrophysics, 525 Davey Lab, The Pennsylvania State University, University Park, PA, 16802, USA}

\begin{abstract}

We present the results of Chandra observations of 13 optically-selected clusters with $0.6 < z < 1.1$, discovered via the Red-sequence Cluster Survey (RCS). All but one are detected at S/N$>3$; though three were not observed long enough to support detailed analysis. Surface brightness profiles are fit to $\beta$-models. Integrated spectra are extracted within $\rm{R}_{2500}$, and $T_X$ and $L_X$ information is obtained. We derive gas masses and total masses within $\rm{R}_{2500}$ and $\rm{R}_{500}$. Cosmologically corrected scaling relations are investigated, and we find the RCS clusters to be consistent with self-similar scaling expectations. However, discrepancies exist between the RCS sample and lower-$z$ X-ray selected samples for relationships involving $L_X$, with the higher-$z$ RCS clusters having lower $L_X$ for a given $T_X$. In addition, we find that gas mass fractions within $\rm{R}_{2500}$ for the high-$z$ RCS sample are lower than expected by a factor of $\sim2$. This suggests that the central entropy of these high-$z$ objects has been elevated by processes such as pre-heating, mergers, and/or AGN outbursts, that their gas is still infalling, or that they contain comparatively more baryonic matter in the form of stars. Finally, relationships between red-sequence optical richness (\bgc) and X-ray properties are fit to the data. For systems with measured $T_X$, we find that optical richness correlates with both $T_X$ and mass, having a scatter of $\sim30\%$ with mass for both X-ray and optically-selected clusters. However, we also find that X-ray luminosity is not well correlated with richness, and that several of our sample appear to be significantly X-ray faint.
\end{abstract}


\section{Introduction \label{s:intro}}

The Extended Medium Sensitivity Survey~\citep[EMSS]{gioia90} sparked renewed interest in the ongoing search for clusters of galaxies at high redshift.  
Since then, numerous high redshift surveys have been carried out, both optically~\citep[and others]{gilbank04,donahue02,postman,bower}, and in the X-ray~\citep[and others]{valtchanov,bauer,wilkes}.  
The motivations for such searches are multifaceted, but the most compelling of these are cosmological in nature.

Clusters of galaxies are an important source of information about the underlying cosmology of the universe.  
They are considered to be essentially ``closed boxes'', meaning that the primordial matter that they were initially assembled from has remained trapped in their deep potential wells since they were formed.  
This makes them ideal objects with which to study galaxy formation and evolution.  
In addition, clusters are the largest virialized objects in the universe.  By virtue of this fact we are able to, through high redshift samples, investigate the growth of large scale structure.  
A firm knowledge of the evolution of the cluster mass function would provide an enormous contribution to constraining cosmological parameters such as $\sigma_8$ (the normalization of the density perturbation spectrum) and $w$~\citep[the dark energy equation of state; e.g.,][]{voit05}.

Two ingredients are required to achieve this goal.  
First, a statistically significant sample of clusters in multiple redshift bins is needed.  
Second, reliable mass estimates of the clusters in that sample must be obtained.  
Difficulties in reaching the first requirement include the vast amount of telescope time required to carry out such a search in the X-ray, and the propensity for false detections due to projection effects in optical surveys.  
The primary challenge in reaching the second part of this goal is again the high cost of observing time to achieve either X-ray or dynamical mass estimates.

The Red-sequence Cluster Survey~\citep[RCS;][]{gladders00,gladders05,yee07} has attempted to evade such difficulties -- RCS is an optical survey which uses the color-magnitude relationship of cluster ellipticals to find galactic overdensities in small slices of redshift space.  
This technique has been estimated to bring false detection rates down to $\sim 5-10\%$~\citep{gladdersphd,blindert07,cohn07}.  
The chosen filters ($R_c$ and $z'$), optimize this finding algorithm for the redshift range $0.2<z<1.2$ and provide photometric redshift information with accuracies of $\sim 10 \%$.  
In addition, optical richness information is immediately available from the survey data and, if sufficiently calibrated, this information could provide a highly efficient way to estimate the masses of cluster candidates.

The first phase of the Red-Sequence Cluster Survey~\citep[RCS-1;][]{gladders05}, from which our cluster sample was drawn, covers 90 square degrees and was performed at CFHT and CTIO.  
RCS-1 has identified 6483 cluster candidates in the redshift range $0.2 <z< 1.2$, over 1000 of which are at least as optically rich as Abell class 0 clusters~\citep{gladders05}.  

The motivations for this work are to characterize high-redshift optically-selected cluster samples, probe cluster evolution, and move forward in attempts to calibrate a robust relationship between optical richness and cluster mass.  
This paper presents a detailed analysis of the {\it{Chandra}} data of thirteen RCS-1 clusters with redshifts in the range $0.6<z<1.1$  
Our analysis investigates the temperatures and gas distributions of ten of these clusters, and provides mass estimates for use in the calibration of relationships between optical richness and cluster mass.  

We also use our results to investigate the X-ray scaling laws of our sample, and thereby probe redshift evolution in these relationships.  
To facilitate comparisons between the RCS clusters and lower-redshift X-ray selected samples, we make use of our previous {\it{Chandra}} analysis of the Canadian Network for Observational Cosmology (CNOC) subsample of the Extended Medium Sensitivity Survey (EMSS)~\citep{hickscnoc,yee96,gioia90}.  
This sample, with redshifts in the range $0.1<z<0.6$, was chosen based on X-ray luminosity \citep[$L_X\geq2\times10^{44}$ erg s$^{-1}$;][]{gioia90}.

This paper is organized as follows:  In Sections~\ref{s:obs} and~\ref{s:sig}, we introduce our sample and describe the basic properties of our data.  In Sections~\ref{s:sb2} and~\ref{s:spec}, we investigate gas distributions and obtain cluster temperatures.  We derive masses for our sample in Section~\ref{s:mass}.  High-$z$ X-ray scaling relationships are examined in Section~\ref{s:laws2}, while correlations between optical richness and cluster X-ray properties are explored in Section~\ref{s:or2}.  In Section~\ref{s:selection} we investigate possible sources of bias in cluster sample selection.  A summary and discussion of our results is presented in Section~\ref{s:sum}.
Unless otherwise noted, this paper assumes a cosmology of $\rm{H}_0=70~\rm{km}~\rm{s}^{-1}~\rm{Mpc}^{-1}$, $\Omega_{\rm{M}}=0.3$, and $\Omega_{\Lambda}=0.7$.  All errors are quoted at 68\% confidence levels.

\section{Cluster Sample \& Observations \label{s:obs}}  

{\it{Chandra}} Advanced CCD Imaging Spectrometer (ACIS) observations of thirteen RCS clusters were taken during the period 10 April 2002 - 23 October 2005.  
Twelve of these clusters were observed with the ACIS-S CCD array, and one was observed with ACIS-I, with an overall range in individual exposures of $10 - 100$ kiloseconds.  
Seven of the clusters in this sample were observed on multiple occasions.  
All multiple observations were merged for imaging analysis to provide higher overall signal-to-noise ratios.  
Those clusters with $\Delta t_{\rm{obs}} \leq 3$ months between observations were merged for spectral analysis as well.  
Each of the observations analyzed in this study possesses a focal plane temperature of $-120^o$ C.  

Aspect solutions were examined for irregularities and none was found.  
Background contamination due to charged particle flares was reduced by removing time intervals during which the background rate exceeded the average background rate by more than $20\%$.  
The quiescent background was further reduced by using VFAINT mode.  
The event files were then filtered on standard grades and bad pixels were removed.  
Table~\ref{table1} provides a list of each of the clusters in our sample, including their precise designation (hereafter shortened for simplicity), redshift, obsid, and corrected exposure information for each observation. 

After the initial cleaning of each data set, 0.3-7.0 keV images, instrument maps, and exposure maps were created using the CIAO 3.3.0.1 tool MERGE\_ALL.  
Data with energies below 0.3 keV and above 7.0 keV were excluded due to uncertainties in the ACIS calibration and background contamination, respectively.  
Ideally, all data below 0.6 keV would have been excluded to minimize low energy uncertainties; however, the combined faintness and high redshifts of our objects require the utilization of lower energy photons as well.  
Point source detection was performed by running the tools WTRANSFORM and WRECON on the flux images.  

Figure~\ref{fig1} contains smoothed 0.3-7.0 keV {\it{Chandra}} flux images of each of the clusters in our sample (produced by the CIAO tool CSMOOTH), including a combined image of the three objects at $z\sim0.9$ which belong to a supercluster in the 23h field~\citep{gilbankprep}.  
As seen in the figure, this sample covers a wide range of cluster X-ray morphology, from very regular objects (e.g., RCS1419+5326), through well-detected clusters with significant substructure (e.g., RCS2318+0034), all the way to very disturbed systems (e.g., RCS2112-6326).  
It is worth noting that in Figure~\ref{fig1}, the brightest part of RCS2318+0034 does not seem to lie at the center of the cluster, indicating that this object may have recently undergone a merger, or could at least possess an appreciable amount of substructure.  
Together, these clusters represent an assembly of some of the richest high-$z$ ($0.6<z<1.1$) clusters in the RCS-1 survey.

\section{Signal-to-Noise Ratios and Cluster Positions\label{s:sig}}

To estimate the significance of RCS cluster detections in the X-ray, we made use of relatively simple statistics.  
Counts were extracted from a 500 $h_{70}^{-1}$ kpc radius region around the aimpoint of each observation in the 0.3-7.0 keV band ($C$), and also from a region far away from the aimpoint on the same chip which served as a background ($B$).  
Obvious point sources were removed from each region.  
Signal-to-noise ratios were calculated based on dividing net counts, $N=C-B$, by the standard deviation, $\sigma=\sqrt{C+B}$.  
Using this method, twelve cluster signals were detected at a signal-to-noise ratio greater than $3$, with the remaining object detected at S/N$=1.1$ (Table~\ref{table2}).

Using adaptively smoothed 0.3-7.0 keV flux images (Figure~\ref{fig1}), we determined the location of the X-ray emission peak of each cluster.  The images of RCS2112-6326 and RCS2156-0448 appear to contain multiple regions of extended emission, therefore we cannot determine a precise X-ray position for these objects.  
In the case of RCS1326+2903, two RCS 13h clusters lie in the field of view.  
The original observation was designed to observe a $z=1.01$ cluster at an RA, Dec of 13:26:29, +29:03:06 (J2000).  
Our astrometry indicates, however, that we are most likely detecting the emission of a lower redshift RCS cluster ($z=0.75$) at an optical position of 13:26:31, +29:03:12.  
Because of the uncertainty surrounding this detection, we have carried both possibilities throughout much of our analysis; however, we include the more likely candidate (at $z=0.75$) in our subsequent fitting and plots.  
All other clusters (with the exceptions of RCS2112-6326 and RCS2156-0448) were found within $31\arcsec$ of their optical positions.  
Table~\ref{table2} lists optical positions, X-ray positions, net counts within 500 $h_{70}^{-1}$ kpc, and signal-to-noise ratios derived from the method described above.

\section{Surface Brightness\label{s:sb2}}

A radial surface brightness profile was computed over the range
0.3-7.0 keV in circular annuli for each cluster.  These profiles
were then fit with $\beta$ models:

\begin{equation}
I(r) = I_B + I_0 \left( 1 + {r^2 \over r_c^2} \right)^{-3\beta+\frac{1}{2}}
\label{sb_eq}
\end{equation} 
where $I_B$ is a constant representing the surface brightness contribution of the background, $I_{0}$ is the normalization and $\rm{r}_{\rm{c}}$ is the core radius.  
The parameters of the best fitting models of the ten clusters for which surface brightness fitting was possible are shown in Table \ref{table3}, and images of these fits are given in Figure~\ref{fig2}.  
Though many of the clusters exhibit hints of substructure, most were reasonably well fit by a $\beta$ model (see Table~\ref{table3} for goodness of fit data).  
Other than somewhat low normalizations, the results of surface brightness fitting are unremarkable ($0.51<\beta<0.72$), except in the two cases of RCS1326+2903 ($\beta=1.04$), which lies at the edge of our detection threshold (Table~\ref{table2}), and RCS2318+0034 which appears to not be completely relaxed, as its brightest emission is slightly offset from the center of its extended emission (Figure~\ref{fig1}).

\section{Spectral Analysis\label{s:spec}}

\subsection{Integrated Spectral Fits and {$\rm{R}_{2500}$}\label{s:integrated}}

Spectra were extracted from each point-source-removed event file in a circular region with
a $300~\rm{h}_{70}^{-1}~\rm{kpc}$ radius.  
In the cases of RCS0224-0002 and RCS1419+5326 ($\Delta t_{\rm{obs}}>2$ yr), individual spectra were extracted from each obsid and fit simultaneously.  
The spectra were analyzed with XSPEC~\citep{arnaud96}, using weighted response matrices (RMFs) and effective area files (ARFs) generated with the CIAO tool SPECEXTRACT and CALDB 3.2.2. 
Background spectra were extracted from the aimpoint chip as far away from the aimpoint as possible. 

Spectra were fitted with single temperature spectral models, inclusive of
foreground absorption.  
Each spectrum was fit with the absorbing column frozen at its measured value~\citep{dickey90}.  
Metal abundances were initially fixed at a value of 0.3 solar~\citep{edge}.  Data with energies below 0.3 keV and above 7.0 keV were excluded from the fits.  

Three of the clusters did not possess enough counts to constrain a spectral fit.  
The results of the ten successful fits, combined with best fitting $\beta$ model parameters from Section~\ref{s:sb2}, were then used to estimate the value of $\rm{R}_{2500}$ for each cluster.
This is accomplished by combining the equation for total gravitating mass~\citep{sarazin88}

\begin{equation}
M_{tot}(<r) = -{{kT(r)r}\over{G \mu m_p}} \left({{\partial~\rm{ln}~\rho}\over{\partial~\rm{ln}~r}} + {{\partial~\rm{ln}~T} \over {\partial~\rm{ln}~r}} \right), 
\label{Mass_eq}
\end{equation}
 
\noindent
(where $\mu m_p$ is the mean mass per particle) with the definition of mass overdensity

\begin{equation}
M_{tot}(r_\Delta) = {{4}\over{3}}\pi\rho_c(z) r^3_\Delta \Delta ,
\end{equation}

\noindent
where $z$ is the cluster redshift, and $\Delta$ is the factor by which the density at $r_\Delta$ exceeds $\rho_c(z)$, the critical density at $z$. 
Here $\rho_c(z)$ is given by $\rho_c(z) = {3{H(z)^{2}/{8 \pi G}}} = {3 {H_0^2 E_z^2}/{8 \pi G}}$, where $E_z=[\Omega_m(1+z)^3 + \Omega_{\Lambda}]^{1/2}$.  
These equations are then combined with the density profile implied from the $\beta$ model (assuming hydrostatic equilibrium, spherical symmetry, and isothermality)

\begin{equation}
\rho_{gas}(r) = \rho_0 \left[1 + {{r^2}\over{r_c^2}}\right]^{-3\beta/2},
\label{eq:dens_eq}
\end{equation}

\noindent
resulting in the equation

\begin{equation}
{{r_\Delta}\over{r_c}} = \sqrt{{\left[{{3\beta k T}\over{G \mu m_p (4/3) \pi \rho_c(z) r_c^2 \Delta}}\right]}-1},
\label{eq:ettori}
\end{equation}

\noindent
\citep{ettori00,ettori04b}.

After the initial estimation of $R_{2500}$, additional spectra were extracted from within that radius, and spectral fitting was performed again.  
This procedure was repeated until temperatures and values of $R_{2500}$ were consistent for a given spectrum.  
Where statistically possible, additional fits were performed allowing abundances to vary also.  
Redshifts were also fit (allowing only $z$, $T_X$, and normalization to vary) for the three clusters in our sample that do not have spectroscopic redshifts.  
We were unable to constrain a redshift for RCS1326+2903.  
The fits of the other two clusters resulted in $z=0.62\pm{0.01}$ for RCS1419+5326, and $z=0.78^{+0.07}_{-0.08}$ for RCS2318+0034, within 10\% and 14\% (respectively) of photometric redshift estimates obtained using the color of the red sequence~\citep{gladders05}.  
Fits with redshift fixed at these values were used in subsequent analysis (Table~\ref{table4}).  The small uncertainties in these values ($\leq10\%$) do not substantially affect our analysis.  

Unabsorbed 2-10 keV luminosities within R$_{2500}$ were calculated using fixed abundance fits.  
These were then converted to bolometric luminosities by scaling, using a thermal emission model in PIMMS.  
For the three clusters for which spectral temperature fitting was impossible (RCS1417+5305, RCS2112-6326, and RCS2157-0448), spectra were extracted within $500~\rm{h}_{70}^{-1}~\rm{kpc}$ radii, and fit in XSPEC with temperatures fixed at 4 keV (slightly lower than the average Tx of the sample) to determine their luminosities.  Temperature uncertainties of $\pm2$ keV were folded into the errors of these estimates.
To estimate $L_X(\Delta={500}$), we extracted spectra from within that radius for the ten clusters in our detailed analysis sample and again used fixed abundance fits, with temperatures also fixed at the R$_{2500}$ value.  
The results of spectral fitting are shown in Table~\ref{table4}, along with 68\% confidence ranges.  Bolometric X-ray luminosities are listed with richness measurements in Table~\ref{table5}.

\subsection{$T_X$-$\sigma$ Comparisons}

Velocity dispersions for three of the clusters in this sample were obtained from~\citet{gilbank07} and~\citet{gilbankprep}, and are listed in Table~\ref{table6}.  
Using the $\sigma-{\rm{T}}_X$ relationship of~\citet{xue00}: $\sigma=10^{2.49} ~T_X^{0.65}$, we find that our temperatures are in agreement with the clusters' measured velocity dispersions in all cases (Table~\ref{table6}).  
This result indicates that these three systems, at least, are not overly disturbed.

\section{Mass Estimates\label{s:mass}}

An isothermal cluster whose surface brightness is well fit by a $\beta$ model can be shown to have a gas density profile which follows Equation~\ref{eq:dens_eq}.  
Using this relationship and the equation of hydrostatic equilibrium (Equation~\ref{Mass_eq}), total mass can be determined via

\begin{equation}
M_{tot}(<r) = {{3\beta}\over{G}}~{{k T r}\over{\mu m_p}}~{{{(r/r_c)}^2}\over{{1+{{(r/r_c)}^2}}}}.
\label{Mass_eq2}
\end{equation}

To estimate gas mass (again assuming hydrostatic equilibrum, isothermality and sphericity), the first step is to obtain a central density ($n_0\equiv\rho_0/m_p$).  
There are two complementary ways to go about this.  One is to use the surface brightness normalization:

\begin{equation}
n_0=\left[{{\Gamma(3\beta)}\over{\pi^{1/2}{\Gamma(3\beta-1/2)}}}\left({{\mu_e}\over{X_H\epsilon_0}}\right)\left({{I_0}\over{r_c}(1+z)^4}\right)\right]^{1/2},
\end{equation}

\noindent where the $\Gamma$ function results from surface brightness integration, $\beta$ comes from the fit to surface brightness, $\mu_e$ is the mean atomic mass per free electron (0.62), $X_H$ is the hydrogen mass fraction (0.707), $\epsilon_0$ is the gas emissivity, $I_0$ is the best fitting surface brightness normalization (corrected for absorption), and $r_c$ is the core radius.

A second method of estimating central density makes use of both imaging and spectral fitting:

\begin{equation}
n_0^2={{4\pi d_{ang}^2~(1+z)^2~K~10^{14}}\over{0.82~4 \pi r_c^3~EI}}~{\rm{cm}}^{-6}.
\end{equation}

\noindent Here K is the normalization of the XSPEC model and $EI$ is the emission integral, estimated by integrating the (spherical) emission from the source out to some radius - in our case we use 10 Mpc following the method of~\citet{ettori03}.  

For the RCS sample we employed both of these methods, as it was crucial to confirm that we were not underestimating central density in these comparatively low luminosity objects.  
We also added the data from our previous {\it{Chandra}} analysis of the moderate redshift CNOC sample, to cover a wider range of redshifts in our comparison.  
We found that the methods agree (on average) to within 10\%, and proceeded in our analysis using the surface brightness normalization method. 

From these equations, along with Equations~\ref{eq:dens_eq} and~\ref{eq:ettori}, and using the results of spectral and surface brightness fitting, gas masses and total masses were determined out to $\rm{R}_{2500}$ and $\rm{R}_{500}$ for the clusters in this sample.  We also calculate gas mass fractions for the RCS clusters and find them to be systematically lower than the gas fractions of lower redshift X-ray selected clusters.  The robustness and implications of this result are explored in detail in Section~\ref{fgas}.  

Gas masses, total masses, and gas mass fractions can be found in Tables~\ref{table7} and~\ref{table8}.  We note that while extrapolations to larger radii are possible using out measured $\beta$-fit parameters, R$_{2500}$ is the radius to which we have confident measures for our entire sample.

\section{Cluster Scaling Relations\label{s:laws2}}

Studying the relationships between global cluster properties (L$_X$, T$_X$, M$_{\rm{tot}}$, etc.) over a broad range in redshift allows us to investigate the influence of non-gravitational processes on cluster formation and evolution.  
On a less grand scale, these relationships can also lead to interesting clues regarding an individual cluster's dynamical state and composition, as well as provide a method of comparison between different cluster samples. 
In this paper we investigate the evolution of scaling relationships over the redshift range $0.1<z<1.0$, and use them to characterize high-$z$ optically-selected RCS clusters ($0.6<z<1.0$)  

To facilitate comparisons between the RCS clusters and lower-redshift X-ray selected samples, we make use of our previous {\it{Chandra}} analysis of the Canadian Network for Observational Cosmology (CNOC) subsample of the Extended Medium Sensitivity Survey (EMSS)~\citep{hickscnoc,yee96,gioia90}.  
The CNOC sample was assembled primarily based on X-ray luminosity, with a cut at $2\times10^{44}$ erg s$^{-1}$ in the original EMSS catalogs~\citep{gioia90}, and covers a redshift range of $0.1<z<0.6$.

This sample is  not well-matched in redshift to our RCS clusters, but  it is one of the best-studied moderate-redshift cluster samples today, with substantial information about both X-ray and optical properties available. Our previous analysis of this sample using the same methodology~\citep{hickscnoc} also allows us to make a confident comparison of our measurements.

All relationships ($L_X$-$T_X$, $L_X-$M$_{\rm{tot}}$, M$_{\rm{tot}}-T_X$, $L_X$-$Y_X$, and M$_{\rm{tot}}-Y_X$) are fit within either R$_{2500}$ or R$_{500}$, have been scaled by the cosmological factor $E_z={H(z)}/{H_0} = \left[{\Omega_m (1+z)^3 + \Omega_\Lambda} \right]^{1/2}$ and are fit with the form

\begin{equation}
{\rm{log}_{10}}~{\rm{Y}}=C1+C2~{\rm{log}_{10}}~{\rm{X}}.
\end{equation}

\noindent In all relationships $T_X$ is in units of 5 keV, $L_X$ in units of $10^{44}$ erg s$^{-1}$, total mass in $10^{14} \msun$, and $Y_X$($\equiv{\rm{M}}_gT_X$) in $4\times10^{13} \msun$ keV.  
Best fitting relationships are determined using the bisector modification of the BCES algorithm in~\citet{akritas}, and we calculate scatter along the Y-axis as $\left[\Sigma_{i=1,N}\left({\rm{log_{10}~Y}}_i-C_1-C_2~{\rm{log_{10}~X}}_i\right)^2/N\right]^{1/2}$, facilitating comparisons to previous work~\citep[e.g.,][]{ettori04}.

In all fits the cluster RCS0439-2904 was left out, due to spectroscopic indications~\citep{gilbank07,cain} that it does not consist of a single virialized mass but two closely spaced objects in projection along the line of sight.  
We perform fits at $\Delta=2500$ on the individual samples; RCS ($0.62<z<0.91$) and CNOC ($0.17<z<0.55$), as well as combined data from all 23 clusters ($0.17<z<0.91$).  
At $\Delta=500$ only the RCS data are fit.  
All fits are then reproduced with the slope fixed at the expected self-similar value.  
The following discussions pertain to fits with two free parameters unless otherwise noted.  
Results from the fitting performed in this section can be found in Table~\ref{table9}, while Table~\ref{table10} provides comparison fits from the literature.

\subsection{The $L_X-T_X$ Relationship\label{s:lxtx}}

In the absence of significant preheating and/or cooling, theory predicts that cluster luminosities should scale as $L_{bol} \propto T^2$.  
However, observational studies have resulted in relationships which fall closer to $L_{bol} \propto T^3$~\citep{white,allen98,markevitch,arnaud99}.  
These departures from theoretically expected self-similar scaling laws indicate the effects of non-gravitational processes, such as galaxy formation~\citep{voit}.  
There is also interest in whether the $L_X$-$T_X$ relationship evolves with redshift~\citep{ettori04}, which we investigate in this section along with the properties of our sample.  

The best fitting relationships and their scatter are given in Table~\ref{table9} and are plotted in Figure~\ref{fig3}.
At both radii (R$_{2500}$ and R$_{500}$), the slope of the RCS fit is found to be consistent with a predicted self-similar slope of 2.  
The CNOC sample, with a slope of $2.31\pm{0.31}$, is only marginally consistent with predicted scaling, but does agree with other low-redshift $L_X$-$T_X$ relationships.  
The main difference between the RCS and CNOC fits, however, is their normalization, which is significantly lower in the case of the RCS fit, translating into $2.3\pm{0.3}\times10^{44}$ erg s$^{-1}$ at 5 keV, compared to $5.5^{+1.1}_{-0.9}\times10^{44}$ erg s$^{-1}$ for CNOC.  
Fits to the combined sample have significantly higher scatter and a much larger slope ($2.90\pm{0.35}$), inconsistent with self-similar evolution.  
This is an interesting result - taken with the individual fits it suggests that redshift evolution in the normalization of $L_X$-$T_X$ could be perceived as evolution in its slope, if the fitted sample covered a broad enough range of redshifts.

This speculation naturally leads us to the important question of whether this trend towards lower luminosity is due to the different selection of the RCS clusters, or to a general evolutionary trend with redshift.  In Table~\ref{table10} we list our fit parameters along with others taken from the literature.   Since a number of these studies use measurements at R$_{500}$, we compare our fits using estimates extrapolated to this radius.
The results of~\citet{allen01}, from a selected sample of $z \sim$ 1 clusters, are consistent in slope with our individual sample fits, but even higher in normalization than the CNOC fit.  
This is perhaps not surprising given that the clusters in their sample are relaxed lensing clusters, many of which have strong cooling cores which can significantly increase the central cluster luminosity. In general, we do not here have enough information to excise cooling cores from the RCS data; however,, in the CNOC data, we did attempt to remove these features from the cluster temperatures and luminosities \citep{hickscnoc}. Discrepancies between CNOC and RCS are thus not likely to be due to a higher incidence of cooling cores in the lower redshift sample.

We also compare our sample to the $0.4 < z < 1.3$  X-ray selected sample  in \citet{ettori04}, which consists of 28 clusters taken from the Chandra archive.
At $\Delta=500$, they find that the slope of the $L_X$-$T_X$ relationship is much steeper than that predicted by self-similar scaling~(slope=3.72 $\pm$ 0.47;  Figure~\ref{fig3}), suggesting a negative redshift evolution in the relationship (i.e., clusters at high-$z$ have lower $L_X$ for a given $T_X$). A similar result is suggested by \citet{ettori04b}, based on simulations guided in part by low-redshift observations.

Figure~\ref{fig3} shows the extrapolation of our data to R$_{500}$. 
Seven of our nine objects lie on the~\citet{ettori04} relationship, suggesting at least some agreement between the properties of their X-ray and our optically-selected samples.
We note that in general their higher redshift clusters also trend towards lower luminosities.
While the slope of our fit to the RCS sample at $\Delta=500$ is inconsistent with theirs, the slope of our combined $\Delta=2500$ sample is in agreement with their slope, another indication that we may be resolving evolution in slope into changes in the normalization of the relationship with redshift.  
Our scatter for the individual fits is significantly lower than theirs ($\sigma_{{\rm{log Y}}}\leq0.20$ vs. $\sigma_{{\rm{log Y}}}=0.35$), whereas our scatter for the combined fits becomes more comparable (0.28). 
Thus, the RCS high-redshift sample appears to be at least qualitatively similar to this high-redshift X-ray selected sample, in support of a trend for samples of clusters at high redshift to have have lower luminosities at a given temperature.

\subsection{The $L_X-$M$_{\rm{tot}}$ Relationship\label{s:lxmass}}

Upon examining the cosmologically corrected $L_X-$M$_{\rm{tot}}$ relationship, we again see disparity between the normalizations of the CNOC and RCS fits.  
This finding provides additional evidence that there is less gas for a given total mass in our high redshift sample.  
Individual slopes at R$_{2500}$ agree with the self-similar value of 1.33, while the slope of the combined sample fit is higher and inconsistent with that value (Table~\ref{table9}; Figure~\ref{fig4}).  
Our RCS fit at R$_{500}$ is again consistent in normalization but not slope with~\citet{ettori04}.  
Likewise, again our combined (R$_{2500}$) sample slope ($1.77\pm{0.15}$) agrees well with theirs ($1.88\pm{0.42}$).  Our scatter ($0.16\leq\sigma_{{\rm{log Y}}}\leq0.33$) is lower in all cases.

\subsection{The M$_{\rm{tot}}-T_X$ Relationship\label{s:masstx}}

The M$_{\rm{tot}}-T_X$ relationship is by far the lowest-scatter ($\sigma_{{\rm{log Y}}}\leq0.10$) relationship in this work (Figure~\ref{fig5}), though this is largely because of the degeneracy between the two parameters, with much of the scatter arising from differences in the spatial distribution of gas.  
All of our fits at R$_{2500}$ have consistent normalizations, and all but the CNOC fit have slopes which agree well with self-similar predictions.  
Though the CNOC slope is higher ($1.83\pm{0.13}$), it is in good agreement with both of our listed R$_{2500}$ comparison fits~\citep{allen01,arnaud05}, as are the fits of both the RCS sample and the combined sample (Table~\ref{table10}).  

At R$_{500}$ the RCS fit is in agreement with~\citet{finoguenov01},~\citet{arnaud05}, and~\citet{kotov05}.  
Our normalization is somewhat higher, however, than all of theirs, and is in disagreement with that of~\citet{ettori04} (Figure~\ref{fig5}).  
The three objects that are most responsible for driving up the normalization all have gas distributions which appear to be more concentrated than average ($\beta \geq 0.72$; Table~\ref{table3}), which would tend to drive up the total mass at higher radius, noting again that at R$_{500}$ our masses are extrapolations. 
In addition, the two most outlying points consist of our least massive cluster (RCS1326+2903, also our weakest detection in the detailed analysis sample), and our most massive cluster (RCS2318+0034).  
It is worth mentioning again that in Figure~\ref{fig1}, the brightest part of RCS2318+0034 does not seem to lie at the center of the cluster's extended emission, indicating that this object may have recently undergone a merger, or could at least possess an appreciable amount of substructure.    

\subsection{Gas Mass Fractions\label{fgas}}

In Section~\ref{s:mass}, we estimate the core (${\rm{R}}_{2500}$) gas mass fractions of our high-$z$ sample, finding values which are significantly lower than both the gas fractions of lower redshift X-ray selected clusters, and the expected universal gas fraction~\citep[$\Omega_b/\Omega_m=0.175\pm{0.012}$;][]{spergel07}.  
Taking a weighted average over our objects results in a gas mass fraction of $4.5\pm{0.2}\%$, in comparison with the CNOC weighted mean of $9.8\pm{0.3}\%$ and values of $\sim9\%$ found within R$_{2500}$ in clusters with $T_X > 5$ keV~\citep{vikhlinin06}.  
Poor clusters and groups are often found to have lower gas mass fractions~\citep{dellantonio,sanderson03}, therefore we may expect this result for the lower temperature objects in our sample.  
This, however, does not explain our findings for the higher temperature objects.
  
To investigate further, we first performed a K-S test on the $f_g$ values of subsets of both samples, choosing the 8 objects in the RCS sample and the 9 objects in the CNOC sample with temperatures between 3.5 and 8 keV.  This test resulted in D=0.875 and P=0.002, indicating that the gas mass fractions of the two samples are different at a confidence level of greater than 99\%.  A histogram showing the $f_g$ distributions of these subsamples is shown in Figure~\ref{fig6}.

We examined the robustnesss of this result by repeating the K-S test after attempting to remove the effects of any possible trend in gas fraction with temperature.  To do so we assumed that the RCS and CNOC samples can be combined
and that the resulting apparent trend of $f_{\rm{gas}}$ with temperature
is physical ~\citep[note that the resulting relation is much steeper than
the one suggested by][; Figure~\ref{fig6}]{vikhlinin06}.  Under these extreme assumptions, the KS-test yields D=0.764 and P=0.007, thus demonstrating the robustness of our earlier results.

%%To do this we first fit a relationship between $f_g$ and $T_X$ for our combined (RCS plus CNOC) 23 object sample, using

%%\begin{equation}
%%{\rm{log_{10}}}~f_g = C1 + C2~{\rm{log_{10}}}~T_X.
%%\end{equation}

%%\noindent The best fit resulted in $C1=-1.94\pm{0.2}$, and $C2=1.0\pm{0.2}$ (Figure~\ref{fig6}).  We note that the slope we find is significantly higher than those reported in studies of lower redshift samples~\citep[e.g.,][; where $C_2({\rm{R}}_{2500})=0.61\pm{0.31}$]{ettori02}.  Using the slope of this fit we projected the $f_g$ values of the 17 clusters in our K-S subset to a temperature of 5 keV and repeated the K-S test with those values.  Here we find D=0.764 and P=0.007, which strongly supports our earlier results.  

Low gas mass fractions have previously been observed in clusters at high redshift by both the XMM-Newton $\Omega$ project~\citep{sadat} and~\citet{lubin02}, and have been predicted in simulations of high redshift objects~\citep{nagai07,ettori06,kravtsov05,ettori04b}.  
In addition, an SZ/WMAP study performed by~\citet{afshordi07} reports that~$\sim35\%$ of expected baryonic mass is missing from the hot ICM in their 193 clusters.  Redshift evolution, however, may not be the only possibility.  
Multiple studies have confirmed that at least some fraction of their optically selected clusters have lower than expected $L_X$~\citep[e.g.,][]{bower,donahue02,gilbank04,popesso07}, therefore sample selection may also contribute to this effect.  We explore selection biases in more depth in Section~\ref{s:selection}.

Given the possibilities present in the literature and a current lack of sufficient data to perform direct comparisons with significant samples matched in both mass and redshift, it is difficult to determine conclusively that the low gas fractions measured here are the result of cluster evolution.  
Possible physical explanations for lower gas fractions are that our clusters have a comparatively higher amount of baryonic matter in the form of stars~\citep{vikhlinin06,nagai07}, that gas is still infalling \citep[i.e., in the process of virialization;][]{popesso07}, or that some mechanism has injected excess energy into the gas (i.e., galaxy formation, mergers, AGN, radio jets;~\citet{nulsen05}), thereby raising its entropy at high-$z$.  Many of these processes occur with relatively higher frequency at high-redshift~\citep[e.g.,][]{lacey93,eastman07}, thus a general trend toward lower gas fractions might easily be expected in high-$z$ clusters 

\subsection{Cluster Entropy\label{s:entropy}}

Cluster entropy can be used as a tool for investigating the energy budget of baryons in clusters~\citep{ponman99}.  Because it may provide insight into f$_{\rm{gas}}$ discrepancies, we investigate it here for our two samples.  The measurable quantity $S\equiv T_X/n_e^{2/3}$ can be related to thermodynamic entropy by $K={\rm{log}}S$.  The canonical radius for measuring this quantity is 0.1R$_{200}$~\citep{ponman99}, so that is the radius at which we present it here.  

Figure~\ref{fig7} shows a plot of cosmologically corrected entropy ($E_z^{4/3} S$) vs. temperature, with the relationship of~\citet{ponman03} overlayed ($S\simeq120~T_X^{0.65}$ keV cm$^{2}$).  The specific entropy of the RCS clusters seems overall to be slightly higher for a given temperature than that of the CNOC sample.  A K-S test on the clusters with $3.5 < T < 8.0$ results in D=0.431 and P=0.208, indicating a difference between the samples at an $\sim80\%$ confidence level.  

As in the case of the gas mass fractions, we perform an additional K-S test after attempting to remove the trend in cluster entropy with temperature.  Using the relationship of~\citet{ponman03} (above), the KS-test yields D=0.764 and P=0.007, indicating a systematic difference in the entropies of the two samples at a $>99\%$ level.  Weighted means of the corrected (5 keV) entropies of the two K-S sample subsets result in $S_{\rm{CNOC}}=297\pm{9}$ keV cm$^{2}$ and $S_{\rm{RCS}}=425\pm{18}$ keV cm$^{2}$, with the RCS clusters having higher entropy on average by a factor of 1.43.  Because $f_g$ is proportional to gas density, at a constant temperature $f_g \propto S^{3/2}$, therefore differences in entropy between the two samples can account for roughly 85\% of their $f_g$ discrepancy, indicating that additional factors may be in effect as well.

It remains difficult to determine the relative contributions of evolution and selection to possible differences in entropy.  Expectations of higher merger and AGN activity at high-$z$~\citep{lacey93,eastman07} suggest that an evolutionary explanation is feasible; however, X-ray surveys that select high central density objects may be prone to preferentially pick out low-entropy systems.
 
\subsection{$Y_X$ Relationships\label{s:yx}}

The product of cluster temperature and gas mass, $Y_X =\rm{M}_{g}T_X$ has been shown to be a reliable, low-scatter proxy for total cluster mass and to be well correlated to X-ray luminosity~\citep{kravtsov06,maughan07}.  
Here we investigate relationships between these quantities and $Y_X$ for our high redshift optically selected sample.  
We adopt self-similar $E_z$ scaling from~\citet{maughan07} and~\citet{kravtsov06} for the $L_X$ and Mass relationships, respectively; and use their best fitting slopes for our constrained slope fits.

Our individual samples can be seen to lie again on two separate relationships in the $Y_X$-$L_X$ plane, of similar slope and differing normalization (Figure~\ref{fig8}).  This can once more be explained as stemming from systematically lower gas mass fractions in the RCS sample.  We cannot make normalization comparisons to~\citet{maughan07} at R$_{2500}$; however, none of the slopes of our $L_X$-$Y_X$ fits are consistent with the slope resulting from fits to his overall sample, and when we fix the slope to his value our scatter increases by a factor of $\sim2$.  At R$_{500}$ our fit to the RCS data agrees neither in slope nor in normalization with his fit (Figure~\ref{fig8}), but again it should be mentioned that our R$_{500}$ values have been extrapolated from data within R$_{2500}$.  

There is an even more significant discrepancy between the normalizations of the CNOC and RCS samples in the $Y_X-$M$_{\rm{tot}}$ relationship.  
This is easily explained as we are already aware that gas mass fractions are lower in the RCS sample, and total mass vs. $Y_X$ ($\propto$~M$_{\rm{gas}}$) highlights this difference.  
We find overall closer agreement with the slope of the $Y_X-$M$_{\rm{tot}}$ relationship modeled by~\citet{kravtsov06} than we did in the case of $L_X$-$Y_X$. 
Here we see marginal agreement at R$_{2500}$ between their R$_{500}$ slope and that of the RCS fit, and consistency with the slope of the CNOC sample relationship.  
At R$_{500}$, though the slope of the RCS fit is still consistent with theirs, the normalizations diagree.  
The reason for this is illustrated nicely in the right panel of Figure~\ref{fig9}.  
The four clusters in our sample which do not lie on their relationship are those with the lowest gas mass fractions.  
And again we see that the three biggest outliers are those with the highest $\beta$ values, and that of these the two most discrepant are RCS1326+2903 and RCS2318+0034.

\section{Correlations with Optical Richness\label{s:or2}}

Optical richness is effectively a measurement of galaxy overdensity within a given aperture, normalized for the evolving galaxy luminosity function and the expected spatial distribution of galaxies in the cluster. 
Our chosen richness measurement, B$_{\rm{gc}}$~\citep{yee99}, represents the galaxy-cluster spatial covariance amplitude~\citep{long},

\begin{equation}
\xi (r) = \left({{r}\over{r_0}}\right)^{-\gamma} = ~{\rm{B}}_{\rm{gc}}~r^{-\gamma}.
\end{equation}

In practice, B$_{\rm{gc}}$ is based on the excess number counts of galaxies within 357 kpc of the cluster optical center, with a normalization applied to correct for the expected spatial distribution (here we assume $\gamma$=1.8, which is in general agreement with actual galaxy distributions at these radii) and for the evolving luminosity function of cluster galaxies. 
Though some uncertainties exist in the evolution of cluster galaxies at redshifts of $z>0.5$, they can be minimized by employing a red-sequence optical richness, \bgc, which is calculated using only the more uniformly evolving red galaxies in a cluster, and which may be better correlated with the underlying cluster mass.
This is the optical richness parameter which will be used throughout this work.  
Values of \bgc~for this sample are given in Table~\ref{table5}. 

It has been shown that B${\rm{gc}}$ correlates well with the X-ray parameters of relaxed clusters~\citep{yee03}, and in~\citet{hickscnoc} we have derived relationships for correlations of X-ray properties with \bgc. 
These relationships, however, were calibrated for X-ray selected clusters at moderate redshift, and therefore may not accurately describe our current sample.  
Here we test these correlations for optically selected clusters at high redshift.  
In the following we will assume that \bgc~behaves similar to the X-ray
temperature when comparing to the X-ray properties. The rationale for
this choice is that for a cluster with a fixed density profile, both
the temperature and \bgc~do not change with redshift, whereas
for instance M$_{2500}$, $L_X$, etc. do change (following self-similar
evolution). 

Our actual data do not extend much beyond $R_{2500}$, so most of our fitting is performed within that radius. 
Cluster properties included in our fits are $L_X$, $T_X$, and total mass.
Again we employ the BCES algorithm of~\citet{akritas}.  
For each of our fits we adopt the form

\begin{equation}
{\rm{log_{10}~Y}} = C_1 + C_2~{\rm{log_{10}}}~{\rm{B}_{\rm{gc,red}}}
\label{eq:powerlaw2}
\end{equation}

\noindent where Y represents the particular property being fit.  
For $L_X$, $T_X$, and total mass, units of $10^{44}~\rm{erg}~\rm{s}^{-1}$, 5 keV, and $10^{14}\msun$ were used, respectively.  
RCS0439-2904 was again removed from fitting procedures, as
it has been confirmed to be two closely-spaced systems in projection along the line of sight~\citep{gilbank07,cain}.
Best fitting parameters and scatters are given in Table~\ref{table11}, along with comparison fits from the literature.

Figure~\ref{fig10} shows the relationship between temperature and richness.  Here, there is little evidence for a systematic difference between the samples, with both the CNOC and the combined sample showing statistical agreement with the expected slope of $2/\gamma$=1.11~\citep{yee03}.  The RCS sample is on average slightly cooler at a given \bgc, a tendency that might stem from sample selection (see Section~\ref{s:selection}, below).
Note that both X-ray and optically-selected samples contain a few outliers, scattering towards higher temperature or lower richness.
Figure~\ref{fig11} shows relationships between richness, M$_{2500}$ and M$_{200}$.
At R$_{2500}$ only the CNOC sample shows agreement with the expected slope of 3/$\gamma$. 
All R$_{200}$ fits are consistent with the values obtained for the CNOC sample by~\citet{yee03} using galaxy dynamics, and the fit reported in~\citet{blindert07} for a sample of 33 RCS clusters ($0.2 < z < 0.5$).
The consistency between these fits indicates a general agreement between both the samples and the different mass estimators; however, it should also be noted that error bars on the fit parameters are quite large for this relationship.

The $L_X$-\bgc~plot (Figure~\ref{fig12}) in contrast, shows quite a bit of scatter for both samples, and a significant offset between the RCS and CNOC samples. This offset is expected for the RCS sample, based on the results of Section~\ref{s:lxtx}, but here we also include the additional clusters with low X-ray luminosity for which T$_X$ could not be derived.  RCS0439-2904, the object which is spectroscopically confirmed to be a projection of two less massive systems, is the cross second from the right.  The significantly higher amount of scatter that we see in this relationship when compared to any of the other $L_X$ relationships suggests that \bgc~is a less reliable predictor of X-ray luminosity than $T_X$, total mass, or $Y_X$.

Of all the richness relationships we investigate, \bgc~is best correlated to X-ray temperature, with an average scatter (all fits) of only $\sigma_{\rm{log Y}}\sim0.16$ for the objects with measured $T_X$ (minus RCS0439-2904, which is not included in fitting).
Interestingly, this is less than the average scatter of our $L_X$-$T_X$ relationship for the same objects (all fits; $\sigma_{\rm{log Y}}\sim0.22$), probably again due to the issue of missing ICM baryons in the high-$z$ RCS sample.
Since $T_X$ is closely related to total mass, whereas $L_X$ is intimately tied to the gas density, differing gas mass fractions will shift the normalization in $L_X$-$T_X$ significantly, whereas \bgc~
may be fairly independent of  the amount of X-ray emitting gas.

When fitting mass to richness, overall scatters average (all fits) to $\sigma_{\rm{log Y}}=0.28$ for M$_{2500}$, and $0.32$ for M$_{200}$.  
The scatter in the RCS sample at R$_{200}$ is particularly large, due in part to the two objects with high $\beta$ values (RCS1326+2903 and RCS2318+0034), and RCS2320+0033 which has a lower than expected \bgc.  
In its role as a mass estimator, \bgc~produces on average $0.07-0.19$ more scatter in $\sigma_{\rm{log Y}}$ than the T$_X$-based mass proxies investigated here, and may suffer from a fraction of objects whose richnesses are affected by projection. 
However, the comparative speed and ease with which it can be obtained still recommend it as a potentially useful tool for mass estimations of large high redshift cluster samples.

 \section{Sample Selection and Biases\label{s:selection}}
 
 The RCS sample is effectively selected by richness (\bgc), whereas the CNOC sample was compiled from objects with high X-ray luminosity~\citep{yee96}. 
Discrepancies in the relations between X-ray and optical properties for these two samples may thus partially be caused by sample selection, especially if the underlying distribution of X-ray-to-optical properties is intrinsically broad~\citep[e.g.,][]{gilbank04}.
Both X-ray and optical surveys would then be expected to deliver biased samples of clusters, with the degree of bias based on the level of scatter in the selection criterion. Here we discuss three sources of selection bias in optical and X-ray cluster samples: optical projection effects, Eddington bias due to observational uncertainty, and sample bias for both optical and X-ray samples.

%%\subsection{Projection Effects}

One important difference in the cluster samples stems from the RCS cluster-finding process. 
While all but one of our objects were confirmed (to S/N$>3$) as extended X-ray sources, the RCS sample is expected to also include a small fraction of  objects whose richness is boosted by the superposition of other structures having galaxy colors similar to the cluster's red sequence. 
While these projections are much less problematic than in monochromatic cluster searches, they may still add systems into the RCS catalog with true richnesses significantly lower than the measured \bgc. ~\citet{gladdersphd} performed a series of simulations which suggest that the fraction of RCS clusters composed of significant projections is on the order of 5-10\%.  This estimate has been confirmed at $z \sim 0.3$ via extensive spectroscopy of 33 RCS clusters~\citep{blindert07}; and at $z \sim 0.8$ from a sample of 12 clusters~\citep{gilbank07}. Additional spectroscopy as well as weak lensing estimates of additional clusters is underway.
Recently,~\citet{cohn07} examined the effects of local structures on the red sequence using the Millennium cosmological simulations. 
They found that in the simulation, the frequency of significant projection increases at higher redshift, to $\sim$ 20\% at $z=1$.  
However, their cluster-finding algorithm and richness estimate were significantly different from the RCS algorithm in many aspects (galaxy magnitudes and colors, radial extent and background corrections), so this may not be directly comparable to the samples discussed here.

Our {\it{Chandra}} observations suggest that perhaps 3 of 13 observed clusters may have X-ray luminosities which are significantly lower than expected from the RCS $L_X$-\bgc~relationship.
All three of the outliers in Figure~\ref{fig12} have been observed in detail spectroscopically, and two of these were found to have at least some degree of overlap with additional structures in the line of sight.  RCS0439-2904 was found spectroscopically to consist of two objects in such close proximity that they may be interacting~\citep{gilbank07,cain}.  RCS1417+5305 is a similar case, though here the overlapping systems are different enough in redshift that they might be unrelated~\citep{gilbank07}.  RCS2112-6326 exhibits a single spectroscopic peak at $z\sim1.1$~\citep{barrientosprep}.  It is not clear whether the highest richness systems in this sample might be subject to a higher contamination rate than the RCS-1 survey as a whole. 

%%\subsection{Eddington Bias}

Because there is significant observational uncertainty  in our richness estimates,  and the number of clusters declines rapidly with increasing richness, it is also necessary to evaluate effects of a possible Eddington bias in the X-ray/optical relationships.  
We calculate this possible bias by using the observed distribution of \bgc~in the RCS-1 sample, which falls as \bgc$^{\rm{N}}$, where N~$\sim -4$.   
Uncertainties in \bgc~are calculated based on the statistics of galaxy counts in the clusters and in the statistical foreground/background galaxy distribution~\citep{yee99}, and tend to increase modestly with increasing richness. 
We model the typical gaussian 1-$\sigma$ uncertainty in \bgc~as a function of \bgc~from an empirical fit to the observed distribution in RCS-1:

\begin{equation}
{\rm{log}_{10}}~(\sigma ) = 0.899+0.535~{\rm{log}_{10}}~({\rm{B}_{\rm{gc,red}}})
\end{equation}

This relationship predicts that the uncertainty will be $\sim 180$  at \bgc=300 $h_{50}^{-1}$ Mpc$^{1.77}$, at the lower end of our cluster richness distribution, and $\sim 320$ at 1000 $h_{50}^{-1}$ Mpc$^{1.77}$ for very rich clusters.  
(Note that these errorbars are not identical to the detection significance for the cluster, but instead reflect the uncertainty in the measurement of the cluster's richness).  
Convolving these relations predicts that the true distribution in richness for a measured \bgc~is skewed to lower values, with a mean value that is $\sim 80-90\%$ of the measured value.
We then use our observed relationship between X-ray temperature derived from the CNOC clusters and B$_{gc}$ to calculate that the mean observed temperature for RCS clusters should be about 10\% lower than the expected relationship at \bgc=1000 $h_{50}^{-1}$ Mpc$^{1.77}$ and about 20\% lower at \bgc=500.  
These decrements will also tend to steepen the logarithmic slope of the $T_X$-B$_{gc}$ relationship by about 0.15. 
Varying the distribution parameters within reasonable limits produces corrections on the order of 10-30\% in normalization at a given \bgc, and a systematic increase of 0.1-0.3 in the slope.  
Comparison with the $T_X$-\bgc~relation shown in Figure~\ref{fig10} suggests that a correction for this Eddington bias would ameliorate the discrepancies between the RCS and CNOC samples, likely resulting in statistical agreement between their respective fits. 

A similar calculation for X-ray luminosities was performed, with decrements in the X-ray luminosity of about 40\% and 25\% beneath predicted values at \bgc~500 and 1000 $h_{50}^{-1}$ Mpc$^{1.77}$, steepening the logarithmic slope of the $L_X$-\bgc~relationship by about 0.3. 
This correction is not, however, sufficient to create agreement between the RCS and CNOC samples once bias in richness measurements has been accounted for, as is also indicated by their differing L$_X$-T$_X$ relations. Note these calculations assume that observational uncertainty is the primary source of scatter in the correlations.

%%\subsection{Sample Bias}

A final consideration in comparing X-ray and optically-selected samples is the possibility that both selection methods produce biases when selecting clusters from a population with a significant intrinsic variation in X-ray or optical properties. 
If there is a significant intrinsic scatter in the properties of gas in cluster cores, systematic differences in X-ray characteristics between optically and X-ray selected samples may naturally arise. 
The ROXS survey, a joint X-ray/optical survey for clusters~\citep{donahue02} found that some of their optically selected clusters had lower than expected $L_X$, suggesting that  selection effects could be culpable. ~\citet{gilbank04} also performed an independent X-ray/optical survey for clusters, using the red-sequence as well as the monochromatic matched-filter technique. They found that the red-sequence methodology significantly out-performs monochromatic techniques in discovering and characterizing clusters.  Even so, they also found a significant difference in the X-ray luminosities of X-ray versus optically-selected clusters, with several examples of spectroscopically-confirmed low $L_X$ clusters. More recently~\citet{stanek07} report that Malmquist bias may be responsible for a higher (by a factor of $\sim2$) average $L_X$ in X-ray flux-limited samples.
These studies all suggest that for a given cluster mass or temperature  there is a significant intrinsic scatter in X-ray luminosity or optical richness, or possibly both.  

If this is the case, then both X-ray and optically-selected clusters may be prone to bias. We first consider the effects of such bias on our observed $L_X$-$T_X$ relations, where we found the RCS clusters to be systematically lower in luminosity for a given temperature, and in addition calculate lower core gas fractions.  This discrepancy could be interpreted as evidence for evolution in the properties of the ICM, and the loose agreement of the RCS data with the high redshift X-ray selected sample of~\citet{ettori04} supports this conclusion.  
In addition, our X-ray luminous CNOC comparison sample may include a significant bias. These clusters were chosen from the wide-area EMSS survey~\citep{gioia90} primarily  based on their X-ray luminosities and may indeed represent a sample of particularly luminous clusters.
We summarize by noting that these selection biases in both X-ray and optical samples can be significant, but can be evaluated quantitatively, given additional independent information about the underlying cluster mass. In general, variations in the X-ray properties of clusters can be inferred most robustly from optically-selected clusters, and vice-versa, to minimize these effects. 

\section{Summary and Discussion \label{s:sum}}

We have performed an in-depth X-ray investigation of 13 high redshift ($0.6<z<1.1$) optically selected clusters of galaxies from the Red-sequence Cluster Survey (RCS; Table~\ref{table1}).  
All but one of these clusters was detected by {\it{Chandra}} at a signal-to-noise ratio of greater than 3 (Table~\ref{table2}), though two additional clusters in our sample (RCS1417+5305 and RCS2112-6326), did not posess enough signal to support further analysis. 
Initial imaging of the objects reveals that the RCS sample spans a wide range in cluster morphology (Figure~\ref{fig1}), from very regular objects (e.g., RCS1419+5326) to more disturbed systems (e.g., RCS2112-6326).

Surface brightness profiles were extracted for ten clusters in $1-2\arcsec$ annular bins, and were reasonably well fit by single $\beta$ models (Figure~\ref{fig2} and Table~\ref{table3}).  
Cluster emission was modeled with XSPEC, beginning with a spectral extraction region of 300 kpc radius.  
The results of single temperature spectral fits, combined with best fit $\beta$ models, were used to determine $\rm{R}_{2500}$.  
Spectra were re-extracted from regions of that radius for further temperature fitting and $\rm{R}_{2500}$ luminosity estimates, until extraction regions and R$_{2500}$ estimates were in agreement.  
Results of this process are given in Table~\ref{table4}.  
We have also used the $\sigma-T_X$ relationship to compare the X-ray temperatures of three of our objects to currently available velocity dispersions~\citep{gilbank07,gilbankprep}.  
We find consistency in all cases (Table~\ref{table6}), suggesting that these three objects are at least relatively undisturbed.

Using the results of both spectral fitting and surface brightness modeling, X-ray masses were calculated for ten clusters in our sample out to $\rm{R}_{2500}$ (Table~\ref{table7}), with extrapolation to $\rm{R}_{500}$ (Table~\ref{table8}).  
Canonical X-ray scaling laws were investigated for nine clusters and compared to those of the moderate redshift ($0.1<z<6$) CNOC sample (Tables~\ref{table9} and~\ref{table10}).  
For the $L_X-T_X$ relationship (Figure~\ref{fig3}), both RCS and CNOC fits have slopes consistent with self-similar predictions; however, their normalizations disagree.   
Interestingly, the slope of our combined RCS-CNOC sample agrees with that of X-ray selected clusters at similar redshift~\citep{ettori04}, suggesting that evolution in the normalization of the $L_X-T_X$ relationship may lie behind the observed steeper slopes.  
Results from $L_X-$M$_{\rm{tot}}$ fits are qualitatively very similar to those of the $L_X$-$T_X$ relationship (Figure~\ref{fig4}), RCS and CNOC slopes are consistent with self-similarity at R$_{2500}$, but disagree in normalization due to differing ICM densities.  

The most notable outcome of our mass estimations is that the $\Delta=2500$ gas mass fractions of RCS clusters are lower than expected by a factor of $\sim2$~\citep{vikhlinin06}.  
Low gas mass fractions are also reported in the findings of other high redshift cluster studies, both observation and theory~\citep{lubin02,ettori04b,kravtsov05,sadat,ettori06,nagai07,afshordi07}.  
Physical explanations for low gas fractions include suggestions that much of the baryonic mass has been converted into stars, that the gaeous component of these clusters is still infalling, or that some mechanism (i.e., galaxy formation, AGN, mergers, radio jets) is responsible for raising the entropy of the gas.  Though we do see some evidence for higher entropy in the RCS sample (Figure~\ref{fig7}), it may not be enough to explain the entire f$_{\rm{gas}}$ discrepancy.
Further study will have to be undertaken to determine whether the overall lower gas fractions that permeate this sample are ubiquitous at high redshift, or an outcome of sample selection.

Explanations aside, the growing evidence that massive ($T_X \sim6$ keV) clusters may have an evolving or broad range of central gas mass fractions may have important consequences for the interpretation of future cluster surveys in the microwave and X-ray bands, which select in part on the basis of central gas density.  Scatter in this parameter will tend to reduce completeness and, if not properly accounted for, inject bias in the cluster samples such surveys produce.  In addition, since SZ mass determinations tend to rely on the assumption of a constant gas mass fraction~\citep{reid06}, complimentary data may be required to avoid systematic errors in SZ total mass estimates.

Using red-sequence optical richness measurements of both samples the relationships between \bgc~and global cluster properties ($T_X$, $L_X$, $\rm{M}_{2500}$, and M$_{200}$) were investigated (Table~\ref{table11}).  
We find that \bgc~is poorly correlated to X-ray luminosity, with average scatter in the relationships reaching $\sigma_{\rm{log Y}}\sim0.36$.  
Temperature, however, is nicely predicted by optical richness. 
 The scatter in the $T_X-$\bgc~relationship is lower even than the scatter we find in $L_X-T_X$ $(\sigma_{{\rm{log Y}}}\sim0.16$ compared to 0.22), though this measurement may exclude several of the strongest outliers in our sample. 
 This can be explained by the lack of hot baryons in RCS clusters, which would affect cluster $L_X$, but not temperature or \bgc. 
 Richness-mass relationships are generally consistent with one another and with previous studies~\citep{yee03,blindert07}, and we find an average scatter of $\sigma_{{\rm{log Y}}}\sim0.30$ for these relationships (Figure~\ref{fig11}). 
 Though this scatter is on average somewhat higher than the other mass proxies investigated here (by $0.07-0.19$ in $\sigma_{\rm{logY}}$), the comparative speed and ease with which it can be obtained still recommend it as a promising tool for mass estimations of large high redshift cluster samples.

\acknowledgements 

Support for this work was provided by the National Aeronautics and Space Administration through a Graduate Student Research Program (GSRP) fellowship, NGT5-140, and Chandra Awards GO0-1079X and GO0-1063B, issued by the Chandra X-ray Observatory Center, which is operated by the Smithsonian Astrophysical Observatory for and on behalf of the National Aeronautics Space Administration under contract NAS8-03060.  EE acknowledges NSF grant AST 02-06154.  MB and BC were supported by subcontract 2834-MIT-SAO-4018 of contract  SV74018  issued by the Chandra X-ray Center on behalf of NASA under contract NAS8-08060.  The RCS is supported by grants to HKCY from the National Science and Engineering Research Council of Canada and the Canada Research Chair Program.  We would also like to thank Phil Armitage, Monique Arnaud, Webster Cash, John Houck, Andisheh Mahdavi, Richard Mushotzky, and Craig Sarazin for their contributions and input.

%\appendix

%% If you wish to include an acknowledgments section in your paper,
%% separate it off from the body of the text using the \acknowledgments
%% command.

%% Included in this acknowledgments section are examples of the
%% AASTeX hypertext markup commands. Use \url without the optional [HREF]
%% argument when you want to print the url directly in the text. Otherwise,
%% use either \url or \anchor, with the HREF as the first argument and the
%% text to be printed in the second.

%We are grateful to V. Barger, T. Han, and R. J. N. Phillips for
%doing the math in section~\ref{bozomath}.
%More information on the AASTeX macros package are available at
%\url{http://www.aas.org/publications/aastex} or the
%\anchor{ftp://www.aas.org/pubs/}{AAS ftp site}.
%For technical support, please write to
%\email{aastex-help@aas.org}.

%% Appendix material should be preceded with a single \appendix command.
%% There should be a \section command for each appendix. Mark appendix
%% subsections with the same markup you use in the main body of the paper.

%% Each Appendix (indicated with \section) will be lettered A, B, C, etc.
%% The equation counter will reset when it encounters the \appendix
%% command and will number appendix equations (A1), (A2), etc.

%\appendix

%\section{Appendicial material}

\clearpage

%% FIGURES

\begin{figure}
\includegraphics[width=6in,clip=true]{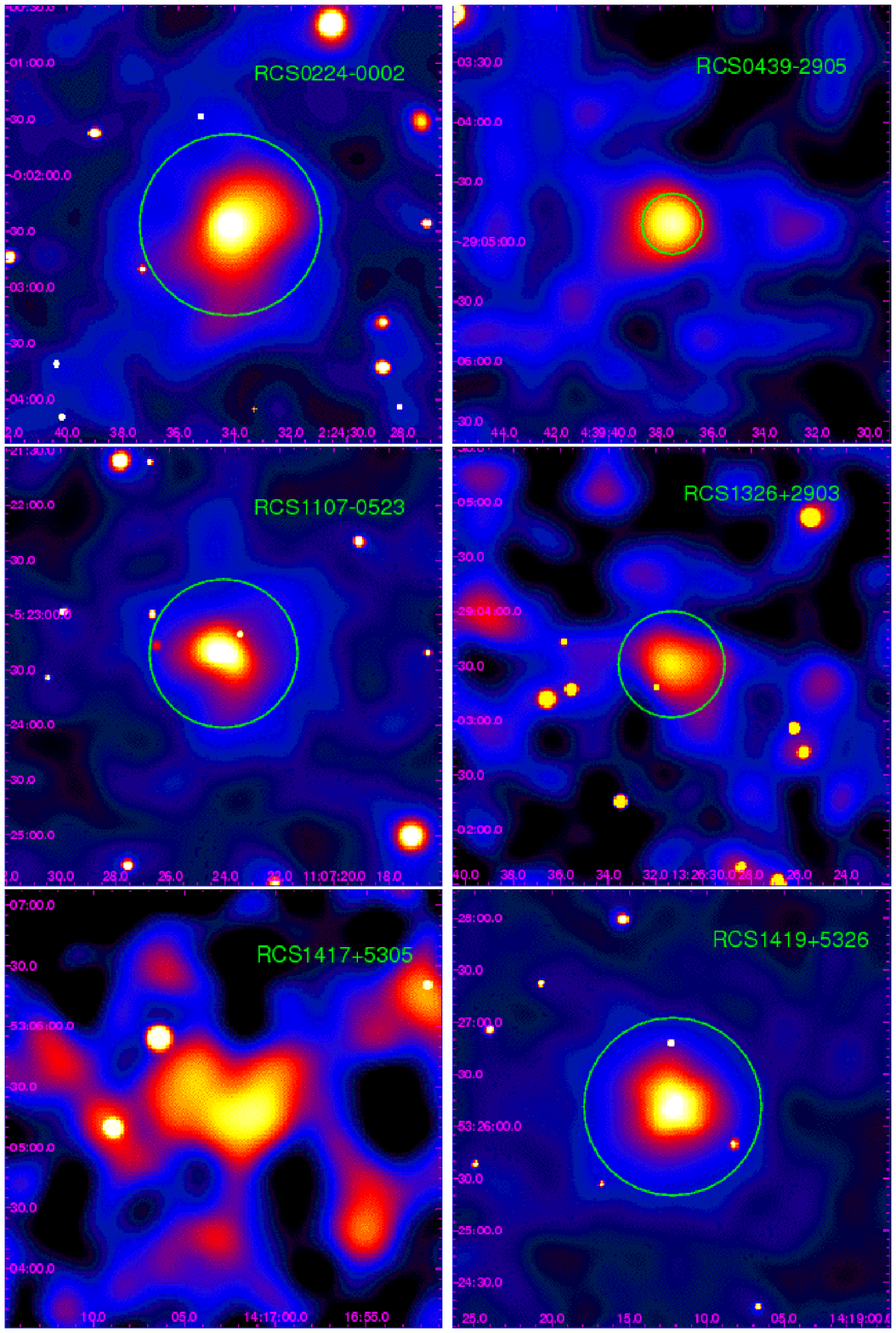}
\end{figure}
\begin{figure}
\includegraphics[width=6in,clip=true]{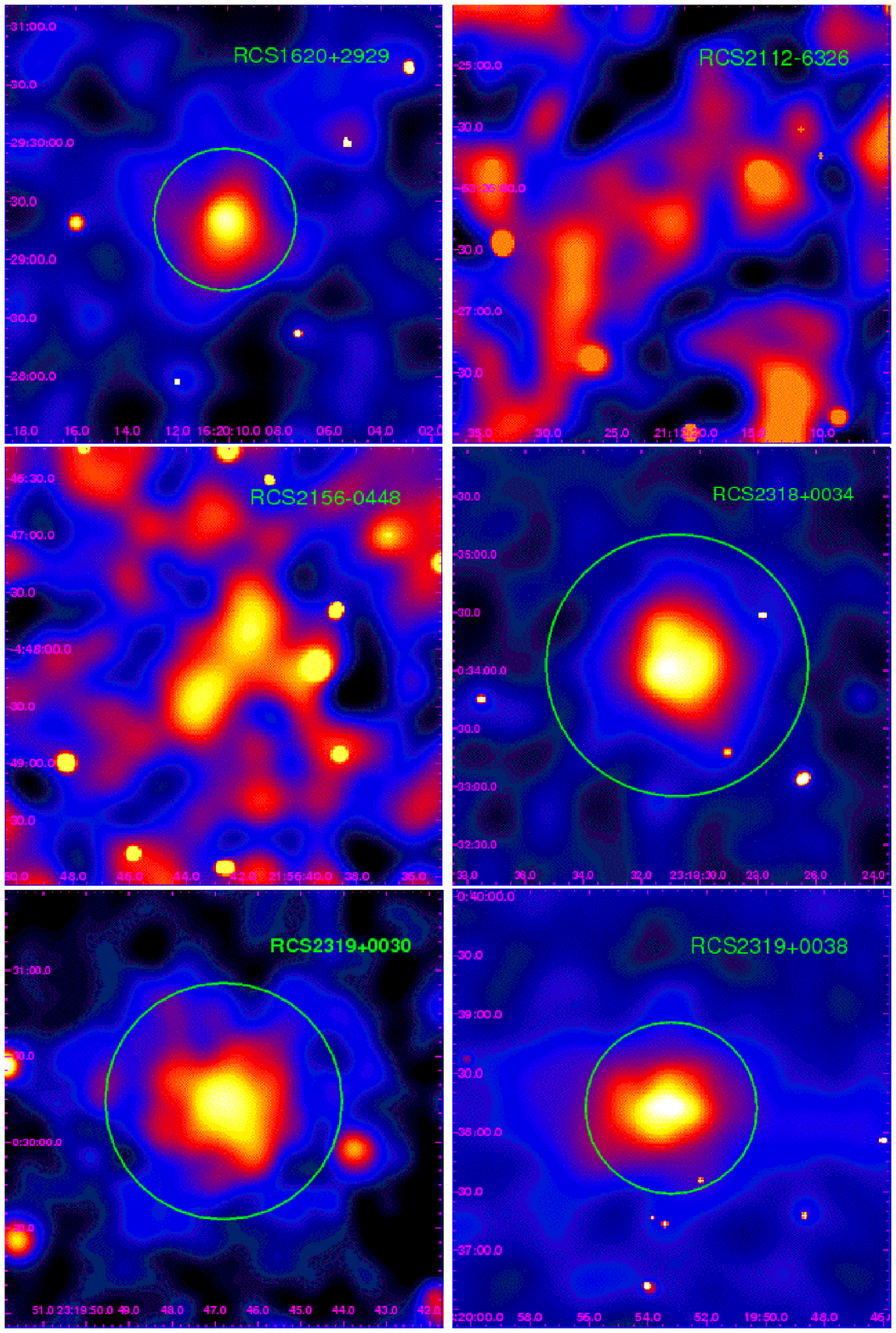}
\end{figure}
\begin{figure}
\includegraphics[width=6in,clip=true]{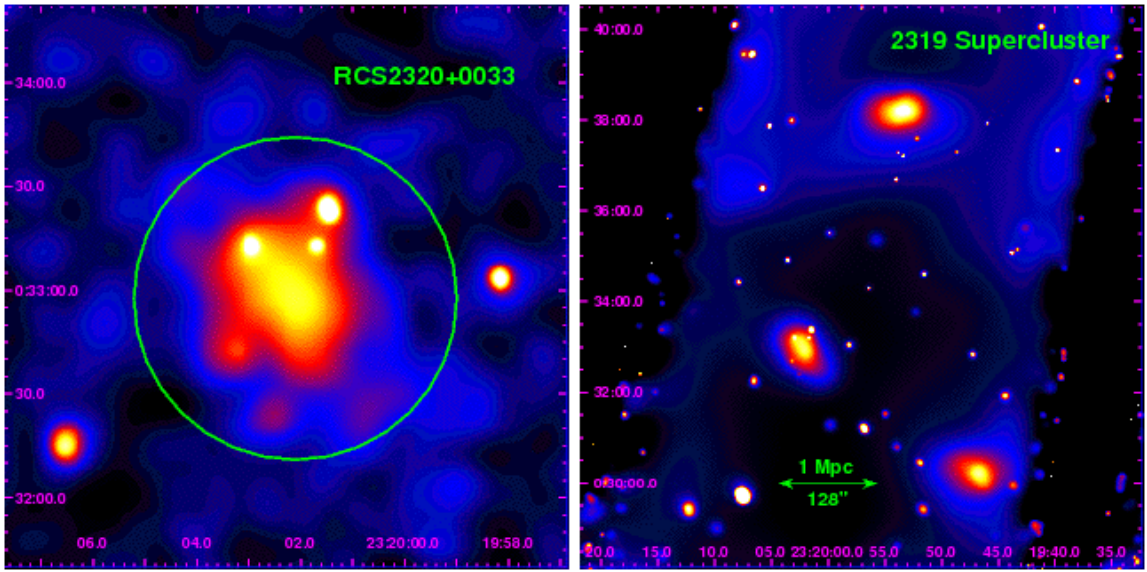}
\caption{{\bf{Smoothed Flux Images.}}  Adaptively smoothed X-ray flux images of our sample in the 0.3-7.0 keV band.  
Circles denote calculated values of $\rm{R}_{2500}$ for each cluster.  
The three single cluster images which lack circles did not contain enough cluster signal to constrain a $\beta$ model or a temperature, and thus lack estimates of $\rm{R}_{2500}$.  
In each image, north is up and east is to the left.  
The last image shows the three clusters which make up the $z=0.9$ supercluster in the 23h field.  
The aimpoint cluster (RCS2319+0038) lies at the top of the image on the backside illuminated CCD ACIS-S3, and the other two clusters (RCS2319+0030 and RCS2320+0033) lie on the frontside illuminated CCD, ACIS-S2. 
Instrumental differences in the two chips cause their respective backgrounds to have slightly different values in the image.\label{fig1}}
\end{figure}

\clearpage

\begin{figure}
\centerline{\includegraphics[width=2.5in, angle=90]{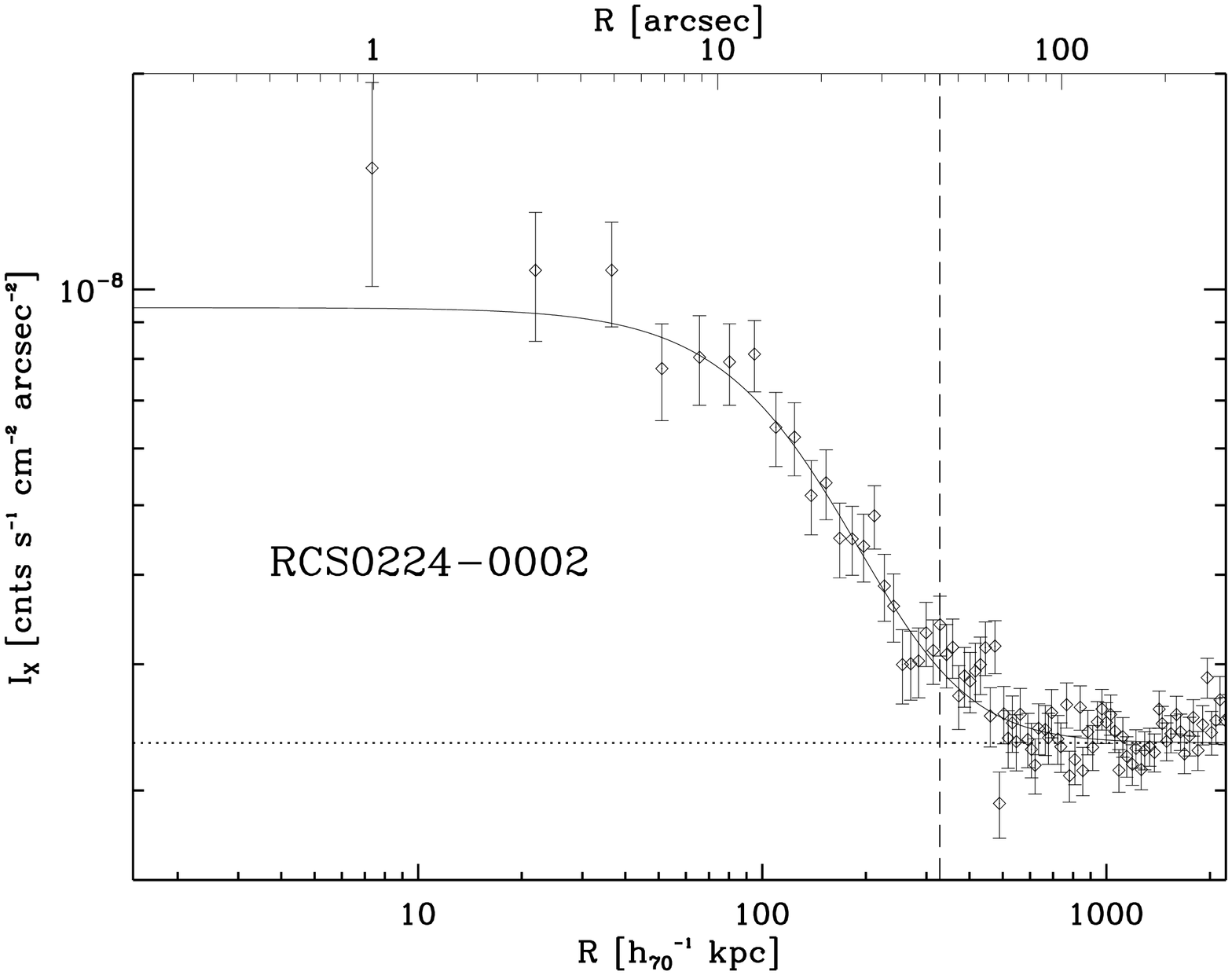}
\includegraphics[width=2.5in, angle=90]{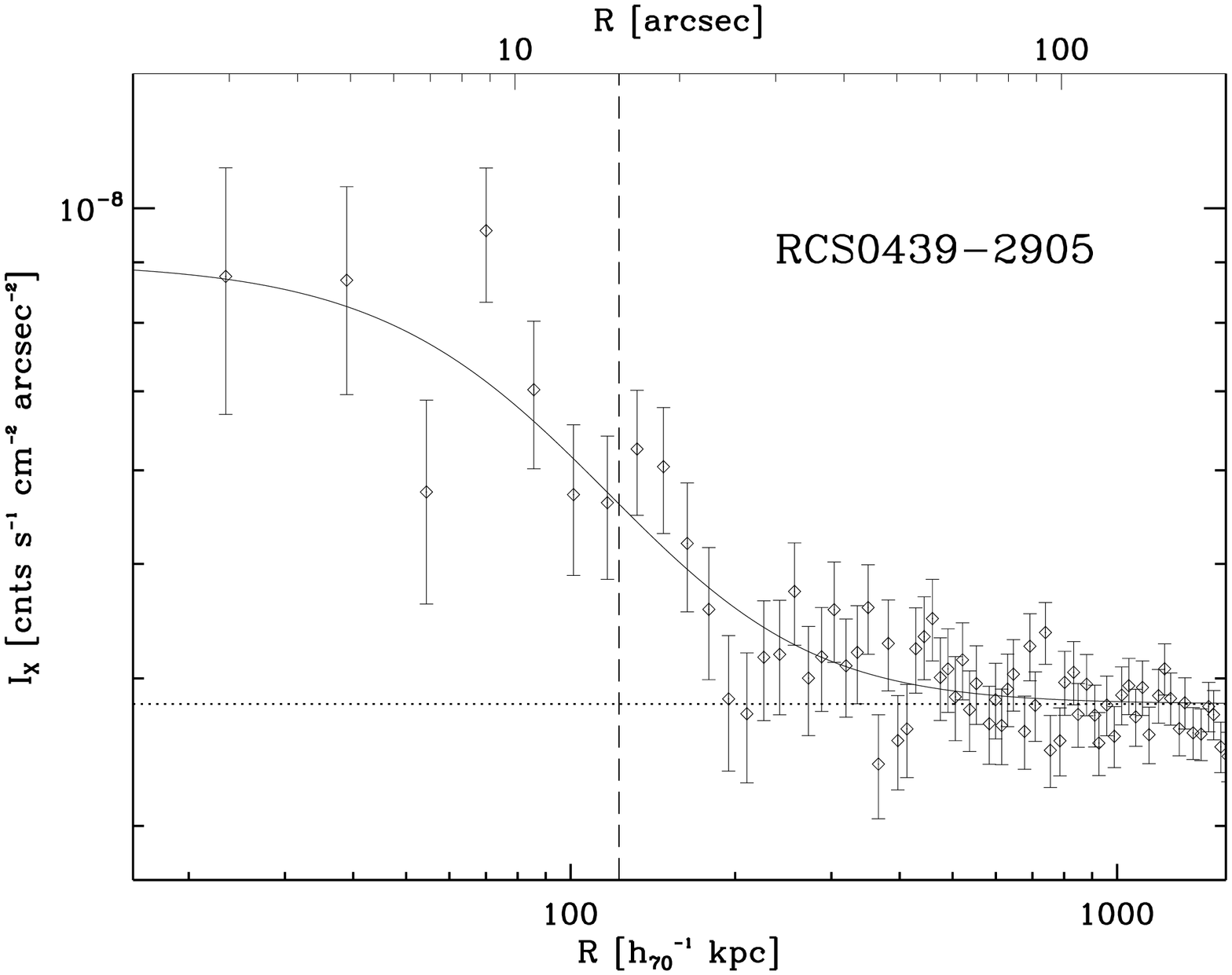}}
\centerline{\includegraphics[width=2.5in, angle=90]{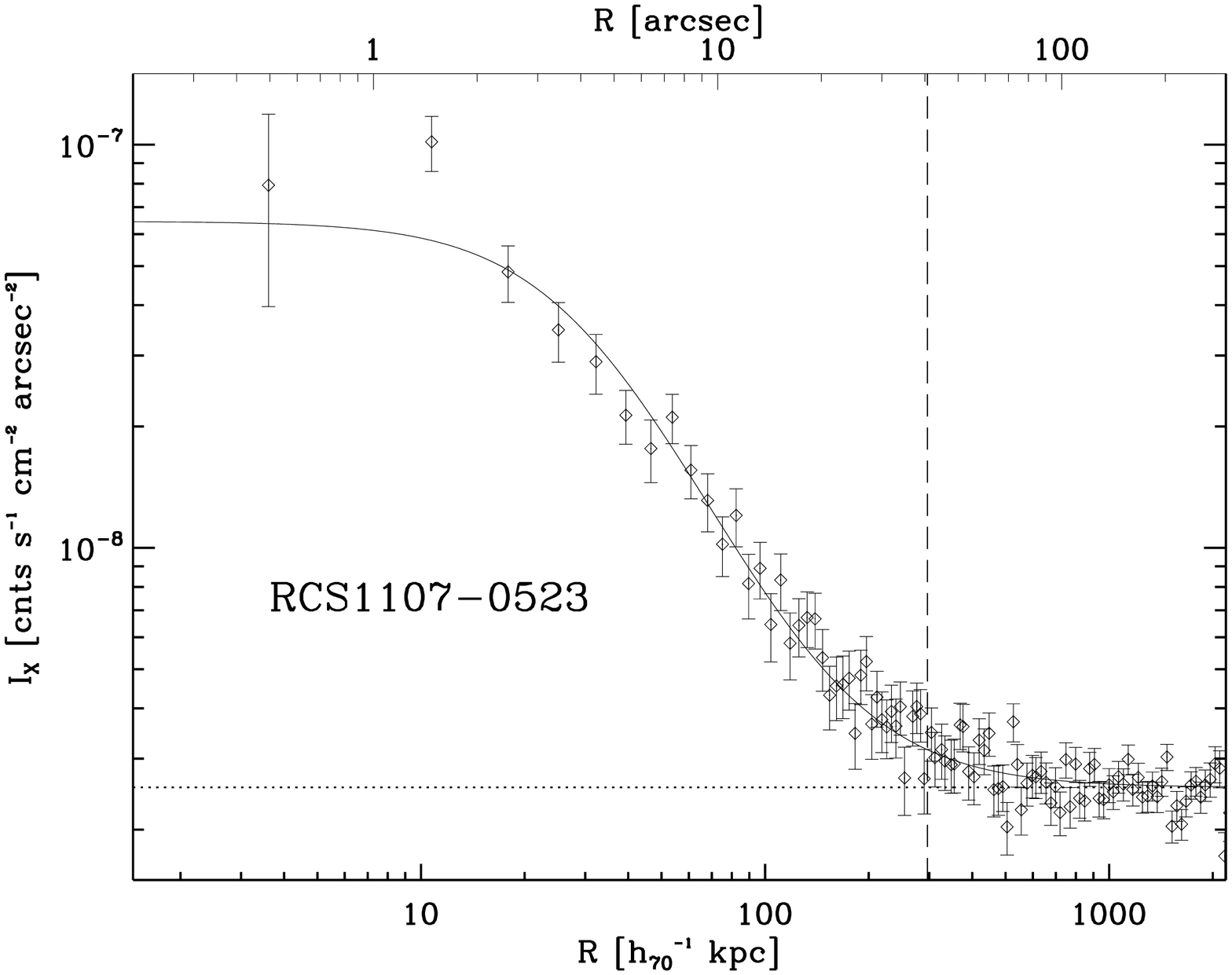}
\includegraphics[width=2.5in, angle=90]{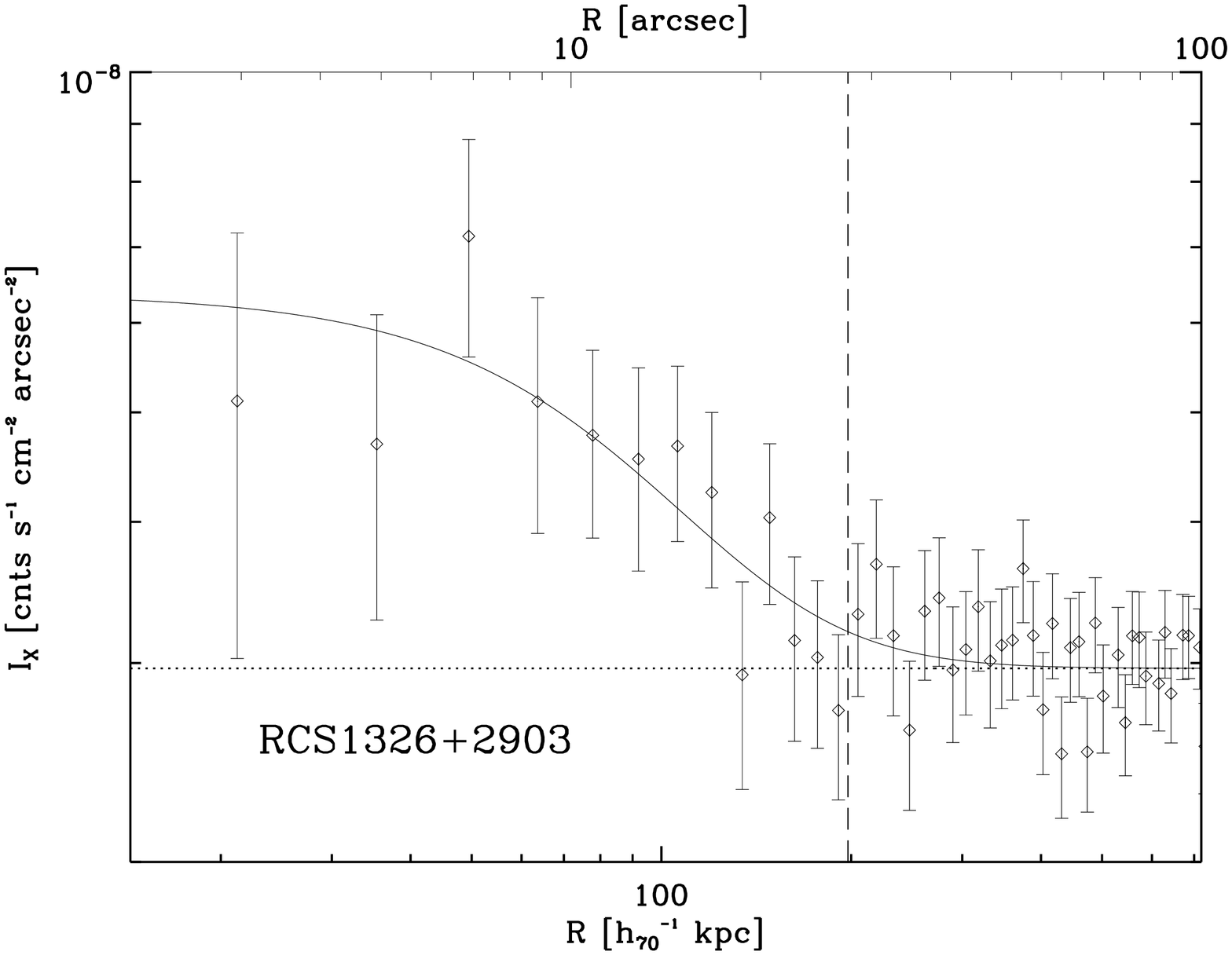}}
\centerline{\includegraphics[width=2.5in, angle=90]{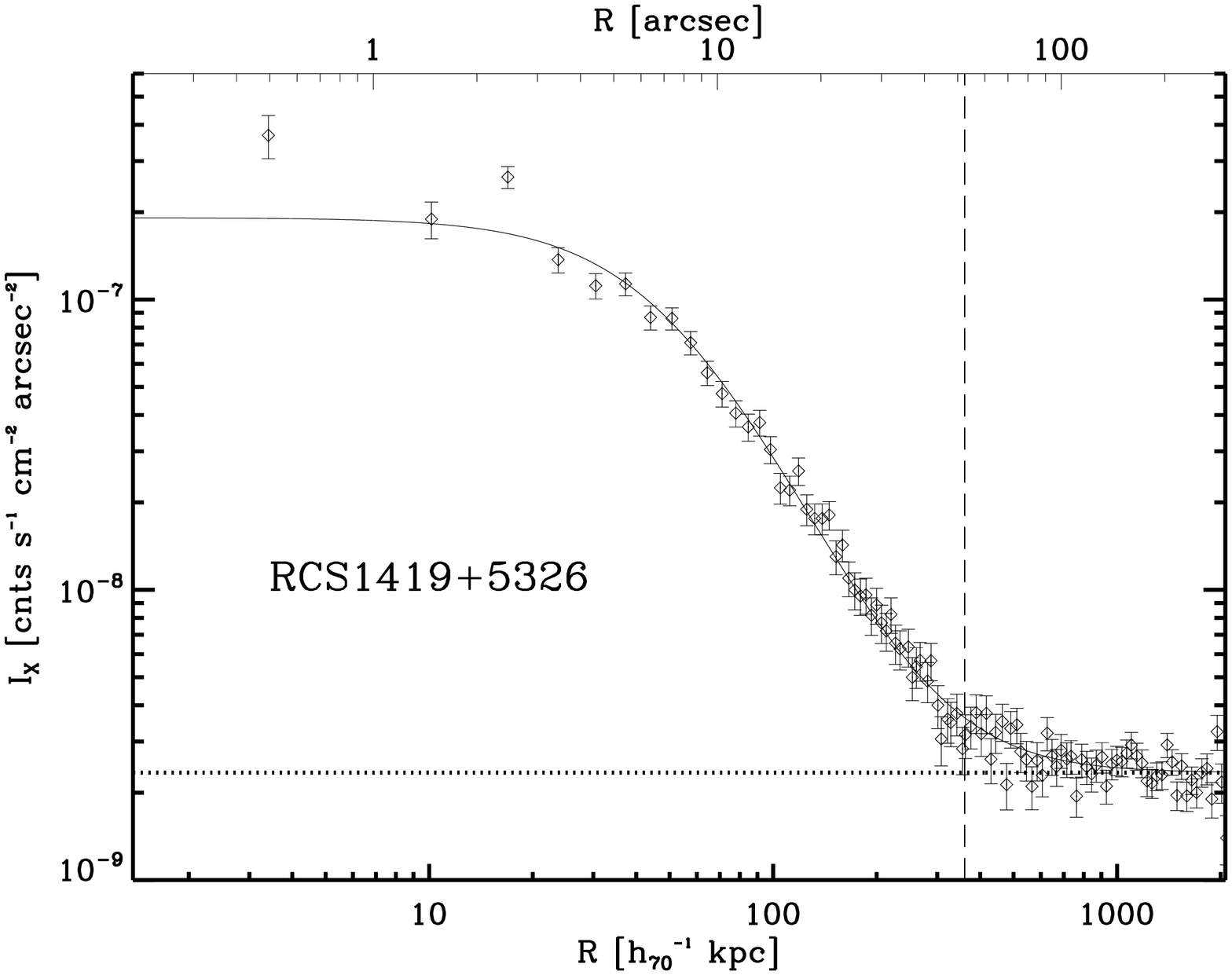}
\includegraphics[width=2.5in, angle=90]{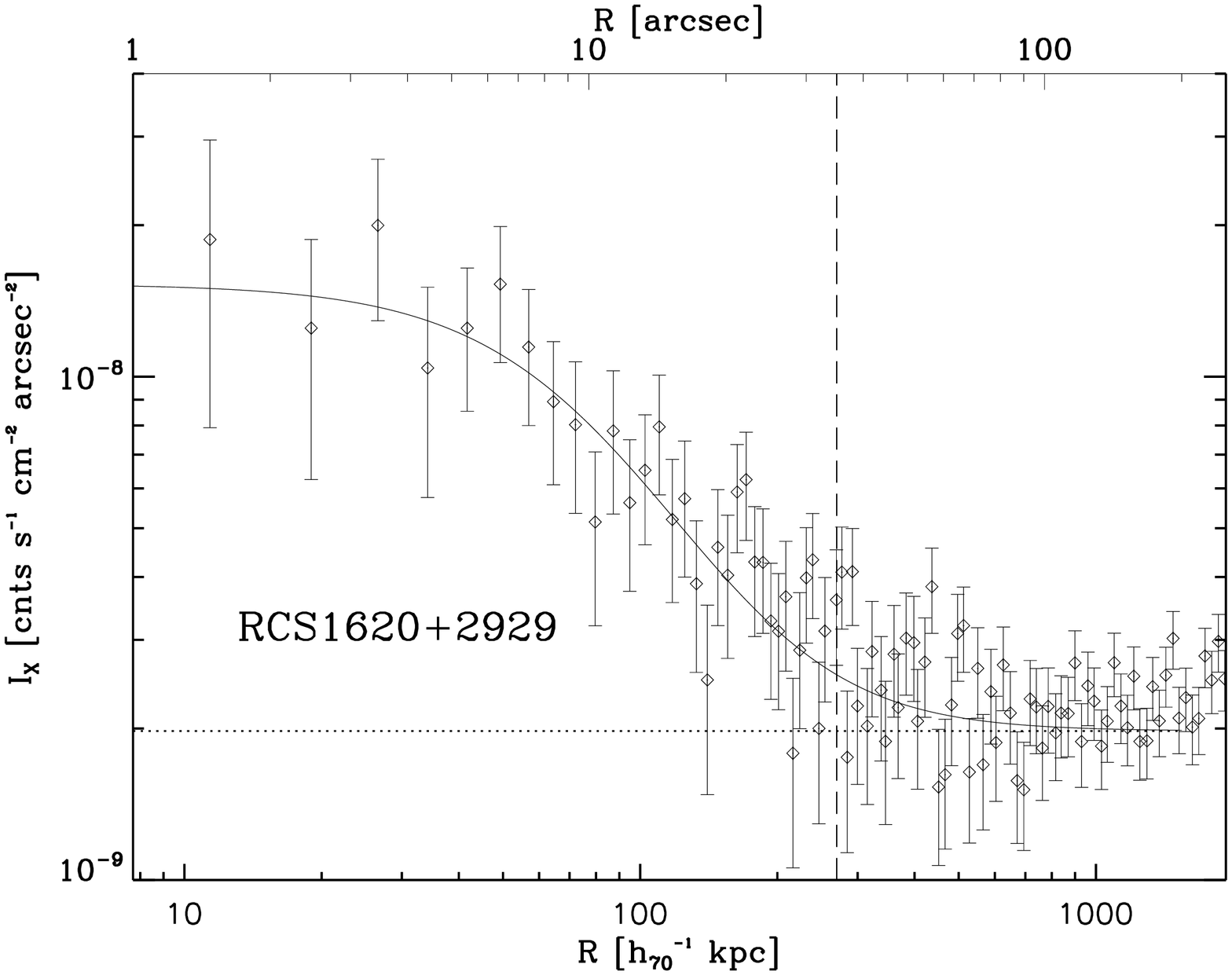}}
\end{figure}
\begin{figure}
\centerline{\includegraphics[width=2.5in, angle=90]{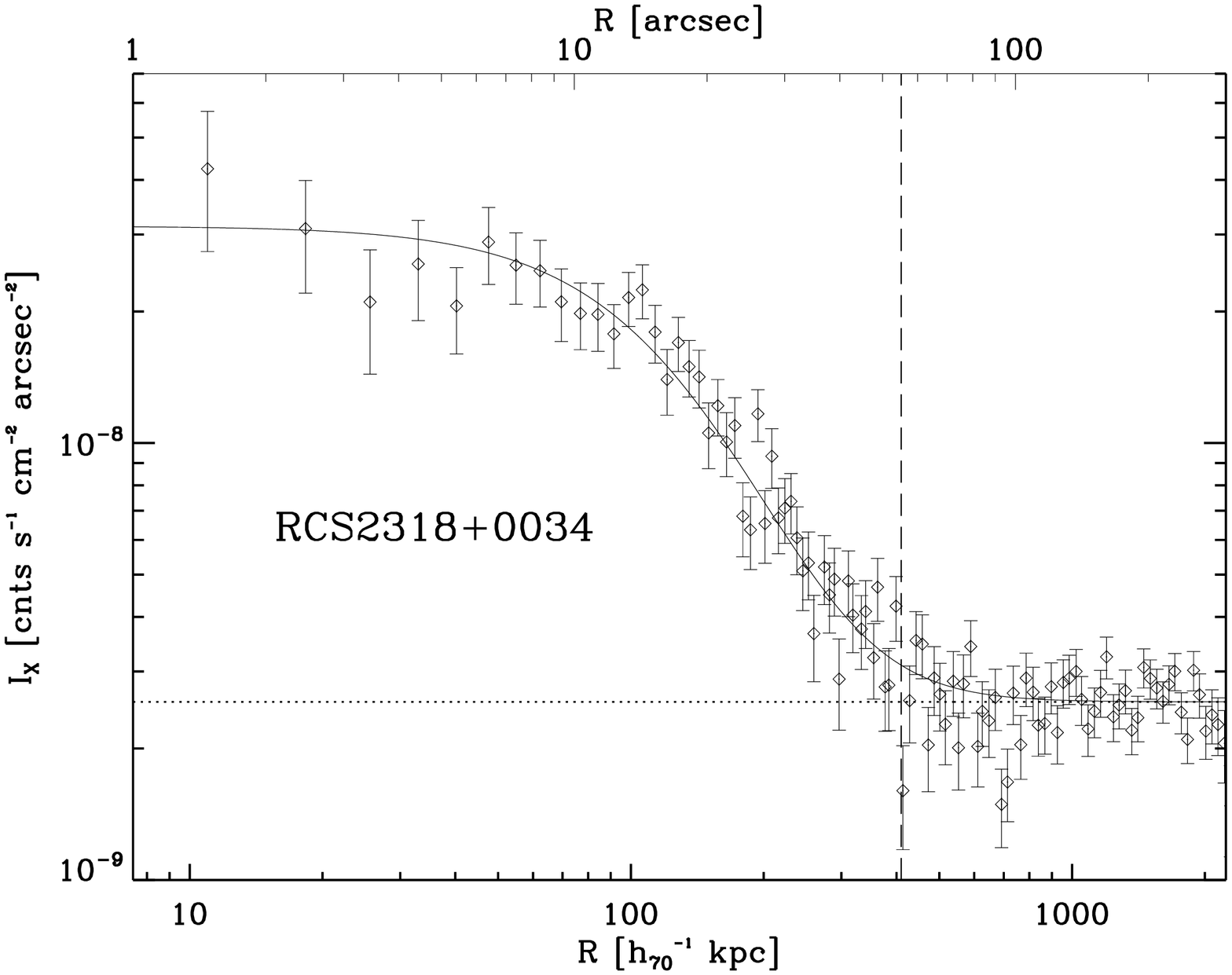}
\includegraphics[width=2.5in, angle=90]{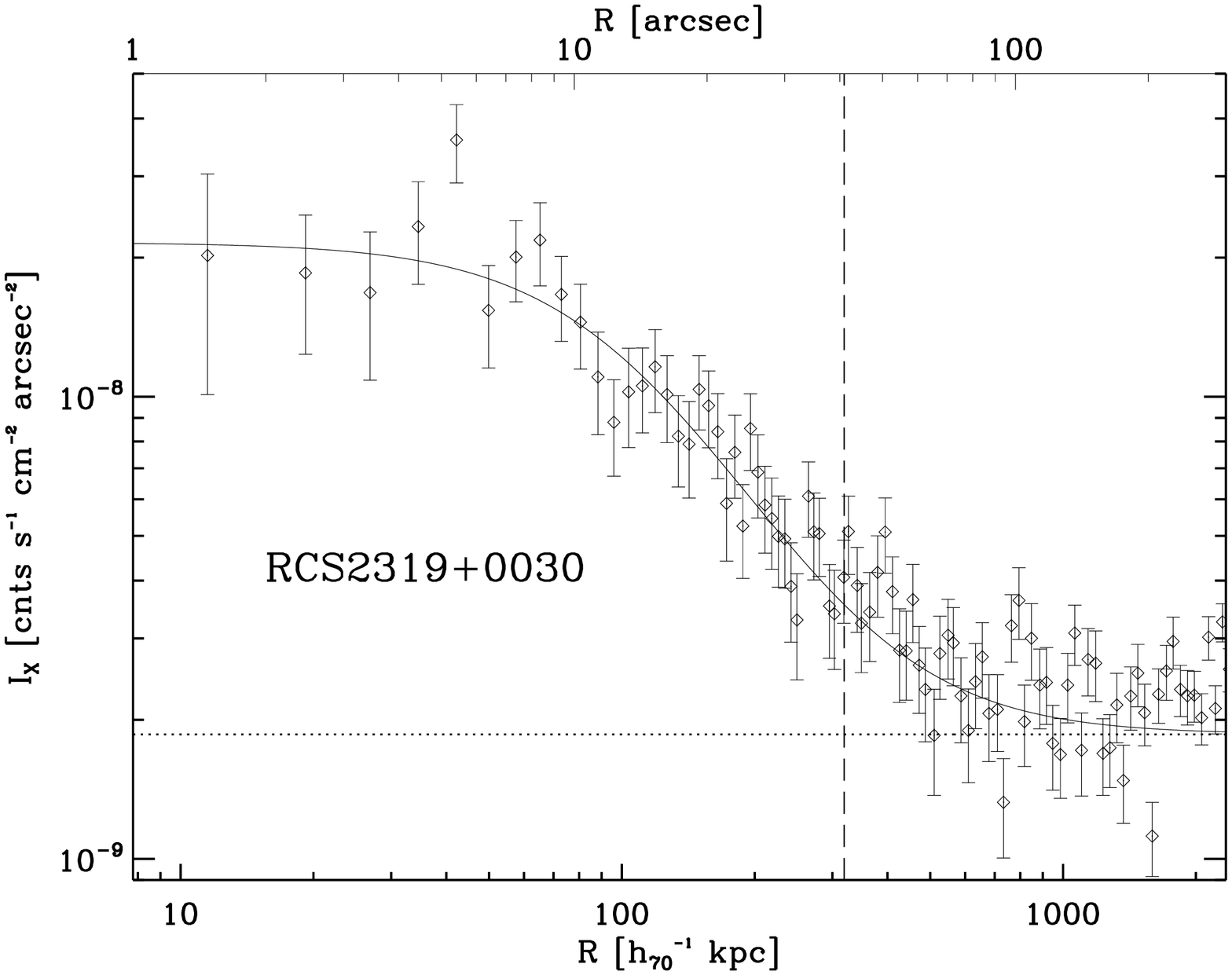}}
\centerline{\includegraphics[width=2.5in, angle=90]{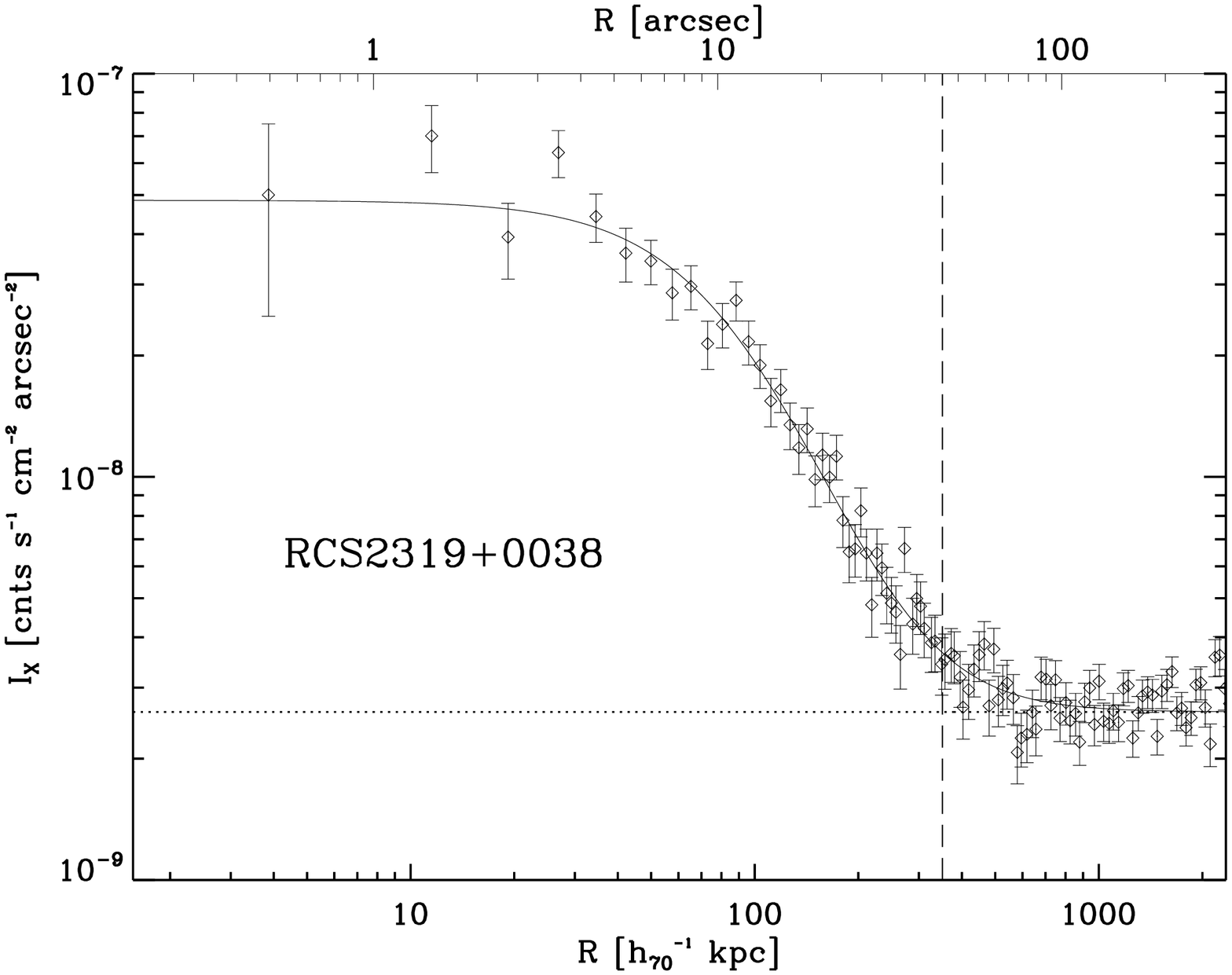}
\includegraphics[width=2.5in, angle=90]{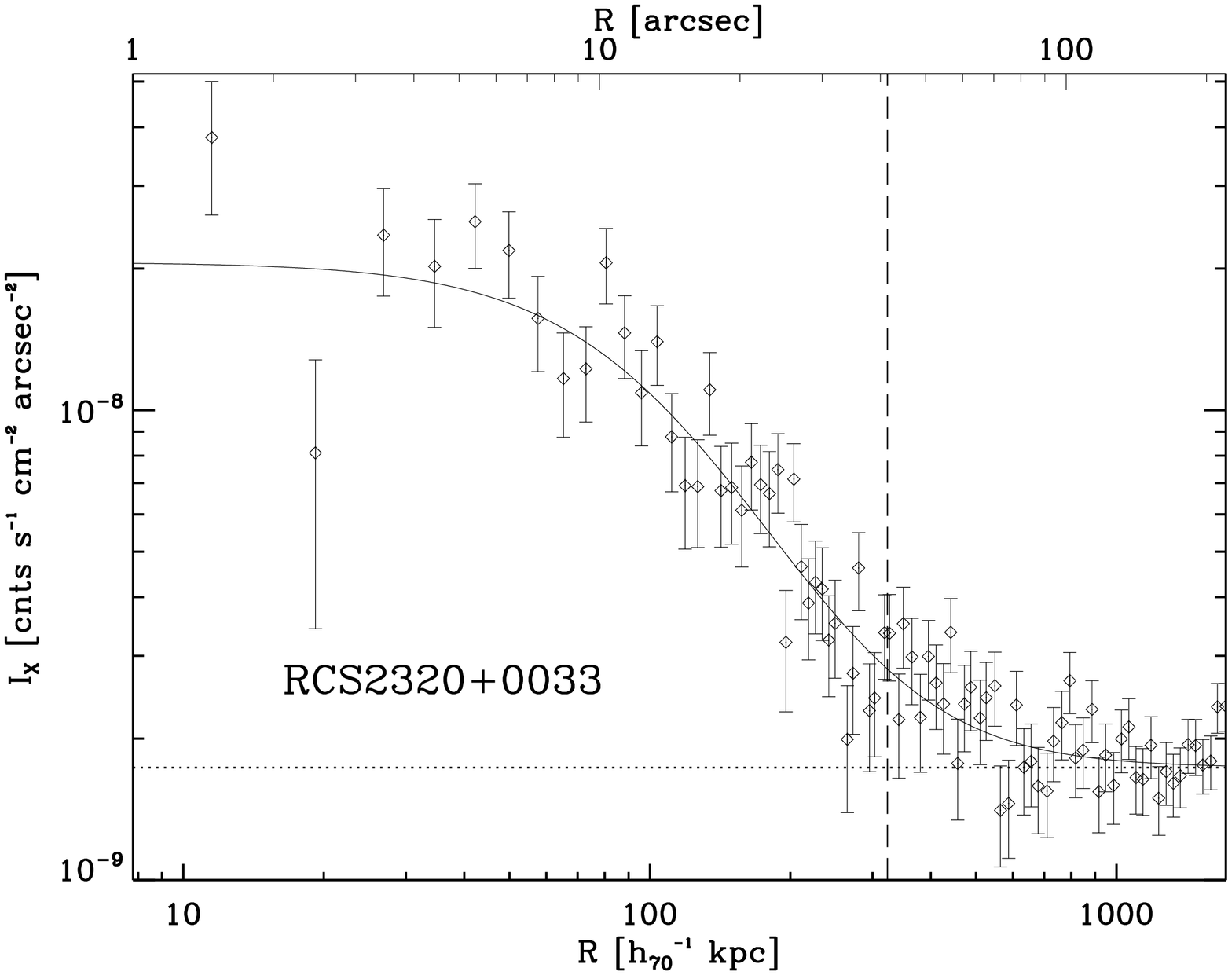}}
\caption{{\bf{Surface Brightness Profiles.}}  Radial surface brighness profiles for the 0.3-7.0 keV band accumulated in annular bins for ten clusters in our sample.  
A solid line traces the best fitting $\beta$ model of each cluster.  
Horizontal dotted lines represent best fit background values, and vertical dashed lines indicate R$_{2500}$.  
Many of the profiles exhibit some substructure; however, most were reasonably well fit by a standard $\beta$ model (see Table~\ref{table3} for goodness of fit data).\label{fig2}}
\end{figure}

\clearpage

\begin{figure}
\centerline{\includegraphics[width=2.5in, angle=90]{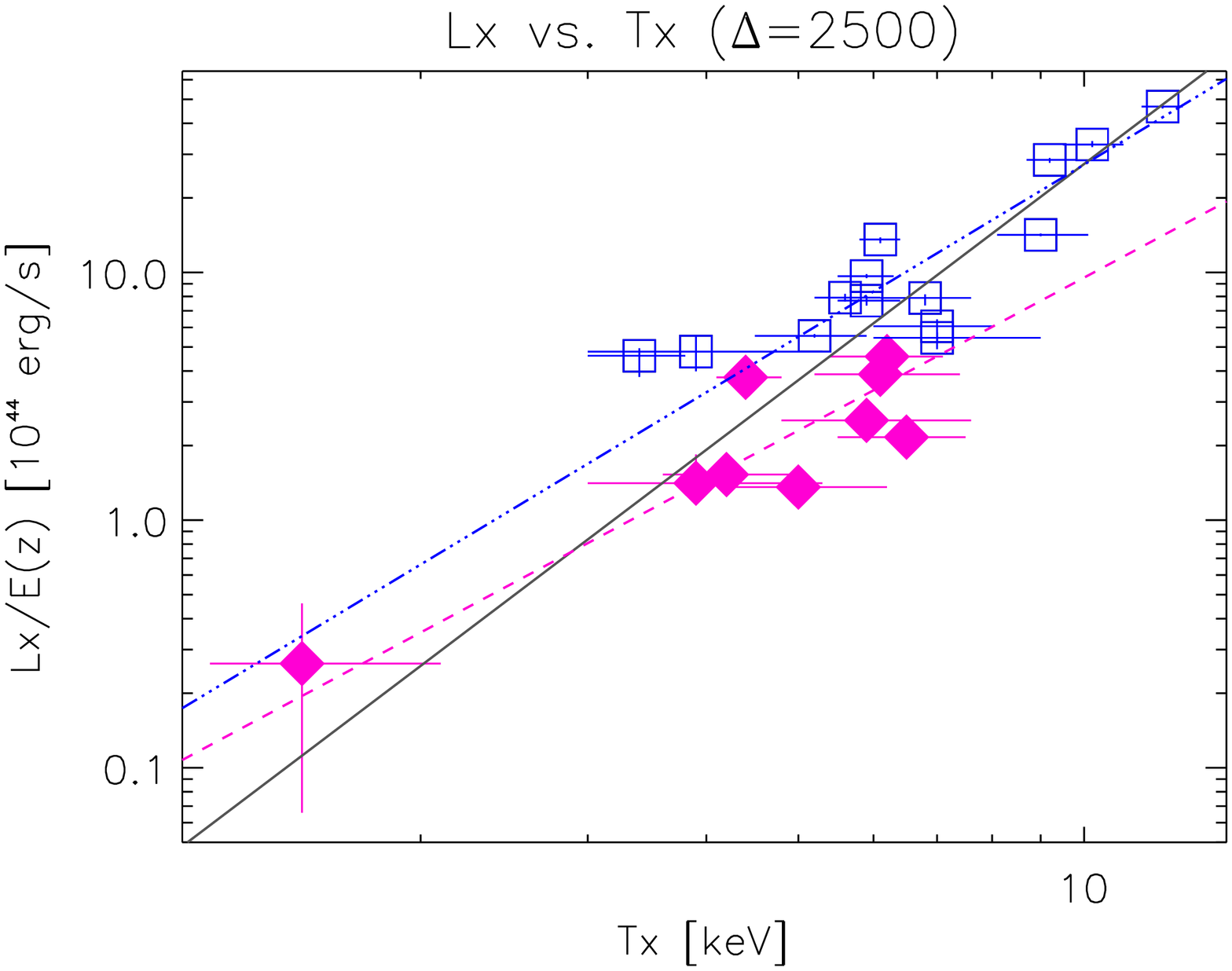}
\includegraphics[width=2.5in, angle=90]{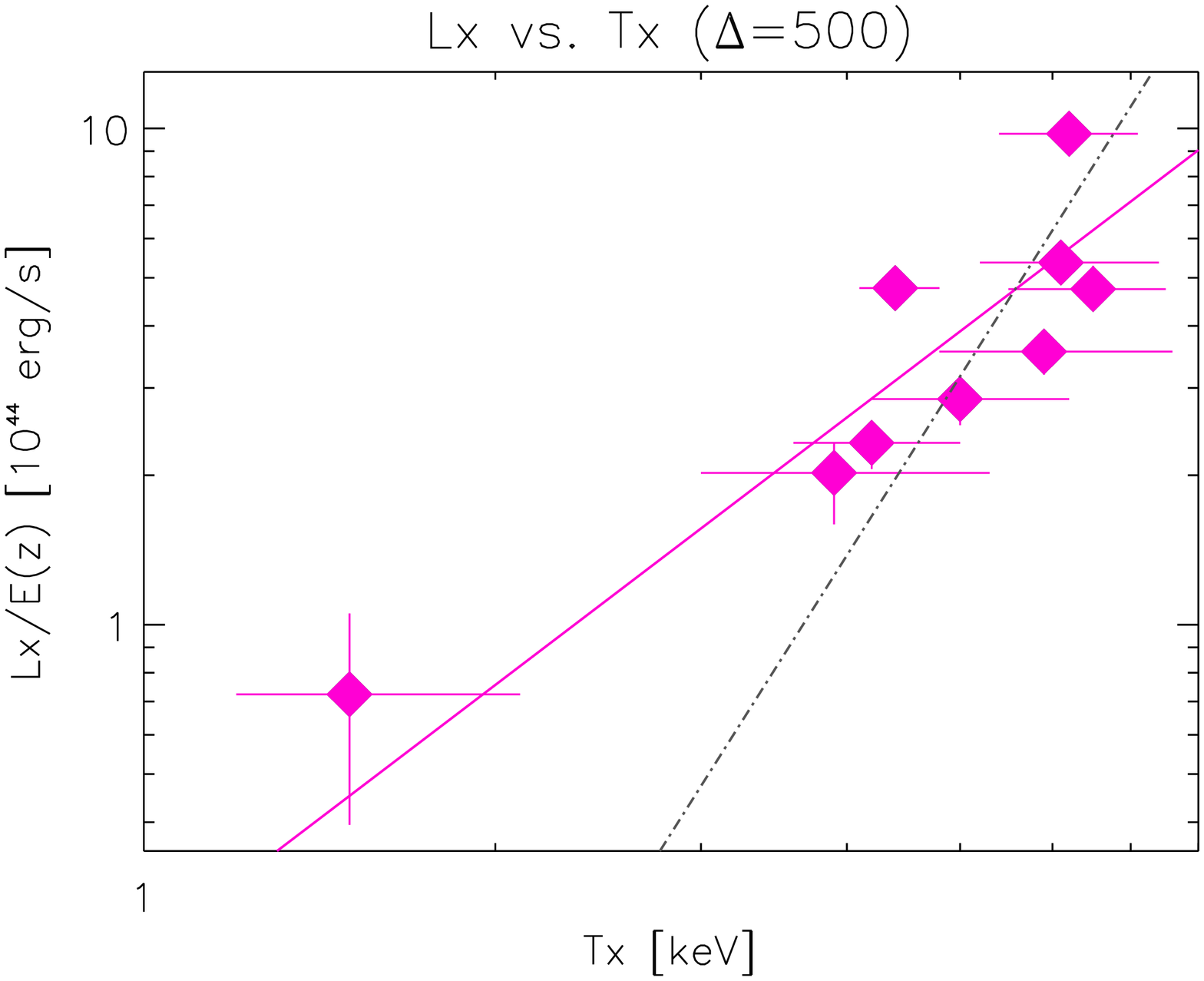}}
\caption{{\bf{Lx-Tx Relationships.}}  {\it{Left panel:}}  X-ray temperatures are plotted against cosmologically corrected unabsorbed bolometric luminosities within $\rm{R}_{2500}$.  
Squares designate moderate redshift CNOC clusters ($z_{\rm{avg}}=0.32$), and diamonds represent high-$z$ RCS clusters ($z_{\rm{avg}}=0.80$).  
The dashed line traces the best fitting relationship for only the RCS clusters, which has a slope of $2.05\pm{0.3}$ and the dot-dash line denotes the best fit to the CNOC data with a powerlaw slope of $2.3\pm{0.3}$, both in agreement with self-similar expectations.  
The solid line indicates the best fitting relationship for the entire sample, with a slope of $2.9\pm{0.3}$, inconsistent with the self-similar value, but in marginal agreement with~\citet{ettori04} who find $3.7\pm{0.5}$ for a cluster ensemble with $0.4<z<1.3$. 
{\it{Right panel:}}  $L_X$ vs. $T_X$ at $\Delta=500$.  The solid line denotes our best-fitting relationship for the RCS clusters with slope $1.8\pm{0.4}$, again consistent with self-similar scaling.  
The dot-dash line shows the fit of~\citet{ettori04}, which was also measured within R$_{500}$.  Seven of our nine objects lie on their relationship, suggesting at least some agreement between the properties of their X-ray and our optically-selected samples.\label{fig3}}
\end{figure}

\clearpage
\begin{figure}
\centerline{\includegraphics[width=2.5in, angle=90]{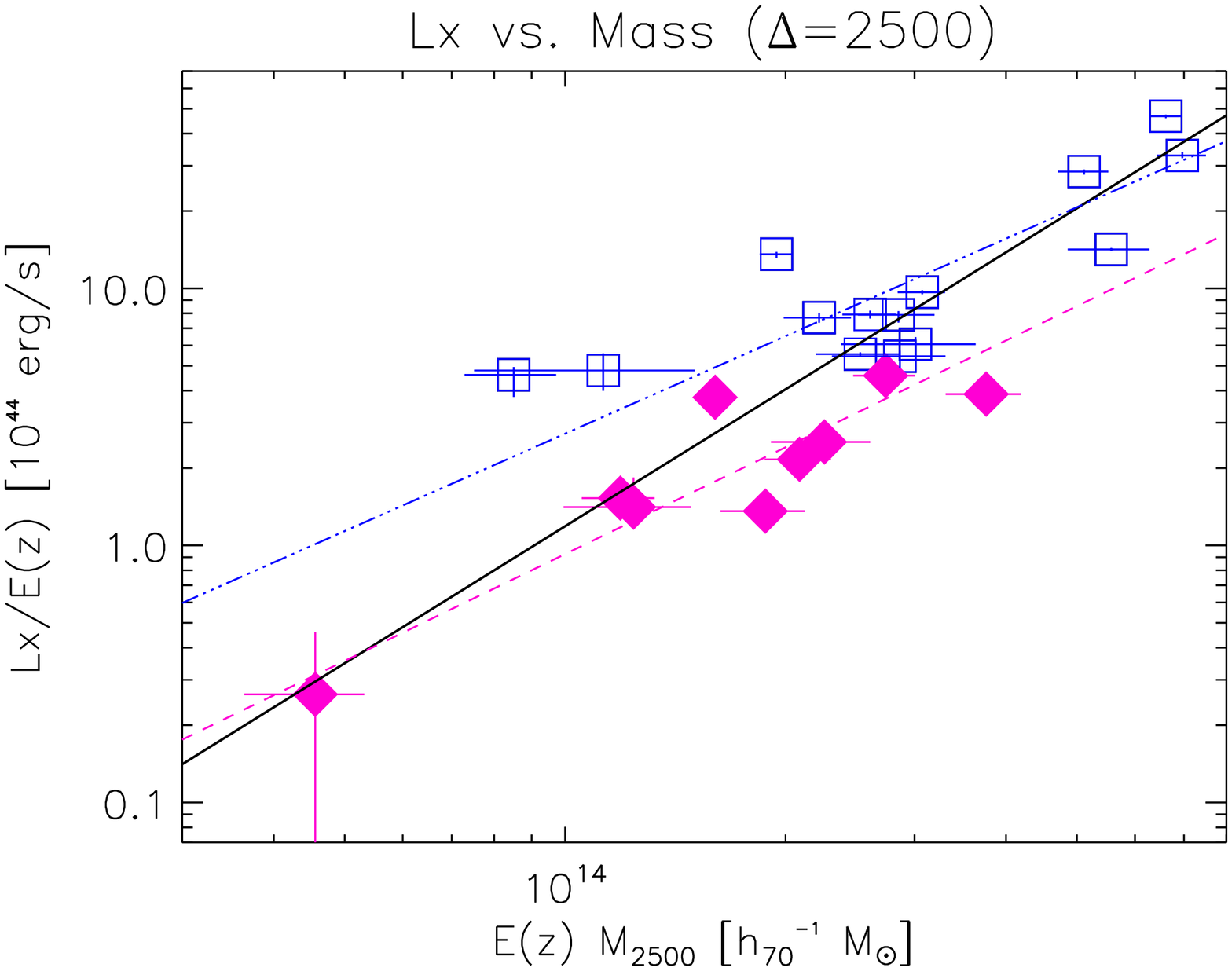}
\includegraphics[width=2.5in, angle=90]{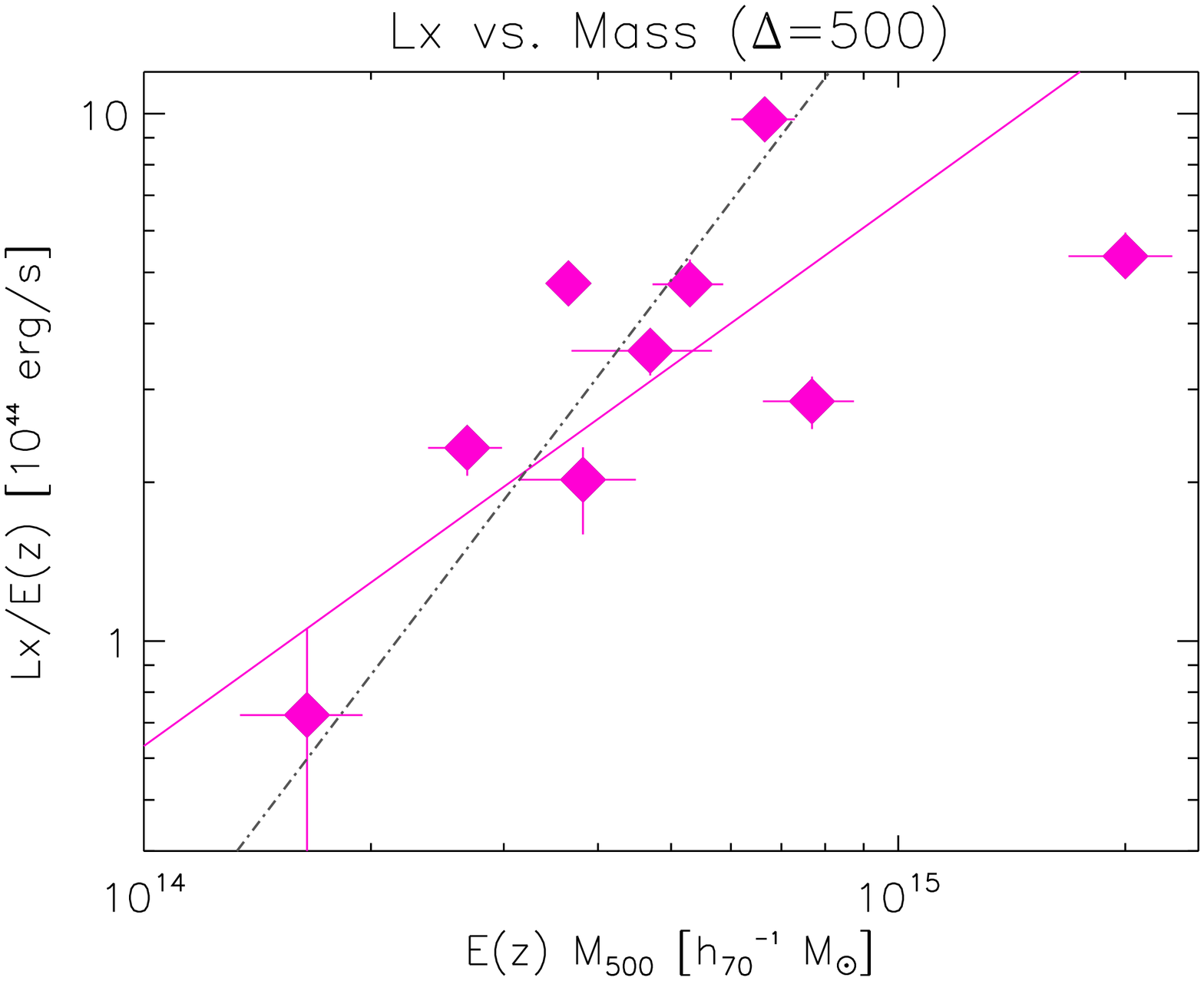}}
\caption{{\bf{$L_X-$M$_{\rm{tot}}$ Relationships.}}  {\it{Left panel:}}  X-ray mass is plotted against cosmologically corrected unabsorbed bolometric luminosity within $\rm{R}_{2500}$.  
Diamonds designate high redshift RCS clusters, and squares represent moderate redshift CNOC clusters.  
The dashed line traces the best fitting relationship for only the RCS clusters, which has a slope of $1.4\pm{0.1}$, while the dot-dash line denotes the best fit to the CNOC data with a powerlaw slope of $1.3\pm{0.2}$, both again in excellent agreement with the self-similar slope of 1.33.  
The solid line indicates the best fitting relationship for the combined sample, which again has a higher slope of $1.77\pm{0.15}$.  
{\it{Right panel:}}  Our $L_X$-M$_{500}$ data is plotted with both our relationship (solid line; slope $1.03\pm{0.28}$) and that of~\citet{ettori04} (dot-dash; slope $1.88\pm{0.42}$) overlayed.\label{fig4}}
\end{figure}

\clearpage
\begin{figure}
\centerline{\includegraphics[width=5in, angle=90]{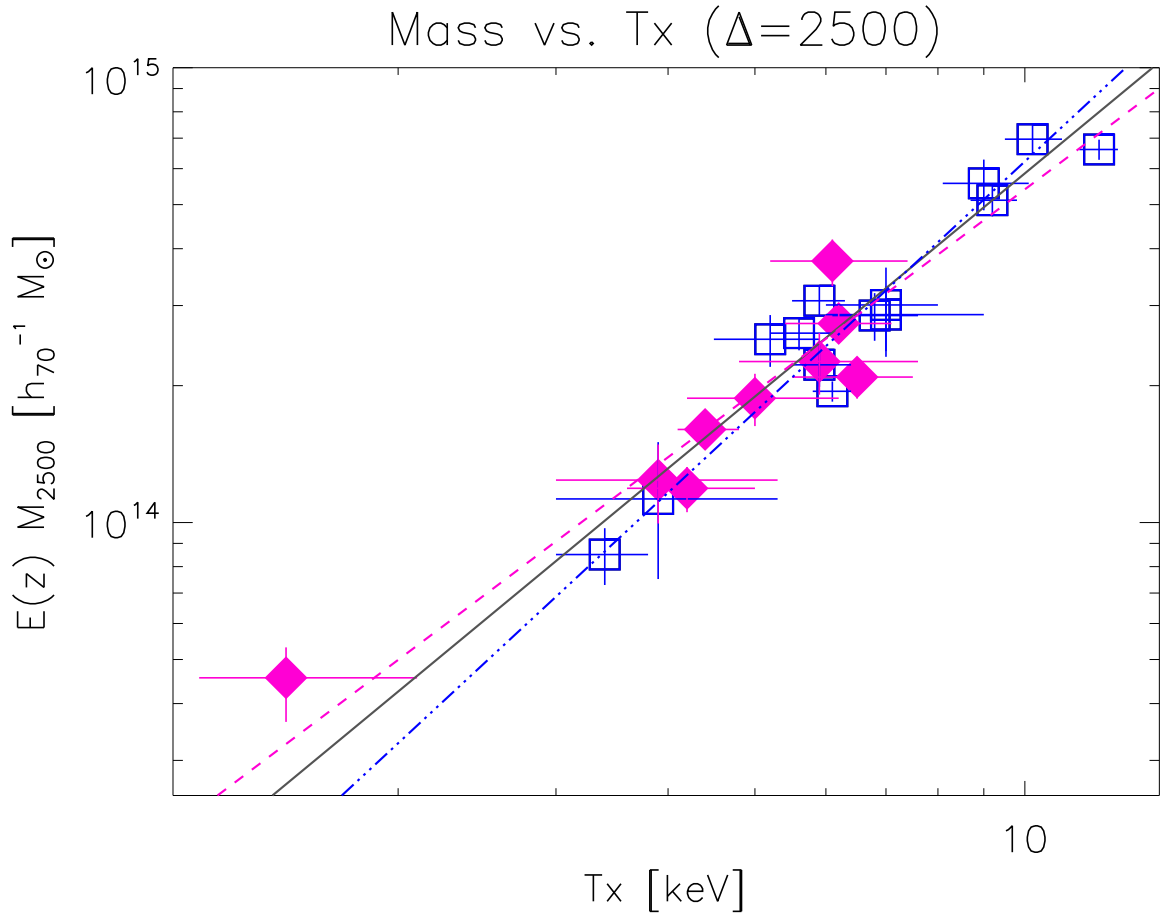}
}
\caption{{\bf{M$_{\rm{tot}}-T_X$ Relationship.}}  
%{\it{Left panel:}}  
X-ray temperatures are plotted against cosmologically corrected mass estimates from Section~\ref{s:mass}.  
Squares designate the CNOC clusters ($0.1<z<0.6$), and diamonds represent RCS clusters ($0.6<z<1.0$).  
The dashed line indicated the RCS fit, with a slope of $1.5\pm{0.3}$.  
Fits to the entire sample (solid line) are also conisistent with self-similar evolution (slope $1.6\pm{0.2}$).  
The CNOC fit has a higher slope (dot-dash; $1.83\pm{0.13}$), but is in good agreement with those from the literature~\citep{allen01,arnaud05} \label{fig5}}

%{\it{Right panel:}}  RCS M$_{500}$-$T_X$ data with best fit (solidd line) overlayed.  The dot-dash line shows the $\Delta=500$ fit of~\citet{ettori04}.  
%Though the slopes are in agreement, the RCS normalization is significantly higher (though still in agreement with other fits in the literature (Table~\ref{table10}).  
%The high normalization is primarily due to two clusters in our sample with comparatively higher $\beta$ values, possibly indicating that they are less relaxed systems.\label{fig5}}
\end{figure}

\clearpage

\begin{figure}
\centerline{\includegraphics[width=2.5in, angle=90]{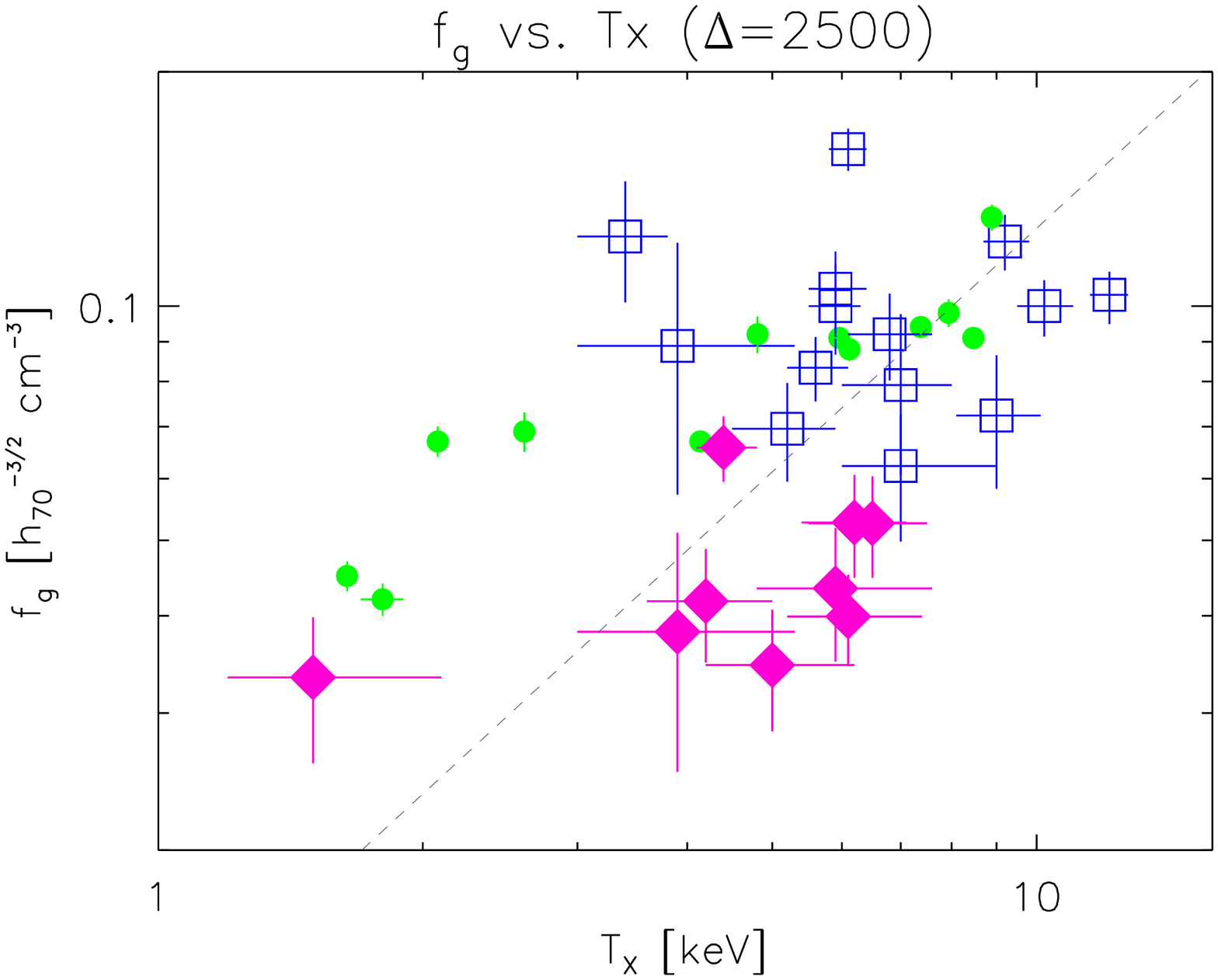}
\includegraphics[width=2.5in, angle=90]{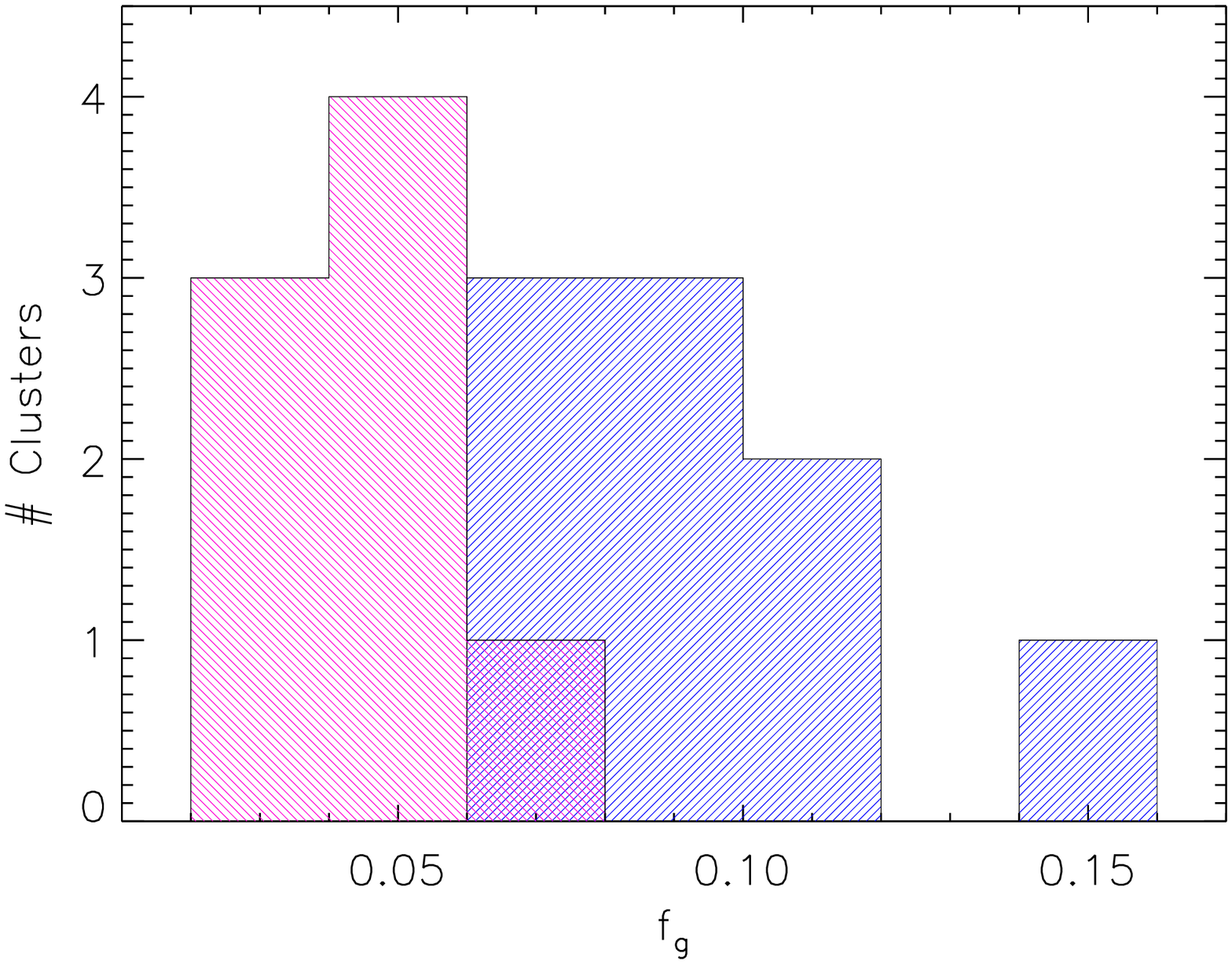}}
\caption{{\bf{Gas Mass Fractions.}}
{\it{Left panel:}}
 $T_X$ is plotted against gas mass fractions within ${\rm{R}}_{2500}$.  
Squares designate moderate redshift CNOC clusters ($0.1<z<0.6$), and diamonds represent higher-$z$ RCS clusters ($0.6<z<1.0$).  
The solid line indicates the best fitting relationship for the entire sample, with 
%%a normalization of $C_1=-1.94\pm{0.2}$,
 a slope of $1.0\pm{0.2}$.  Circles indicate points taken from~\citet{vikhlinin06}.
%%, and $\sigma_{{\rm{log}}{fg}}=0.30$.  
%%This slope is noticeably higher than that reported in studies of lower redshift samples~\citep[where $C_2({\rm{R}}_{500})=0.34\pm{0.22}$ and $C_2({\rm{R}}_{2500})=0.61\pm{0.31}$, respectively]{mohr99,ettori02}, possibly indicating redshift evolution in the $f_g$-$T_X$ relationship.
{\it{Right panel:}}  ~Histogram of gas mass fractions for the eight RCS (left side) and nine CNOC clusters (right side) with $3.5<T_X<8$ keV. A K-S test performed on these two samples resulted in D=0.875 and P=0.002, indicating that the gas mass fractions of the samples are different at $> 99\%$ confidence.
 \label{fig6}}
\end{figure}

\clearpage
\begin{figure}
\centerline{\includegraphics[width=5in, angle=90]{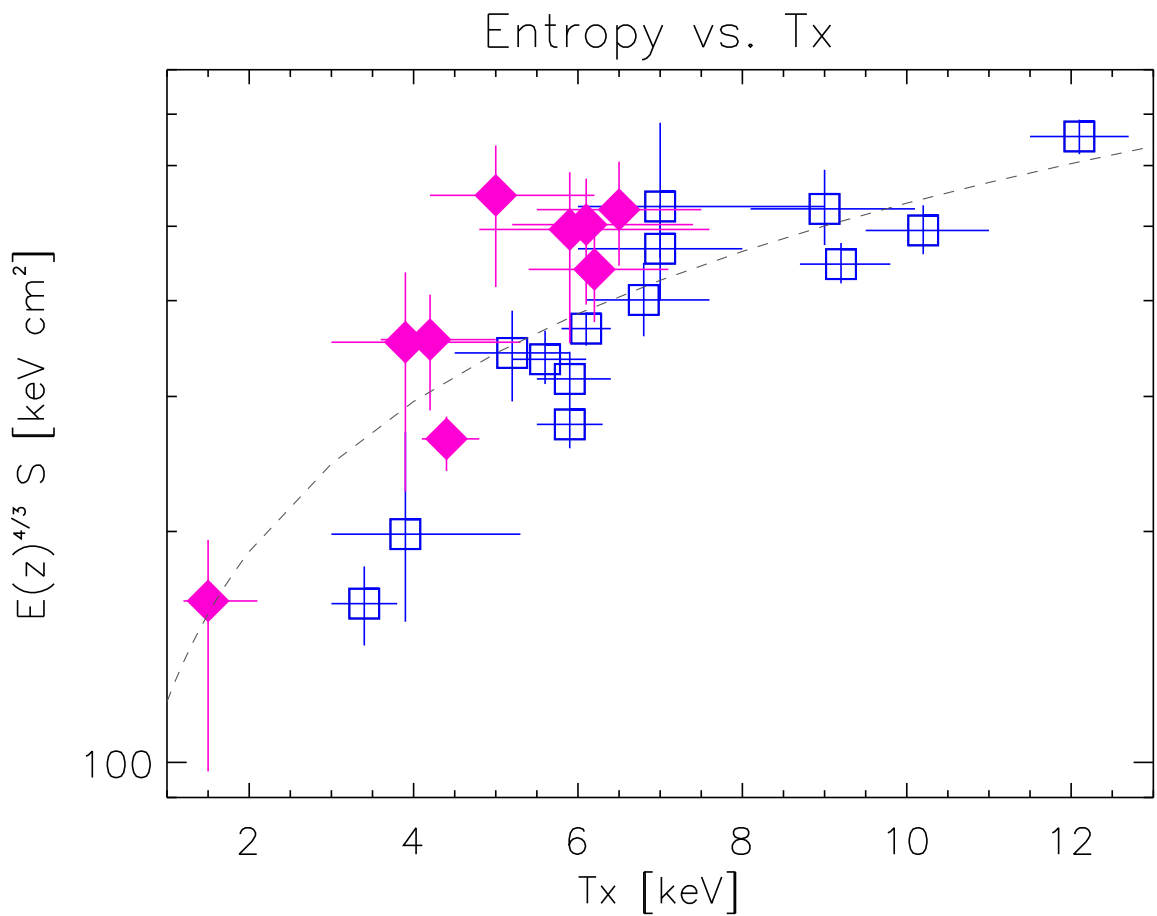}
}
\caption{{\bf{Cluster Entropy.}} 
%{\it{Left panel:}}  
Entropy measured at 0.1R$_{200}$ is plotted against X-ray temperature for the clusters in this study.
RCS clusters (diamonds) have slightly higher entropies for a given $T_X$ than CNOC clusters (squares), though probably not enough to account for the whole of the discrepancies seen in gas fractions between the samples.
The dashed line indicates the relationship of~\citet{ponman03}.\label{fig7}}
\end{figure}

\clearpage
\begin{figure}
\centerline{\includegraphics[width=5in, angle=90]{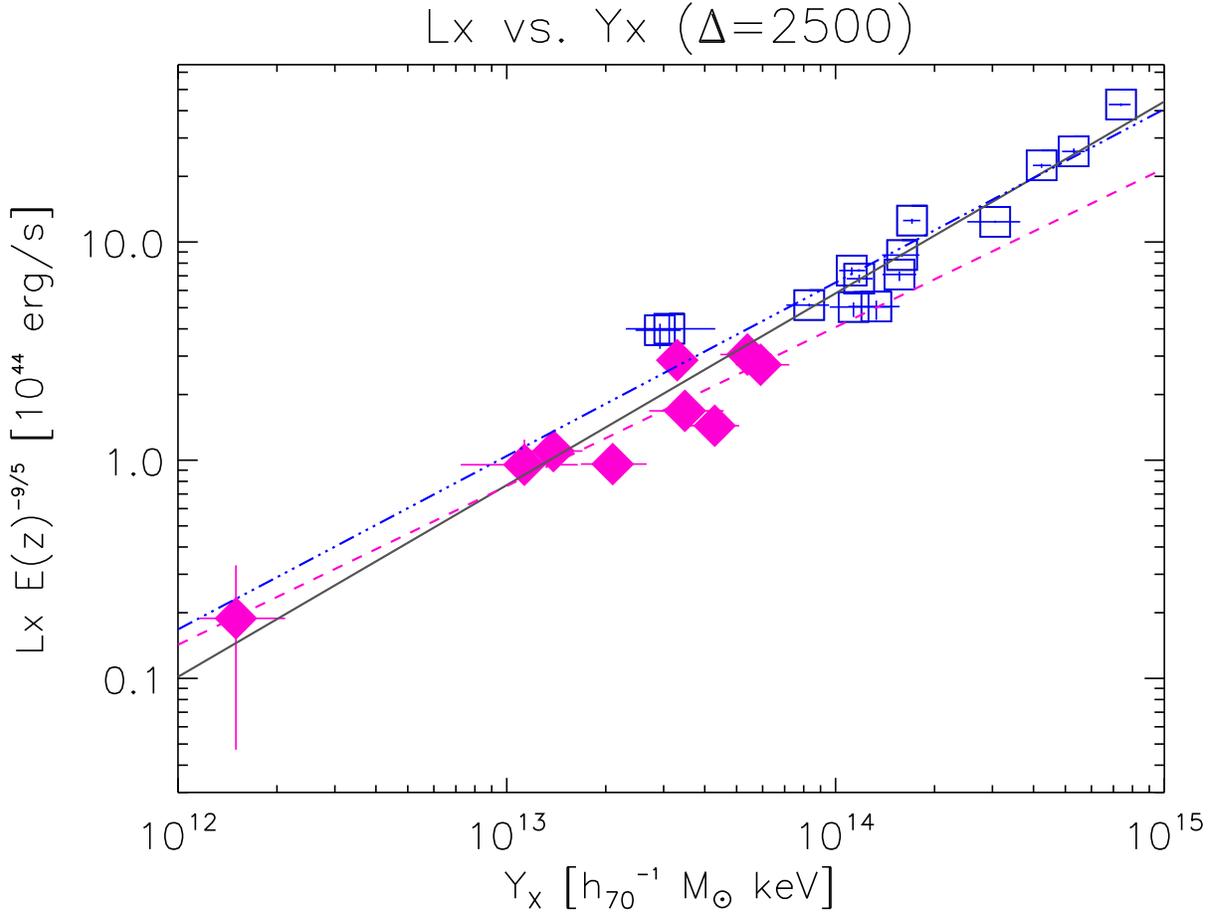}
}
\caption{{\bf{$L_X$-$Y_X$ Relationship.}} 
%{\it{Left panel:}}  
$Y_X$ is plotted against X-ray luminosity within $\rm{R}_{2500}$.  
Though CNOC (squares; dash-dot line) and RCS (diamonds; dashed line) slopes agree, their normalizations are inconsistent, probably due to differences in gas mass fractions between the two samples.  
The solid line indicates the best fitting relationship for the entire sample.  Overall this is the tightest relationship involving $L_X$ that we investigate in this work.\label{fig8}}
%{\it{Right panel}}  $L_X$ vs. $Y_X$ for R$<$R$_{500}$.  
%Our best fit relationship to the RCS sample (solid line) is plotted against that%of~\citet{maughan07} (dot-dash line).  
%Our fit agrees with theirs in neither slope nor normalization.\label{fig8}}
\end{figure}

\clearpage

\begin{figure}
\centerline{\includegraphics[width=2.5in, angle=90]{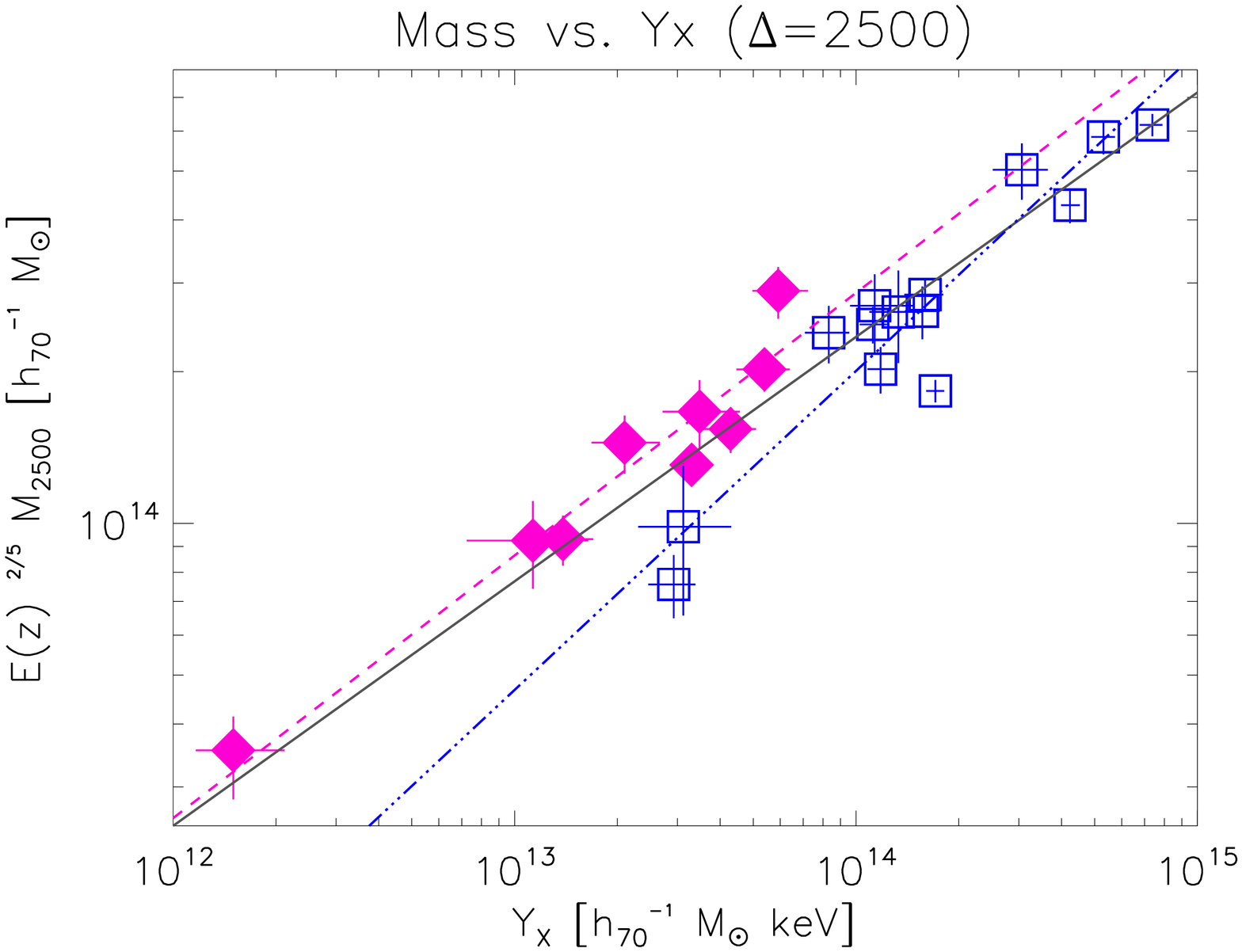}
\includegraphics[width=2.5in, angle=90]{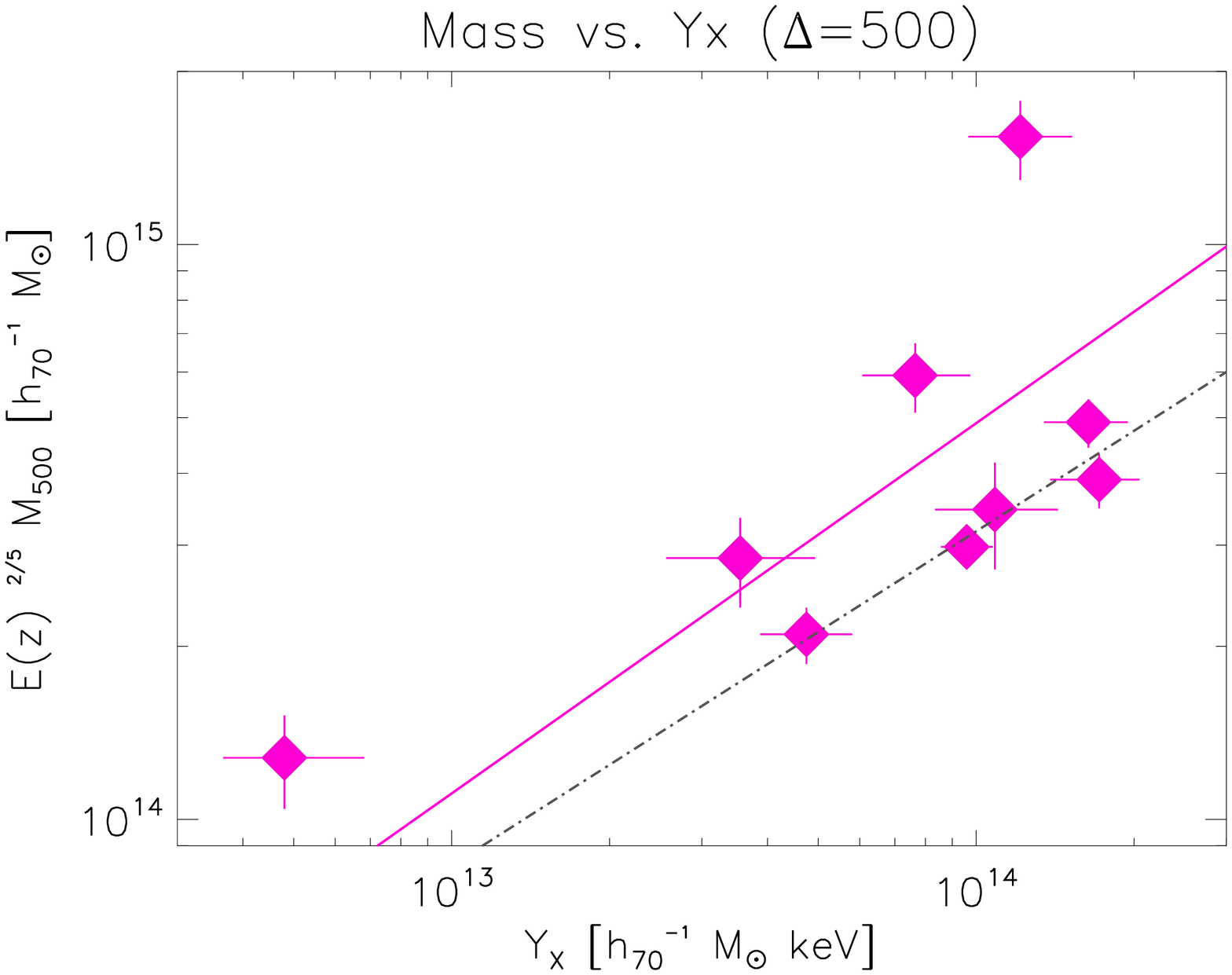}}
\caption{{\bf{M$_{\rm{tot}}-Y_X$ Relationship.}}  {\it{Left panel:}}  $Y_X$ is plotted against mass estimates within R$_{2500}$.  
Here we find marginal agreement between the slopes of our individual fits: CNOC (squares; dot-dash line) and RCS (diamonds; dashed line), and the slope of~\citet{kravtsov06}.  
{\it{Right panel:}}  The $Y_X-$M$_{\rm{tot}}$ relationship is fit for RCS clusters at $\Delta=500$ (solid line).  
In a direct comparison with~\citet{kravtsov06} (dot-dash), our slopes are in agreement but normalizations are inconsistent.  The points that do not lie on their relationship
are also the clusters  that have the lowest gas mass fractions in our sample.
Bear in mind that we are extrapolating to get out to R$_{500}$, as our data mostly lie within R$_{2500}$.\label{fig9}}
\end{figure}

\clearpage
\begin{figure}
\centerline{\includegraphics[angle=90,width=6in]{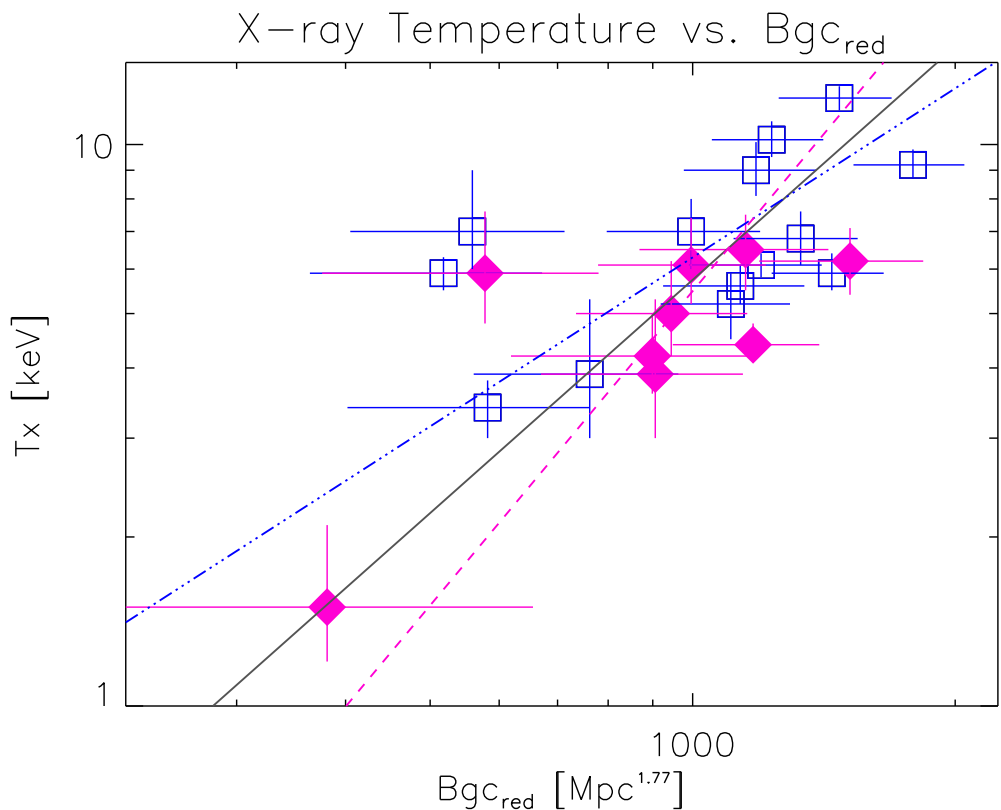}}
\caption{{\bf{$T_X$ vs. $\rm{B}_{gc,red}$.}}  A log-log plot of $T_X$ vs. $\rm{B}_{gc,red}$ within R$_{2500}$ for the CNOC (squares) and RCS (diamonds) samples.  Error bars represent 68\% confidence intervals.  
The fit to the combined sample (solid line) is in agreement with both the CNOC (dot-dash) and RCS (dashed) fits.  
Both the CNOC and combined fits are also consistent in slope with the expected value of 1.11~\citep{yee03}.  
These relationships, on average, show the lowest scatter of any richness relationships investigated in this work, with a lower average scatter even than $L_X$-$T_X$.  
The scatter of the RCS fit is mostly driven by the outlying point (RCS2320+0033) which has a very low \bgc for its mass.\label{fig10}}
\end{figure}

\clearpage

\begin{figure}
\centerline{\includegraphics[width=2.5in,angle=90]{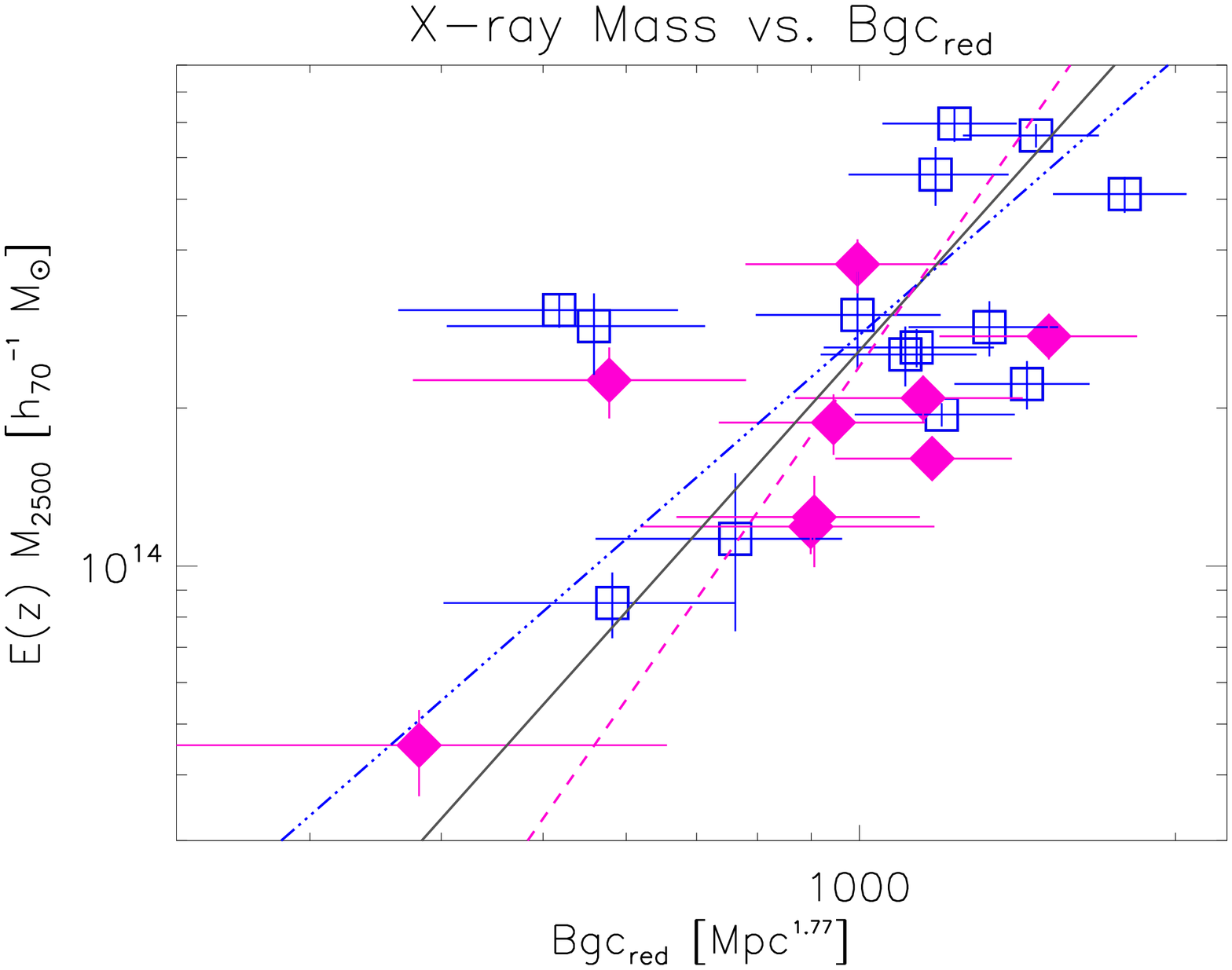}
\includegraphics[width=2.5in,angle=90]{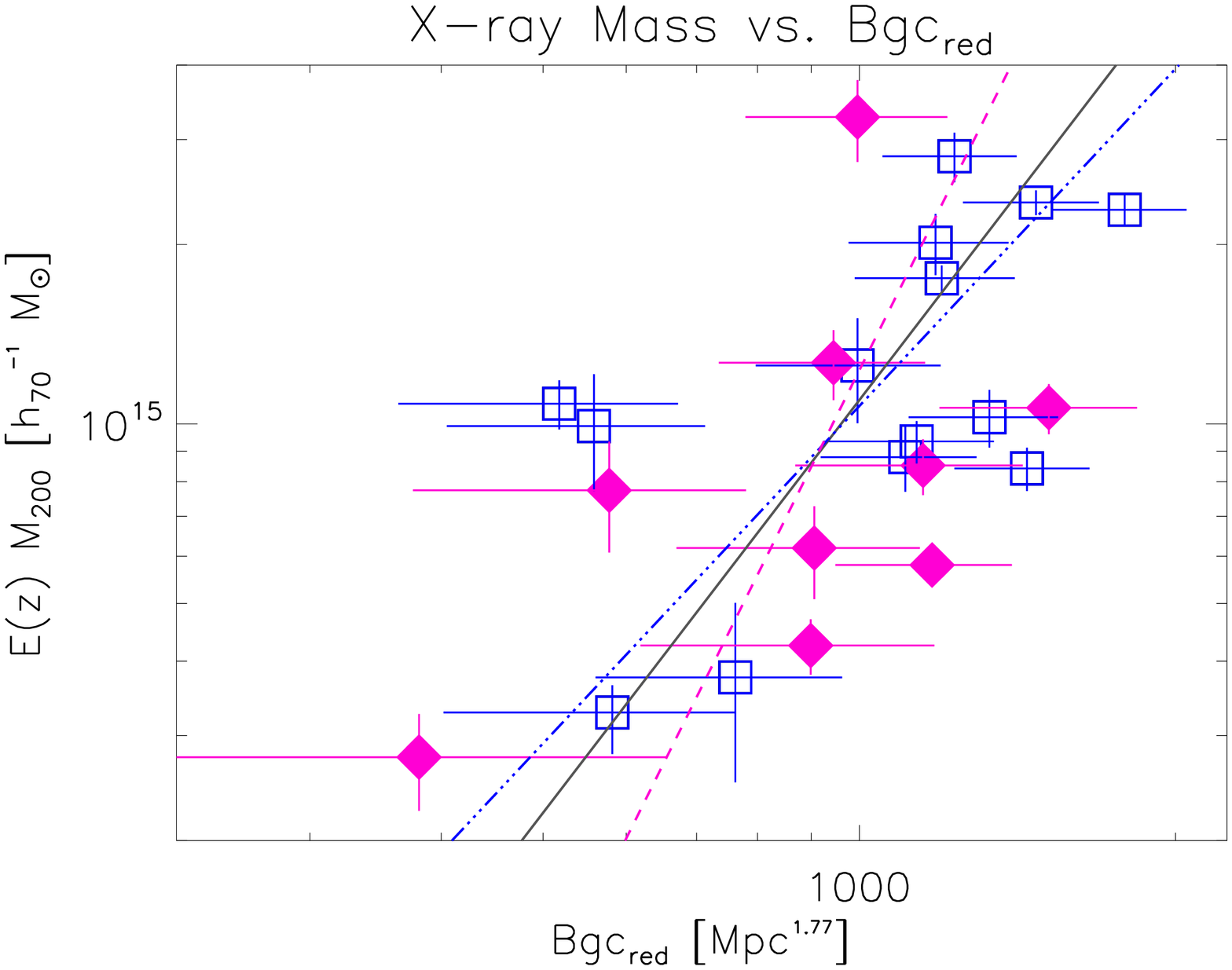}}
\caption{{\bf{X-ray Mass vs. $\rm{B}_{gc,red}$.}}  {\it{Left panel:}}  Total mass is plotted against $\rm{B}_{gc,red}$ for R$<$R$_{2500}$.  
Error bars represent 68\% confidence intervals.  
CNOC clusters are shown as squares, and diamonds designate the RCS sample.  
The dashed line shows the best relationship for the RCS sample, while dot-dash and solid lines indicate fits to the CNOC and combined samples, respectively.  
The average scatter in the relationships is $\sigma_{\rm{log Y}}=28\%$; however, all fits are consistent with one another due to large errors.  
{\it{Right panel:}}  \bgc~vs. M$_{200}$.  X-ray masses were extrapolated to R$_{200}$ for comparison with the relationship of~\citet{blindert07}.  
All three of our fits are consistent with their relationship, which was determined via dynamical investigations of 33 moderate redshift RCS clusters.  
Scatter in our fits averages to $32\%$\label{fig11}}
\end{figure}

\clearpage

\begin{figure}
\centerline{\includegraphics[angle=90,width=6in]{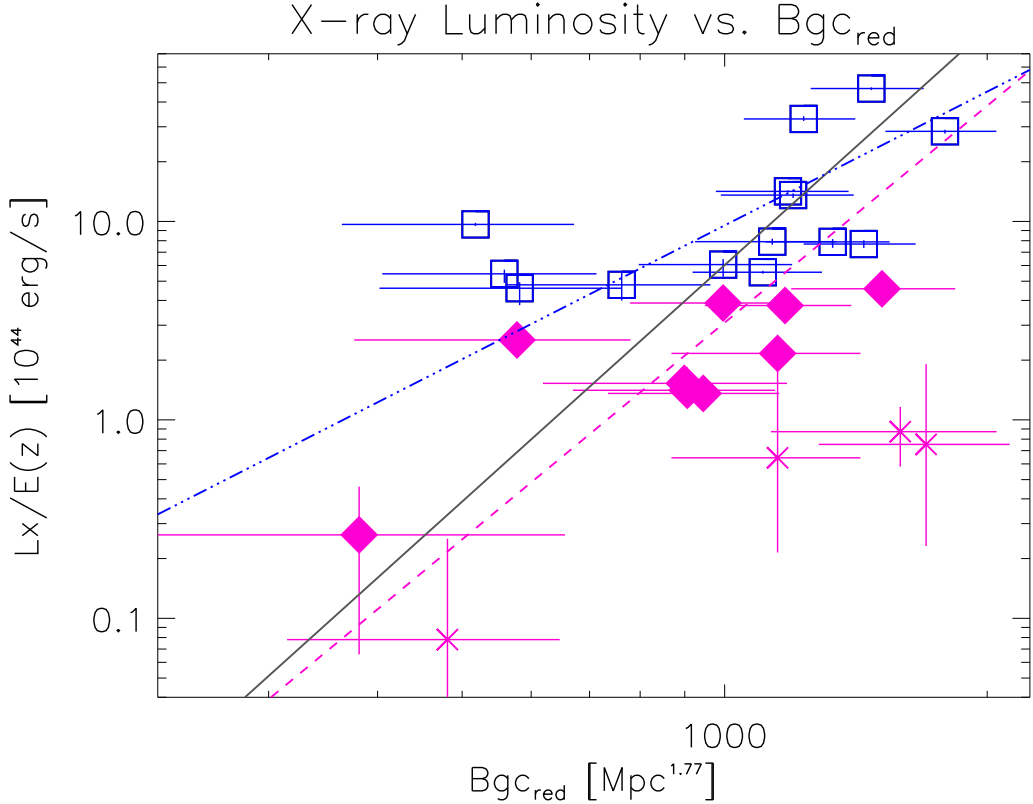}}
\caption{{\bf{$L_X$ vs. $\rm{B}_{gc,red}$.}}  A log-log plot of $L_X$ vs. $\rm{B}_{gc,red}$ for the combined CNOC/RCS sample.  
CNOC clusters are shown as squares, and diamonds designate RCS clusters that are included in fitting.  
Asterixs indicate the four clusters that were not included in fitting due to non-detection (point farthest to the left), known superposition (RCS0439-2904; point second from the right), or insufficient counts (remaining two points).  
Luminosities for these objects (excluding RCS0439-2904) were determined in XSPEC assuming a temperature of $4\pm{2}$ keV.  
Error bars represent 68\% confidence intervals.  
A dashed line shows the RCS sample fit, while dash-dot and solid lines indicate CNOC and combined sample fits, respectively.  
On average, these relationships show the most scatter of all those that we investigate in this work.\label{fig12}}
\end{figure}

\clearpage

%% TABLES

\begin{deluxetable}{ccccccc}
\tablecolumns{7}
\tablewidth{0pt}
\tablecaption{Cluster Sample\label{table1}}
\tablehead{
\multicolumn{2}{c}{Cluster} & 
\colhead{{\rm{z}}} & 
\colhead{$1\arcsec$} &
\colhead{obsid} &
\colhead{Individual Exposure} &
\colhead{Total Exposure}\\
\multicolumn{2}{c}{} & 
\colhead{} &
\colhead{[$h_{70}^{-1}$ kpc]} &
\colhead{} & 
\colhead{[seconds]} &
\colhead{[seconds]} 
}
\startdata
\multicolumn{2}{l}{RCS022434-0002.5} & 0.778\tablenotemark{a} &  7.44  & 3181 &  12051 & 100844\\
\multicolumn{2}{l}{} & {} &{}  & 4987  & 88793 & \\
\multicolumn{2}{l}{RCS043938-2904.7} &  0.960\tablenotemark{b}&  7.93    & 3577 &  64507 & 93263\\
\multicolumn{2}{l}{} & {} &{}  & 4438 & 28756 & \\
\multicolumn{2}{l}{RCS110723-0523.3} & 0.735\tablenotemark{c} & 7.28     & 5825 & 49466 &  94058\\
\multicolumn{2}{l}{} & {} &{}  & 5887  & 44592 \\
\multicolumn{2}{l}{RCS132631+2903.1} &  0.75\tablenotemark{d}& 7.34    & 3291  & 30907 &65499 \\
\multicolumn{2}{l}{} & {} &{} & 4362 &  34592 & \\
\multicolumn{2}{l}{RCS141658+5305.2} &  0.968\tablenotemark{c}&  7.95   & 3239\tablenotemark{e}  & 62824  & 62824\\
\multicolumn{2}{l}{RCS141910+5326.2} &  0.62\tablenotemark{f}&  6.79   & 3240  & 9904 & 57307\\
\multicolumn{2}{l}{} & {} &{}  & 5886  & 47403 & \\
\multicolumn{2}{l}{RCS162009+2929.4} &  0.870\tablenotemark{c}&  7.71   & 3241  & 35953 & 35953 \\
\multicolumn{2}{l}{RCS211223-6326.0} &  1.099\tablenotemark{g}&  8.17   & 5885  & 70520 & 70520\\
\multicolumn{2}{l}{RCS215641-0448.1} &  1.080\tablenotemark{g,h}& 8.14    & 5353  & 36558 & 71259\\
\multicolumn{2}{l}{} & {}&{}  & 5359 & 34701 & \\
\multicolumn{2}{l}{RCS231831+0034.2} &  0.78\tablenotemark{f}& 7.44    & 4938  & 50454 & 50454\\
\multicolumn{2}{l}{RCS231953+0038.0} &  0.900\tablenotemark{i}& 7.79    & 5750  & 20902 & 74539\\
\multicolumn{2}{l}{(RCS231948+0030.1)} & (0.904)\tablenotemark{i} & (7.80)  & 7172  & 17947 & \\
\multicolumn{2}{l}{(RCS232002+0033.4)} & (0.901)\tablenotemark{i} & (7.79)  & 7173  & 20899 & \\
\multicolumn{2}{l}{} & {} &{}  & 7174  & 14791 & \\
\enddata

\tablenotetext{a}{\citet{hicks07}}
\tablenotetext{b}{\citet{cain}}
\tablenotetext{c}{\citet{gilbank07}}
\tablenotetext{d}{From photometric data~\citep{gladders05}.  This cluster may be at $z\sim1.01$ (see text for explanation).}
\tablenotetext{e}{ACIS-I observation.}
\tablenotetext{f}{From X-ray spectra (this work), see text.}
\tablenotetext{g}{\citet{barrientosprep}}
\tablenotetext{h}{ID uncertain~\citep[see][]{barrientosprep}.}
\tablenotetext{i}{\citet{gilbankprep}}

\end{deluxetable}

\clearpage

\begin{deluxetable}{ccllllccc}
\tabletypesize{\small}
\tablecolumns{9}
\tablewidth{0pt}
\tablecaption{Cluster Positions and Detection Details\label{table2}}
\tablehead{
\multicolumn{2}{c}{Cluster} & 
\multicolumn{2}{c}{Optical Position\tablenotemark{a}} &
\multicolumn{2}{c}{X-ray Position\tablenotemark{a}} &
\colhead{Separation} &
\colhead{Net Counts\tablenotemark{b}} &
\colhead{S/N Ratio}   \\
\multicolumn{2}{c}{} & 
\colhead{RA} &
\colhead{Dec} &
\colhead{RA} &
\colhead{Dec} &
\colhead{[$\arcsec$]}&
\colhead{} &
\colhead{} 
}
\startdata
\multicolumn{2}{l}{RCS0224-0002} & 02:24:34.1 & -00:02:30.9 & 02:24:34.2 & -00:02:26.4 &  4.7 & 1102 &16.2 \\
\multicolumn{2}{l}{RCS0439-2904} & 04:39:38.0 & -29:04:55.2 & 04:39:37.6 & -29:04:50.3 &   7.2 & 461 &6.5    \\
\multicolumn{2}{l}{RCS1107-0523} & 11:07:23.4 & -05:23:13.7 &  11:07:24.0& -05:23:20.7 &  11.4 & 1056 &15.5   \\
\multicolumn{2}{l}{RCS1326+2903\tablenotemark{c}} & 13:26:31 & +29:03:12 &  13:26:31.3 &  +29:03:31.0&  19.9 & 181 & 3.1   \\
\multicolumn{2}{l}{RCS1417+5305} & 14:16:59.8 & +53:05:12.2 &  14:17:01.5&  +53:05:16.2& 15.8  & 138 &4.7 \\
\multicolumn{2}{l}{RCS1419+5326} & 14:19:12.1 & +53:26:11.0 &  14:19:12.1&  +53:26:11.6& 0.6  & 2903 &40.2 \\
\multicolumn{2}{l}{RCS1620+2929} & 16:20:10.0 &  +29:29:21.5&  16:20:10.1& +29:29:20.8& 1.5  & 257 &7.1  \\
\multicolumn{2}{l}{RCS2112-6326} & 21:12:23.1 & -63:25:59.5 & \nodata& \nodata & \nodata  & 232 &4.8  \\
\multicolumn{2}{l}{RCS2156-0448} & 21:56:41.2 & -04:48:13.3 &  \nodata & \nodata & \nodata  & 54 &1.1  \\
\multicolumn{2}{l}{RCS2318+0034} & 23:18:31.5 & +00:34:18.0 &  23:18:30.8 & +00:34:02.5 & 19.9  & 1161 &21.5  \\
\multicolumn{2}{l}{RCS2319+0030} & 23:19:48.7 & +00:30:08.5 &  23:19:46.8 & +00:30:14.3 & 29.1  &  780 &17.8 \\
\multicolumn{2}{l}{RCS2319+0038} & 23:19:53.9 & +00:38:11.6 &  23:19:53.2 & +00:38:12.5 & 10.5  & 1742  &26.2 \\
\multicolumn{2}{l}{RCS2320+0033} & 23:20:03.0 & +00:33:25.1 &  23:20:02.1 & +00:32:57.6 & 30.6  & 725 &16.8 \\
\enddata

\tablenotetext{a}{All positions are given for equinox J2000.} 
\tablenotetext{b}{0.3-7.0 keV band, within ${\rm{R}} < 500 ~h^{-1}_{70}~{\rm{kpc}}$}
\tablenotetext{c}{There is also an RCS cluster at 13:26:29, +29:03:06, which is $39.2\arcsec$ from the X-ray centroid (see text for details)}
\end{deluxetable}

\clearpage

\begin{deluxetable}{ccrrrrr}
\tablecolumns{7}
\tablewidth{0pc}
\tablecaption{$\beta$-Model Fits\label{table3}}
\tablehead{ 
\multicolumn{2}{c}{Cluster} & \colhead{$r_{\rm c}$} [$h_{70}^{-1}$ kpc] &
\colhead{$\beta$}  & \colhead{$I_{\rm 0}$\tablenotemark{a}} &
\colhead{$I_{\rm B}$\tablenotemark{a}}  & \colhead{$\chi^2/\rm{DOF}$}}
\startdata 
 \multicolumn{2}{l}{RCS0224-0002}& $180_{-12}^{+13}$ & $0.72_{- 0.04}^{+ 0.04}$ & $7.1_{-0.4}^{+0.4}$ & $2.33_{-0.02}^{+0.02}$ & 210.7/199 \\ 
\multicolumn{2}{l}{RCS0439-2904}& $108_{-7}^{+8}$ & $0.59_{- 0.04}^{+ 0.04}$ & $5.2_{-0.3}^{+0.4}$ & $3.80_{-0.04}^{+0.03}$ & 67.8/64  \\ 
\multicolumn{2}{l}{RCS1107-0523}& $31_{-2}^{+2}$ & $0.51_{- 0.01}^{+ 0.01}$ & $62_{-4}^{+5}$ & $2.5_{-0.1}^{+0.1}$ & 216.1/191   \\ 
\multicolumn{2}{l}{RCS1326+2903}& $148_{-9}^{+11}$ & $1.04_{-0.06}^{+0.08}$ &  $3.4_{-0.2}^{+0.3}$ & $2.97_{-0.03}^{+0.03}$  & 54.9/64   \\ 
\multicolumn{2}{l}{RCS1419+5326}& $52_{-2}^{+2}$ & $0.60_{- 0.01}^{+ 0.01}$ & $189_{-7}^{+9}$ & $2.34_{-0.02}^{+0.02}$ & 170.6/155 \\
\multicolumn{2}{l}{RCS1620+2929}& $85_{-12}^{+14}$ & $0.60_{- 0.04}^{+ 0.05}$ & $13.3_{-1}^{+1}$ & $1.98_{-0.04}^{+0.04}$ & 147.7/148  \\ 
\multicolumn{2}{l}{RCS2318+0034}& $171_{-4}^{+7}$ & $0.86_{- 0.02}^{+ 0.04}$ & $29_{-1}^{+1}$ & $2.56_{-0.02}^{+0.02}$ & 304.8/300  \\
\multicolumn{2}{l}{RCS2319+0030}& $113_{-7}^{+8}$ & $0.54_{- 0.02}^{+ 0.03}$ & $19.7_{-0.7}^{+1.5}$ & $1.86_{-0.02}^{+0.03}$ & 209.2/155  \\
\multicolumn{2}{l}{RCS2319+0038}& $100_{-6}^{+7}$ & $0.65_{- 0.02}^{+ 0.03}$ & $46_{-2}^{+3}$ & $2.61_{-0.02}^{+0.02}$ & 153.6/155  \\
\multicolumn{2}{l}{RCS2320+0033}& $117_{-8}^{+8}$ & $0.61_{- 0.02}^{+ 0.03}$ & $18.9_{-1.0}^{+1.0}$ & $1.74_{-0.03}^{+0.03}$ & 162.8/147  \\
\enddata
\tablenotetext{a}{Surface brightness $I$ in units of $10^{-9}$ photons sec${}^{-
1}$ cm${}^{-2}$ arcsec${}^{-2}$}
\end{deluxetable}

\clearpage

\begin{deluxetable}{ccccccc}
\tablecolumns{7}
\tablewidth{0pt}
\tablecaption{Integrated Spectral Fits ($\Delta = 2500$)\label{table4}}
\tablehead{
\multicolumn{2}{c}{Cluster} & \colhead {$\rm{R}_{2500}$} & \colhead{kT} & \colhead{Z}& \colhead{${\rm{N}}_{\rm{H}}$}  & \colhead{$\chi^2/\rm{DOF}$}  \\
\multicolumn{2}{c}{} & \colhead{[$\rm{h}_{70}^{-1}$ kpc]} & \colhead{[keV]} &\colhead{[Z$_\sun$]} & \colhead{[${10}^{20}~{\rm{cm}}^{-2}$]} & \colhead{} 
}
\startdata
\multicolumn{2}{l}{RCS0224-0002} & {${329}^{+52}_{-35}$} & {${5.0}^{+1.2}_{-0.8}$} &  0.3 & 2.91   &{48.7/65} \\
\multicolumn{2}{l}{} &     \nodata & {${5.1}^{+1.2}_{-0.9}$} &    {${0.2}^{+0.4}_{-0.2}$} &    &{48.7/64} \\
\multicolumn{2}{l}{RCS0439-2904} & {${123}^{+24}_{-17}$} & {${1.5}^{+0.3}_{-0.2}$} &  0.3& 2.63  &{8.0/10} \\
\multicolumn{2}{l}{RCS1107-0523} & {${296}^{+29}_{-22}$} & {${4.2}^{+0.8}_{-0.6}$}  & 0.3& 4.24   & {52.2/61}\\
\multicolumn{2}{l}{} &   \nodata & {${4.2}^{+0.6}_{-0.5}$}  &    {${0.7}^{+0.5}_{-0.3}$}&    & {50.1/60}\\
\multicolumn{2}{l}{RCS1326+2903}  & {${202}^{+65}_{-34}$} & {${1.5}^{+0.6}_{-0.3}$} & 0.3 & 1.16  & {9.3/11} \\
\multicolumn{2}{c}{($z=1.01$)}& {${128}^{+79}_{-37}$} & {${1.6}^{+0.7}_{-0.3}$} & 0.3 &   & {8.9/11} \\
\multicolumn{2}{l}{RCS1419+5326} &  {${356}^{+17}_{-13}$} & {${4.5}^{+0.4}_{-0.3}$}  &  0.3& 1.18   & {126.3/125}\\
\multicolumn{2}{l}{} &   \nodata & {${4.6}^{+0.4}_{-0.3}$}  &   {${0.3}^{+0.1}_{-0.1}$} &    & {126.2/124}\\
\multicolumn{2}{l}{RCS1620+2929} & {${270}^{+53}_{-34}$} & {${3.9}^{+1.3}_{-0.9}$} & 0.3 & 2.72  &{14.1/20} \\
\multicolumn{2}{l}{} &    \nodata & {${3.9}^{+1.2}_{-0.8}$} &    {${0.3}^{+0.8}_{-0.3}$} &   &{14.1/19} \\
\multicolumn{2}{l}{RCS2318+0034} & {${410}^{+49}_{-37}$} & {${6.1}^{+1.3}_{-0.9}$}  & 0.3 & 4.13  &{48.9/68} \\
\multicolumn{2}{l}{} &   \nodata & {${5.8}^{+1.2}_{-0.8}$}  &  {${0.6}^{+0.3}_{-0.3}$} &   &{47.4/67} \\
\multicolumn{2}{l}{RCS2319+0030} & {${319}^{+55}_{-28}$} & {${6.5}^{+1}_{-1}$} & 0.3 & 4.13  &{44.8/34} \\
\multicolumn{2}{l}{} &    \nodata & {${6}^{+1}_{-1}$} &     {${0.6}^{+0.4}_{-0.4}$} &   &{44.3/33} \\
\multicolumn{2}{l}{RCS2319+0038} & {${351}^{+29}_{-25}$} & {${6.2}^{+0.9}_{-0.8}$} & 0.3 & 4.16  &{72.4/81} \\
\multicolumn{2}{l}{} &    \nodata & {${5.9}^{+0.8}_{-0.7}$} &     {${0.5}^{+0.2}_{-0.2}$} &   &{70.9/80} \\
\multicolumn{2}{l}{RCS2320+0033} & {${323}^{+53}_{-34}$} & {${5.9}^{+2}_{-1}$} & 0.3 & 4.14  &{31.2/32} \\
\multicolumn{2}{l}{} &    \nodata & {${6.0}^{+2}_{-1}$} &     {${0.3}^{+0.4}_{-0.3}$} &   &{31.1/31} \\
\enddata
\tablecomments{Single temperature fits within R$_{2500}$.  When possible, a second fit was performed allowing both the temperature and abundance to vary.  These fits are reported in the second line (where there is one) for each cluster.  In the case of RCS1326+2903, the second line indicates the result of fitting the integrated spectrum with a fixed abundance and a redshift of $z=1.01$.}
%\tablenotetext{a}{Bolometric X-ray luminosity within $R_{2500}$}
\end{deluxetable}

\clearpage

\begin{deluxetable}{ccccc}
\tablecolumns{5}
\tablewidth{0pc}
\tablecaption{Cluster Richness and Luminosity\label{table5}}
\tablehead{                
\multicolumn{2}{c}{Cluster}           &
\colhead{${\rm{B}_{\rm{gc,red}}}$}    &
\colhead{${\rm{L}}_{\rm{x}} ({\rm{R}}_{2500})$} &
\colhead{${\rm{L}}_{\rm{x}} ({\rm{R}}_{500})$}   \\
\multicolumn{2}{c}{}                &
  \colhead{[$h_{50}^{-1}~\rm{Mpc}^{1.77}$]}  &
 \colhead{[$10^{44}~{\rm{erg}}~{\rm{s}}^{-1}$]}  &
  \colhead{[$10^{44}~{\rm{erg}}~{\rm{s}}^{-1}$]}   
}
\startdata
\multicolumn{2}{l}      {RCS0224-0002} & $ 945\pm{210}$ & $ 2.1_{ 0.2}^{+ 0.3}$ & $ 4.4_{ 0.5}^{+ 0.5}$ \\
\multicolumn{2}{l}      {RCS0439-2904} & $1590\pm{460}$ & $ 1.5_{ 0.5}^{+ 0.5}$ & $ 4.0_{ 0.8}^{+ 0.7}$ \\
\multicolumn{2}{l}      {RCS1107-0523} & $ 899\pm{280}$ & $ 2.3_{ 0.2}^{+ 0.3}$ & $ 3.5_{ 0.4}^{+ 0.3}$ \\
\multicolumn{2}{l}      {RCS1326+2903} & $ 381\pm{275}$ & $ 0.4_{ 0.3}^{+ 0.3}$ & $ 1.1_{ 0.5}^{+ 0.5}$ \\
\multicolumn{2}{c}        {($z=1.01$)} & $2670\pm{671}$ & $ 1.1_{ 0.5}^{+ 0.7}$ & $ 2.7_{ 1.0}^{+ 1.0}$ \\
\multicolumn{2}{l}      {RCS1417+5305} & $1879\pm{464}$ & $ 1.3_{ 2.0}^{+ 0.9}$\tablenotemark{a} &  \nodata \\
\multicolumn{2}{l}      {RCS1419+5326} & $1173\pm{224}$ & $ 7.0_{ 0.3}^{+ 0.4}$ & $8.4_{ 0.5}^{+ 0.5}$ \\
\multicolumn{2}{l}      {RCS1620+2929} & $ 906\pm{236}$ & $ 2.3_{ 0.3}^{+ 0.7}$ & $ 3.3_{ 0.7}^{+ 0.5}$ \\
\multicolumn{2}{l}      {RCS2112-6326} & $1011\pm{400}$ & $ 1.2_{ 2.5}^{+ 0.8}$\tablenotemark{a} & \nodata \\
\multicolumn{2}{l}      {RCS2156-0448} & $ 481\pm{166}$ & $ 0.1_{ 0.3}^{+ 0.1}$\tablenotemark{a}\tablenotemark{b} & \nodata \\
\multicolumn{2}{l}      {RCS2318+0034} & $ 996\pm{217}$ & $ 6.0_{ 0.4}^{+ 0.7}$ & $ 8.3_{ 0.7}^{+ 0.9}$ \\
\multicolumn{2}{l}      {RCS2319+0030} & $1150\pm{281}$ & $ 3.6_{ 0.4}^{+ 0.6}$ & $ 7.9_{ 0.8}^{+ 0.7}$ \\
\multicolumn{2}{l}      {RCS2319+0038} & $1515\pm{323}$ & $ 7.6_{ 0.4}^{+ 0.6}$ & $16.2_{ 0.8}^{+ 0.6}$ \\
\multicolumn{2}{l}      {RCS2320+0033} & $ 578\pm{202}$ & $ 4.2_{ 0.3}^{+ 0.5}$ & $ 5.9_{ 0.6}^{+ 0.5}$ \\
\enddata
\tablenotetext{a}{{Bolometric X-ray Luminosity} within 500 $h_{70}^{-1}$ kpc, assuming a temperature of 4 keV}
\tablenotetext{b}{ID uncertain~\citep{barrientosprep}}
%obtained with XSPEC and PIMMS.}  
%\tablenotetext{a}{Corrected exposure time (see text for corrections applied).} 
\end{deluxetable}

\clearpage

\begin{deluxetable}{cccc}
\tablecolumns{4}
\tablewidth{0pt}
\tablecaption{Dynamical Comparisons\label{table6}}
\tablehead{
\multicolumn{2}{c}{Cluster} & 
\colhead{$\sigma$} & 
\colhead{$10^{2.49}~{\rm{T}}_X^{0.65}$\tablenotemark{a}}  \\
\multicolumn{2}{c}{} & 
\colhead{[km ${\rm{s}}^{-1}$]} & 
\colhead{[km ${\rm{s}}^{-1}$]}  
}
\startdata
\multicolumn{2}{l}{RCS1107-0523}&    $700\pm{300}$ & $785^{+95}_{-74}$   \\
\multicolumn{2}{l}{RCS1620+2929}&    $1050\pm{340}$ & $748^{+154}_{-117}$    \\
\multicolumn{2}{l}{RCS2319+0038}&    $860\pm{190}$ & $1012^{+93}_{-97}$   \\
\enddata
\tablenotetext{a}{\citet{xue00}}
\end{deluxetable}

\clearpage

\begin{deluxetable}{cccccc}
\tablecolumns{6}
\tablewidth{0pc}
\tablecaption{Mass Estimates ($\Delta=2500$)\label{table7}}
\tablehead{                
\multicolumn{2}{c}{Cluster} & \colhead{${n}_0$} & \colhead{${\rm{M}}_{\rm{gas}}$} & \colhead{${\rm{M}}_{2500}$} & \colhead{$f_{gas}$}  \\
\multicolumn{2}{c}{} & \colhead{[${10^{-2}~\rm{cm}^{-3}}$]} & \colhead{[$10^{13} ~{\rm{M}_\odot}$]} & \colhead{[$10^{13} ~{\rm{M}_\odot}$]} & \colhead{} 
}
\startdata
\multicolumn{2}{l}      {RCS0224-0002} & $ 0.329_{-0.010}^{+ 0.009}$ & $  0.42_{ -0.05}^{+  0.05}$ & $ 12.15_{ -1.62}^{+  1.59}$ & $ 0.035_{-0.005}^{+ 0.005}$ \\
\multicolumn{2}{l}      {RCS0439-2904} & $ 0.545_{-0.016}^{+ 0.016}$ & $  0.07_{ -0.01}^{+  0.01}$ & $  0.86_{ -0.10}^{+  0.10}$ & $ 0.078_{-0.010}^{+ 0.011}$ \\
\multicolumn{2}{l}      {RCS1107-0523} & $ 1.972_{-0.061}^{+ 0.058}$ & $  0.33_{ -0.04}^{+  0.04}$ & $  7.88_{ -0.88}^{+  0.88}$ & $ 0.042_{-0.005}^{+ 0.006}$ \\
\multicolumn{2}{l}      {RCS1326+2903} & $ 0.323_{-0.010}^{+ 0.010}$ & $  0.10_{ -0.01}^{+  0.01}$ & $  2.97_{ -0.56}^{+  0.54}$ & $ 0.034_{-0.006}^{+ 0.008}$ \\
\multicolumn{2}{c}        {($z=1.01$)} & $ 0.432_{-0.014}^{+ 0.013}$ & $  0.06_{ -0.01}^{+  0.01}$ & $  1.28_{ -0.24}^{+  0.24}$ & $ 0.048_{-0.008}^{+ 0.011}$ \\
\multicolumn{2}{l}      {RCS1419+5326} & $ 2.427_{-0.048}^{+ 0.047}$ & $  0.75_{ -0.06}^{+  0.06}$ & $ 11.39_{ -0.65}^{+  0.64}$ & $ 0.065_{-0.005}^{+ 0.005}$ \\
\multicolumn{2}{l}      {RCS1620+2929} & $ 0.675_{-0.035}^{+ 0.038}$ & $  0.29_{ -0.07}^{+  0.08}$ & $  7.60_{ -1.45}^{+  1.51}$ & $ 0.039_{-0.008}^{+ 0.010}$ \\
\multicolumn{2}{l}      {RCS2318+0034} & $ 0.713_{-0.012}^{+ 0.011}$ & $  0.97_{ -0.06}^{+  0.06}$ & $ 24.28_{ -2.92}^{+  2.82}$ & $ 0.040_{-0.005}^{+ 0.005}$ \\
\multicolumn{2}{l}      {RCS2319+0030} & $ 0.698_{-0.019}^{+ 0.019}$ & $  0.66_{ -0.07}^{+  0.07}$ & $ 12.55_{ -1.35}^{+  1.34}$ & $ 0.052_{-0.006}^{+ 0.007}$ \\
\multicolumn{2}{l}      {RCS2319+0038} & $ 1.205_{-0.033}^{+ 0.033}$ & $  0.87_{ -0.10}^{+  0.10}$ & $ 16.54_{ -1.60}^{+  1.59}$ & $ 0.052_{-0.006}^{+ 0.006}$ \\
\multicolumn{2}{l}      {RCS2320+0033} & $ 0.699_{-0.019}^{+ 0.019}$ & $  0.59_{ -0.07}^{+  0.07}$ & $ 13.57_{ -2.10}^{+  2.06}$ & $ 0.044_{-0.007}^{+ 0.008}$ \\
\enddata
\end{deluxetable}
\clearpage

\begin{deluxetable}{ccccccc}
\tablecolumns{7}
\tablewidth{0pc}
\tablecaption{Mass Estimates ($\Delta=500$)\label{table8}}
\tablehead{                
\multicolumn{2}{c}{Cluster} & \colhead{$\rm{R}_{500}$} & \colhead{${\rm{M}}_{\rm{gas}}$} &\colhead{${\rm{M}}_{500}$} & \colhead{$f_{gas}$}  \\
\multicolumn{2}{c}{} & \colhead{[kpc]}  & \colhead{[$10^{14} ~{\rm{M}_\odot}$]} & \colhead{[$10^{14} ~{\rm{M}_\odot}$]} & \colhead{}
}
\startdata
\multicolumn{2}{l}      {RCS0224-0002} & $  819_{-  69}^{+ 103}$ & $ 0.153_{-0.019}^{+ 0.020}$ & $ 4.975_{-0.692}^{+ 0.680}$ & $ 0.031_{-0.005}^{+ 0.006}$ \\
\multicolumn{2}{l}      {RCS0439-2904} & $  350_{-  28}^{+  41}$ & $ 0.047_{-0.006}^{+ 0.006}$ & $ 0.364_{-0.042}^{+ 0.041}$ & $ 0.129_{-0.018}^{+ 0.020}$ \\
\multicolumn{2}{l}      {RCS1107-0523} & $  665_{-  48}^{+  64}$ & $ 0.113_{-0.013}^{+ 0.013}$ & $ 1.782_{-0.200}^{+ 0.198}$ & $ 0.063_{-0.008}^{+ 0.009}$ \\
\multicolumn{2}{l}      {RCS1326+2903} & $  544_{-  61}^{+ 119}$ & $ 0.032_{-0.004}^{+ 0.004}$ & $ 1.083_{-0.203}^{+ 0.198}$ & $ 0.030_{-0.006}^{+ 0.007}$ \\
\multicolumn{2}{c}        {($z=1.01$)} & $  440_{-  50}^{+ 112}$ & $ 0.042_{-0.005}^{+ 0.005}$ & $ 0.912_{-0.174}^{+ 0.170}$ & $ 0.046_{-0.009}^{+ 0.011}$ \\
\multicolumn{2}{l}      {RCS1419+5326} & $  802_{-  28}^{+  37}$ & $ 0.218_{-0.019}^{+ 0.018}$ & $ 2.600_{-0.148}^{+ 0.147}$ & $ 0.084_{-0.006}^{+ 0.007}$ \\
\multicolumn{2}{l}      {RCS1620+2929} & $  627_{-  74}^{+ 115}$ & $ 0.091_{-0.013}^{+ 0.014}$ & $ 2.342_{-0.420}^{+ 0.409}$ & $ 0.039_{-0.008}^{+ 0.009}$ \\
\multicolumn{2}{l}      {RCS2318+0034} & $  979_{-  77}^{+ 103}$ & $ 0.199_{-0.026}^{+ 0.028}$ & $12.936_{-2.060}^{+ 1.988}$ & $ 0.015_{-0.003}^{+ 0.004}$ \\
\multicolumn{2}{l}      {RCS2319+0030} & $  749_{-  59}^{+ 118}$ & $ 0.264_{-0.031}^{+ 0.031}$ & $ 3.181_{-0.343}^{+ 0.339}$ & $ 0.083_{-0.011}^{+ 0.012}$ \\
\multicolumn{2}{l}      {RCS2319+0038} & $  809_{-  55}^{+  62}$ & $ 0.264_{-0.032}^{+ 0.032}$ & $ 4.007_{-0.388}^{+ 0.384}$ & $ 0.066_{-0.008}^{+ 0.009}$ \\
\multicolumn{2}{l}     {RCS2320+0033} & $  760_{-  73}^{+ 112}$ & $ 0.184_{-0.024}^{+ 0.025}$ & $ 2.823_{-0.602}^{+ 0.585}$ & $ 0.065_{-0.014}^{+ 0.018}$ \\

\multicolumn{2}{l}      {RCS0439-2904} & $  569_{-  43}^{+  63}$ & $ 0.100_{-0.013}^{+ 0.013}$ & $ 0.619_{-0.071}^{+ 0.070}$ & $ 0.161_{-0.024}^{+ 0.028}$ \\
\enddata
\end{deluxetable}

\clearpage
\begin{deluxetable}{cccccccc}
\tablecolumns{8}
\tablewidth{0pc}
\tablecaption{Fitting Parameters\label{table9}}
\tablehead{                
\colhead{}          &
\colhead{}    &
\multicolumn{3}{c}{$\Delta=2500$} & 
\multicolumn{3}{c}{$\Delta=500$} \\
\colhead{Fit} &
\colhead{Sample} &
 \colhead{$C_1$} & 
 \colhead{$C_2$} &
 \colhead{$\sigma_{{\rm{log Y}}}$} &
\colhead{$C_1$} &
 \colhead{$C_2$} &
 \colhead{$\sigma_{{\rm{log Y}}}$}
}
\startdata
 {$E_z^{-1}L_X-T_X$}&   {RCS} & $ 0.36\pm{ 0.06}$ & $ 2.05\pm{ 0.34}$&  0.17 & $ 0.59\pm{ 0.05}$ & $ 1.79\pm{ 0.42}$&  0.15 \\
            {}&      {} & $ 0.45\pm{ 0.03}$&2.0 (fixed)&  0.20 & $ 0.65\pm{ 0.03}$    &2.0 (fixed)&  0.19 \\
            {}&  {CNOC} & $ 0.74\pm{ 0.08}$ & $ 2.31\pm{ 0.31}$&  0.18&\nodata&\nodata&\nodata \\
            {}&      {} & $ 0.85\pm{ 0.01}$    &2.0 (fixed)&  0.19&\nodata&\nodata&\nodata \\
            {}& {TOTAL} & $ 0.56\pm{ 0.07}$ & $ 2.90\pm{ 0.35}$&  0.28&\nodata&\nodata&\nodata \\
            {}&      {} & $ 0.81\pm{ 0.01}$    &2.0 (fixed)&  0.33&\nodata&\nodata&\nodata \\
\hline
  {$E_z^{-1}L_X-E_z$M$_{\rm{tot}}$}&   {RCS} & $-0.03\pm{ 0.04}$ & $ 1.38\pm{ 0.12}$&  0.16 & $-0.20\pm{ 0.16}$ & $ 1.03\pm{ 0.28}$&  0.24 \\
            {}&      {} & $ 0.06\pm{ 0.02}$&1.33 (fixed)&  0.19 & $-0.28\pm{ 0.02}$   &1.33 (fixed)&  0.33 \\
            {}&  {CNOC} & $ 0.44\pm{ 0.12}$ & $ 1.26\pm{ 0.21}$&  0.20&\nodata&\nodata&\nodata \\
            {}&      {} & $ 0.48\pm{ 0.01}$   &1.33 (fixed)&  0.23&\nodata&\nodata&\nodata \\
            {}& {TOTAL} & $ 0.07\pm{ 0.10}$ & $ 1.77\pm{ 0.15}$&  0.29&\nodata&\nodata&\nodata \\
            {}&      {} & $ 0.40\pm{ 0.01}$   &1.33 (fixed)&  0.33&\nodata&\nodata&\nodata \\
\hline
 {$E_z$M$_{\rm{tot}}-T_X$}&   {RCS} & $ 0.29\pm{ 0.03}$ & $ 1.48\pm{ 0.27}$&  0.09 & $ 0.76\pm{ 0.08}$ & $ 1.72\pm{ 0.65}$&  0.22 \\
            {}&      {} & $ 0.28\pm{ 0.03}$&1.5 (fixed)&  0.10 & $ 0.69\pm{ 0.03}$    &1.5 (fixed)&  0.23 \\
            {}&  {CNOC} & $ 0.24\pm{ 0.02}$ & $ 1.83\pm{ 0.13}$&  0.07&\nodata&\nodata&\nodata \\
            {}&      {} & $ 0.27\pm{ 0.01}$    &1.5 (fixed)&  0.08&\nodata&\nodata&\nodata \\
            {}& {TOTAL} & $ 0.28\pm{ 0.02}$ & $ 1.63\pm{ 0.18}$&  0.09&\nodata&\nodata&\nodata \\
            {}&      {} & $ 0.27\pm{ 0.01}$    &1.5 (fixed)&  0.08&\nodata&\nodata&\nodata \\
\hline
 {$E_z^{-9/5}L_X-Y_X$}&   {RCS} & $ 0.32\pm{ 0.05}$ & $ 0.73\pm{ 0.05}$&  0.11 & $ 0.22\pm{ 0.04}$ & $ 0.65\pm{ 0.10}$&  0.11 \\
            {}&      {} & $ 0.40\pm{ 0.03}$&1.1 (fixed)&  0.22 & $ 0.08\pm{ 0.03}$    &1.1 (fixed)&  0.27 \\
            {}&  {CNOC} & $ 0.50\pm{ 0.08}$ & $ 0.80\pm{ 0.09}$&  0.12&\nodata&\nodata&\nodata \\
            {}&      {} & $ 0.29\pm{ 0.01}$    &1.1 (fixed)&  0.20&\nodata&\nodata&\nodata \\
            {}& {TOTAL} & $ 0.41\pm{ 0.03}$ & $ 0.88\pm{ 0.04}$&  0.14&\nodata&\nodata&\nodata \\
            {}&      {} & $ 0.30\pm{ 0.01}$    &1.1 (fixed)&  0.22&\nodata&\nodata&\nodata \\
\hline
 {$E_z^{2/5}$M$_{\rm{tot}}-Y_X$}&   {RCS} & $ 0.25\pm{ 0.03}$ & $ 0.52\pm{ 0.05}$&  0.06 & $ 0.43\pm{ 0.08}$ & $ 0.64\pm{ 0.22}$&  0.23 \\
            {}&      {} & $ 0.23\pm{ 0.03}$&0.581 (fixed)&  0.09 & $ 0.37\pm{ 0.03}$  &0.581 (fixed)&  0.25 \\
            {}&  {CNOC} & $ 0.05\pm{ 0.04}$ & $ 0.63\pm{ 0.05}$&  0.08&\nodata&\nodata&\nodata \\
            {}&      {} & $ 0.04\pm{ 0.02}$  &0.581 (fixed)&  0.09&\nodata&\nodata&\nodata \\
            {}& {TOTAL} & $ 0.18\pm{ 0.02}$ & $ 0.49\pm{ 0.03}$&  0.10&\nodata&\nodata&\nodata \\
            {}&      {} & $ 0.04\pm{ 0.02}$  & 0.581 (fixed)&  0.14 &\nodata&\nodata & \nodata \\
\enddata
\tablecomments{Best fits to scaling relations cosmologically corrected by the factor $E_z$.  Temperature is in units of 5 keV; luminosity in $10^{44}$ erg s$^{-1}$; mass in units of $10^{14} \msun$; $Y_X$ in $4 \times 10^{13} \msun$ keV.  Scatter along the Y-axis is calculated as $\left[\Sigma_{i=1,N}\left({\rm{log~Y}}_i-C_1-C_2~{\rm{log~X}}_i\right)^2/N\right]^{1/2}$.}
\end{deluxetable}

\clearpage
\begin{deluxetable}{cccc}
\tablecolumns{4}
\tablewidth{0pc}
\tablecaption{Fitting Comparisons\label{table10}}
\tablehead{                
\colhead{Sample\tablenotemark{a}} &
 \colhead{$C_1$} & 
 \colhead{$C_2$} &
 \colhead{Redshift}
}
\startdata
\multicolumn{4}{c} {\underline{$E_z^{-1}L_X-T_X$}}\\
{\underline{$\Delta=2500$}} \\   

{RCS} & $0.36\pm{0.06}$ & $2.05\pm{0.34}$  &  $0.6<z<1.0$ \\
       {CNOC} & $0.74\pm{0.08}$ & $2.31\pm{0.31}$& $0.1<z<0.6$ \\
         {TOTAL} & $0.56\pm{0.07}$ & $2.90\pm{0.35}$& $0.1<z<1.0$ \\
ASF01 & {$0.98^{+0.09}_{-0.10}$} & {$2.08\pm{0.06}$} & $0.1<z<0.45$\\
{\underline{$\Delta=500$}} \\
{RCS} &$0.59\pm{0.05}$ & $1.79\pm{0.42}$& $0.6<z<1.0$\\
ETB04 & {$0.50\pm{0.11}$} & {$3.72\pm{0.47}$} & $0.4<z<1.3$ \\
%BMS04 & {$0.75^{+0.2}_{-0.08}$} & {$2.5\pm{0.1}$} & theory \\
\hline
 \multicolumn{4}{c}{\underline{$Ez^{-1}L_X-E_z$M$_{\rm{tot}}$}}\\
{\underline{$\Delta=500$}} \\
{RCS} &$-0.20\pm{0.16}$ & $1.03\pm{0.28}$& $0.6<z<1.0$\\
ETB04 & {$-0.63\pm{0.32}$} & {$1.88\pm{0.42}$} &  $0.4<z<1.3$ \\
\hline
\multicolumn{4}{c}{\underline{$E_z$M$_{\rm{tot}}-T_X$}} \\   
{\underline{$\Delta=2500$}} \\
{RCS} & $0.29\pm{0.03}$ & $1.48\pm{0.27}$ & $0.6<z<1.0$\\
         {CNOC} & $0.24\pm{0.02}$ & $1.83\pm{0.13}$& $0.1<z<0.6$  \\
        {TOTAL} & $0.28\pm{0.02}$ & $1.63\pm{0.18}$&  $0.1<z<1.0$ \\
APP05 & {$0.23\pm{0.05}$} & {$1.70\pm{0.07}$} & $z\leq0.15$\\
%SPF03 & {$0.08\pm{0.4}$} & {$1.84\pm{0.14}$} & $0.0<z<0.21$\\
ASF01 & {$0.27\pm{0.34}$} & {$1.51\pm{0.27}$} & $0.1<z<0.45$\\
{\underline{$\Delta=500$}} \\
{RCS} & $0.76\pm{0.08}$ & $1.72\pm{0.65}$ & $0.6<z<1.0$\\
FRB01 & {$0.52\pm{0.45}$} & {$1.78\pm{0.10}$} & $z<0.09$\\
APP05 & {$0.58\pm{0.14}$} & {$1.71\pm{0.09}$} & $z\leq0.15$\\
KV05 & {$0.51\pm{0.31}$} & {$1.79\pm{0.19}$} & $0.4<z<0.7$\\
ETB04 & {$0.59\pm{0.05}$} & {$1.98\pm{0.3}$} & $0.4<z<1.3$\\
\hline
\multicolumn{4}{c}{\underline{$E_z^{-9/5}L_X$-$Y_X$}}\\
{\underline{$\Delta=500$}} \\
{RCS} &$0.22\pm{0.04}$ & $0.65\pm{0.10}$& $0.6<z<1.0$\\
M07 & {$-0.10\pm{0.04}$} & {$1.1\pm{0.04}$} & $0.1<z<1.3$\\
\hline
 \multicolumn{4}{c}{\underline{$E_z^{2/5}$M$_{\rm{tot}}-Y_X$}}\\
{\underline{$\Delta=500$}} \\
{RCS} &$0.43\pm{0.08}$ & $0.64\pm{0.22}$& $0.6<z<1.0$\\
APP07 & {$0.17\pm{0.2}$} & {$0.55\pm{0.03}$} & $z\leq0.15$\\
KVN06 & {$0.27\pm{0.006}$} & {$0.581\pm{0.009}$} & theory \\
\enddata
\tablenotetext{a}{Referenced samples: (ASF01)~\citet{allen01}; (APP05)~\citet{arnaud05}; (APP07)~\citet{arnaud07}; (BMS04)~\citet{borgani04}; (ETB04)~\citet{ettori04}; (FRB01)~\citet{finoguenov01}; (KV05)~\citet{kotov05}; (KVN06)~\citet{kravtsov06}; (M07)~\citet{maughan07}; (SPF03)~\citet{sanderson03}.}
\end{deluxetable}

\clearpage

\begin{deluxetable}{ccccc}
\tablecolumns{5}
\tablewidth{0pc}
\tablecaption{$B_{gc,red}$ Fitting Parameters\label{table11}}
\tablehead{                
\colhead{Fit} &
\colhead{Sample} &
\colhead{$C_1$} &
 \colhead{$C_2$} &
 \colhead{$\sigma_{{\rm{log Y}}}$}
}
\startdata

       {$L_X$}&   {RCS} & $-10.40\pm{ 2.61}$ & $  3.68\pm{ 0.89}$& 0.39 \\
            {}&      {} & $ -6.10\pm{  0.04}$ &2.22 (fixed)&   0.24 \\
            {}&  {CNOC} & $ -5.88\pm{ 1.66}$ & $  2.31\pm{ 0.55}$& 0.32 \\
            {}&      {} & $ -5.65\pm{  0.02}$ &2.22 (fixed)&   0.32 \\
            {}& {TOTAL} & $-10.10\pm{ 2.29}$ & $  3.68\pm{ 0.76}$& 0.49 \\
            {}&      {} & $ -5.65\pm{  0.02}$ &2.22 (fixed)&   0.40 \\
{} & {YE03} & $-4.48\pm{0.75}$ & $  1.84\pm{0.24 }$& \nodata \\
\hline
       {$T_X$}&   {RCS} & $ -5.54\pm{ 1.56}$ & $  1.86\pm{ 0.54}$& 0.21 \\
            {}&      {} & $ -3.35\pm{  0.04}$ &1.11 (fixed)&   0.14 \\
            {}&  {CNOC} & $ -2.90\pm{ 0.75}$ & $  1.00\pm{ 0.24}$& 0.14 \\
            {}&      {} & $ -3.26\pm{  0.02}$ &1.11 (fixed)&   0.16 \\
            {}& {TOTAL} & $ -4.08\pm{ 0.94}$ & $  1.38\pm{ 0.31}$& 0.18 \\
            {}&      {} & $ -3.28\pm{  0.02}$ &1.11 (fixed)&   0.16 \\
{} & {YE03} & $-2.29\pm{0.4 }$ & $0.78\pm{0.13 }$& \nodata \\
%\hline
%       {$Y_X$}&   {RCS} & $-15.20\pm{ 4.12}$ & $  5.06\pm{ 1.41}$& 0.57 \\
%            {}&  {CNOC} & $ -7.75\pm{ 2.34}$ & $  2.77\pm{ 0.76}$& 0.38 \\
%            {}& {TOTAL} & $-14.20\pm{ 3.24}$ & $  4.83\pm{ 1.07}$& 0.65 \\
\hline
        {M$_{2500}$}&   {RCS} & $ -8.20\pm{ 3.11}$ & $  2.86\pm{ 1.06}$& 0.36 \\
            {}&      {} & $ -4.93\pm{  0.04}$ &1.67 (fixed)&   0.22 \\
            {}&  {CNOC} & $ -4.81\pm{ 1.51}$ & $  1.75\pm{ 0.49}$& 0.26 \\
            {}&      {} & $ -4.66\pm{  0.02}$ &1.67 (fixed)&   0.27 \\
            {}& {TOTAL} & $ -6.31\pm{ 1.36}$ & $  2.24\pm{ 0.45}$& 0.30 \\
            {}&      {} & $ -4.74\pm{  0.02}$ &1.67 (fixed)&   0.28 \\
\hline
        {M$_{200}$}&  {RCS} & $ -9.53\pm{ 6.20}$ & $  3.54\pm{ 2.11}$& 0.51 \\
            {}&      {} & $ -4.27\pm{  0.04}$ &1.67 (fixed)&   0.30 \\
            {}&  {CNOC} & $ -4.61\pm{ 1.53}$ & $  1.88\pm{ 0.50}$& 0.26 \\
            {}&      {} & $ -4.06\pm{  0.02}$ &1.67 (fixed)&   0.26 \\
            {}& {TOTAL} & $ -5.86\pm{ 1.43}$ & $  2.30\pm{ 0.48}$& 0.32 \\
            {}&      {} & $ -4.12\pm{  0.02}$ &1.67 (fixed)&   0.28 \\
	    {} & {YE03} & $-4.55\pm{0.89 }$ & $  1.64\pm{0.28 }$& \nodata \\
{} & {B07} & $-5.70\pm{3.4 }$ & $  2.1\pm{1.2 }$& \nodata \\
\enddata
\tablecomments{Fits to richness scaling relationships.  Luminosity is given in units of $10^{44}$ erg s$^{-1}$, temperature in units of 5 keV, and mass in $10^{14}\msun$.  Parameters for the present work are measured within $\Delta=2500$ unless otherwise noted. Referenced samples are (ASF01) ~\citet{yee03}, and (B07)~\citet{blindert07}.  Scatter ($\sigma_{\rm{log Y}}$) is given as $\left[\Sigma_{i=1,N}\left({\rm{log~Y}}_i-C_1-C_2~{\rm{log X}}_i\right)^2/N\right]^{1/2}$.}
\end{deluxetable}

\clearpage

%% Use the figure environment and \plotone or \plottwo to include 
%% figures and captions in your electronic submission.

%% Tables may also be prepared as separate files. See the accompanying
%% sample file table.tex for an example of an external table file.
%% To include an external file in your main document, use the \input
%% command. Uncomment the line below to include table.tex in this
%% sample file. (Note that you will need to comment out the \documentclass,
%% \begin{document}, and \end{document} commands from table.tex if you want
%% to include it in this document.)

%% \input{table}

%% The following command ends your manuscript. LaTeX will ignore any text
%% that appears after it.

\end{document}